\documentclass[fleqn,usenatbib]{mnras}

\usepackage{newtxtext,newtxmath}
\usepackage[T1]{fontenc}

\DeclareRobustCommand{\VAN}[3]{#2}
\let\VANthebibliography\thebibliography
\def\thebibliography{\DeclareRobustCommand{\VAN}[3]{##3}\VANthebibliography}

\usepackage{graphicx}	% Including figure files
\usepackage{amsmath}	% Advanced maths commands
\usepackage{color}
\def\Msun{{\rm ~M}_{\odot}}

\def\Zsun{{\rm ~Z}_{\odot}}

\definecolor{darkgreen}{rgb}{0.0, 0.4, 0.22}
\definecolor{orange}{rgb}{0.9, 0.5, 0.}

\title[Low-Z cosmic SFH: impact of FMR \& starbursts]{
The impact of the FMR and starburst galaxies on the (low-metallicity) cosmic star formation history}

\author[M. Chruslinska et al.]{
Martyna Chru{\'s}li{\'n}ska,$^{1}$\thanks{E-mail: m.chruslinska@astro.ru.nl}
Gijs Nelemans,$^{1,2,3}$
Lumen Boco$^{4,5,6}$
and Andrea Lapi$^{4,5,6,7}$
\\
$^{1}$Department of Astrophysics/IMAPP, Radboud University, PO Box 9010, 6500 GL Nijmegen, The Netherlands\\
$^{2}$Institute for Astronomy, KU Leuven, Celestijnenlaan 200D, 3001 Leuven, Belgium\\
$^{3}$SRON, Netherlands Institute for Space Research, Sorbonnelaan 2, 3584 CA Utrecht, The Netherlands\\
$^{4}$SISSA, Via Bonomea 265, 34136 Trieste, Italy\\
$^{5}$IFPU-Institute for Fundamental Physics of the Universe, Via Beirut 2, 34014 Trieste, Italy\\
$^{6}$INFN-Sezione di Trieste, via Valerio 2, 34127 Trieste, Italy\\
$^{7}$IRA-INAF, Via Gobetti 101, 40129 Bologna, Italy\\
}

\date{Accepted XXX. Received YYY; in original form ZZZ}

\pubyear{2021}

\begin{document}
\label{firstpage}
\pagerange{\pageref{firstpage}--\pageref{lastpage}}
\maketitle

\begin{abstract}
The question how much star formation is occurring at low metallicity throughout the cosmic history appears crucial for the discussion of the origin of various energetic transients, and possibly - double black hole mergers. 
We revisit the observation-based distribution of birth metallicities of stars (f$_{\rm SFR}$(Z,z)), focusing on several factors that strongly affect its low metallicity part: (i) the method used to describe the metallicity distribution of galaxies (redshift-dependent mass metallicity relation - MZR, or redshift-invariant fundamental metallicity relation - FMR), (ii) the contribution of starburst galaxies and (iii) the slope of the MZR. 
We empirically construct the FMR based on the low-redshift scaling relations, which allows us to capture the systematic differences in the relation caused by the choice of metallicity and star formation rate (SFR) determination techniques and discuss the related f$_{\rm SFR}$(Z,z) uncertainty.\\
We indicate factors that dominate the f$_{\rm SFR}$(Z,z) uncertainty in different metallicity and redshift regimes.
The low metallicity part of the distribution is poorly constrained even at low redshifts (even a factor of $\sim$200 difference between the model variations) 
The non-evolving FMR implies a much shallower metallicity evolution than the extrapolated MZR, however, its effect on the low metallicity part of the f$_{\rm SFR}$(Z,z) is counterbalanced by the contribution of starbursts (assuming that they follow the FMR). 
A non-negligible fraction of starbursts in our model may be necessary to satisfy the recent high-redshift SFR density constraints.
\end{abstract}


%%%%%%%%%%%%%%%%%%%%%%%%%%%%%%%%%%%%%%%%%%%%%%%%%%

%%%%%%%%%%%%%%%%% BODY OF PAPER %%%%%%%%%%%%%%%%%%
\section{Introduction}
The observed populations of stars, their remnants and related transients consist of objects formed at different times and with different metallicities.
To model those populations and to correctly interpret observations, it is necessary to know the metallicity dependent star formation history (SFH) of the observed galaxy, galaxies in the probed volume, or even of the entire Universe.
Knowledge of the latter - which is the subject of interest of this study - becomes increasingly important in the era of gravitational wave (GW) astrophysics.
That is because the time between the formation of the progenitor stars and merger of stellar black holes or neutron stars observed in GW can be comparable to the age of the Universe \citep[e.g.][]{Belczynski16}. Moreover, the efficiency of formation of merging binaries may show a strong metallicity dependence. In particular, it has been suggested that double black hole mergers may form much more efficiently in low metallicity environments \citep[$\lesssim$0.1 solar metallicity; e.g.][]{Klencki18,Giacobbo18}.
This makes the modelled properties of the population of such systems particularly sensitive to the assumed  distribution  of  the  cosmic  star  formation  rate  density (SFRD)  at  different  metallicities  and  redshifts, f$_{\rm SFR}$(Z,z) 
\citep[e.g.][]{ChruslinskaNelemansBelczynski19,Neijssel19,Broekgaarden21} and requires knowledge of f$_{\rm SFR}$(Z,z) even beyond the peak of the cosmic SFH.
To confront the model predictions with observations and draw correct conclusions from such a comparison, it is necessary to take into account the f$_{\rm SFR}$(Z,z) uncertainty,
which may be substantial - especially at high redshifts \citep{ChruslinskaNelemans19,Chruslinska20,Boco21}.
Given the many uncertain pieces of information that need to be combined to estimate f$_{\rm SFR}$(Z,z), an (observation-based) determination of this distribution presents a challenge in itself.
In this study we take a closer look at two of those pieces:
the empirical correlation between the star formation rate
and metallicity of star forming galaxies (the so-called fundamental metallicity relation, e.g. \citealt{Ellison08,Mannucci10}) and the contribution of starburst galaxies.
We build on the observation-based f$_{\rm SFR}$(Z,z) model from \citet{ChruslinskaNelemans19} (briefly introduced in Sec. \ref{sec: framework intro}) 
and expand the discussion presented in \citet{Boco21}, aiming to evaluate the uncertainty (or - find realistic, observationally allowed extremes) of the metallicity dependent cosmic SFH in view of those factors.
Where  appropriate  we  adopt  a  standard  flat  cosmology  with $\Omega_{M}$=0.3, $\Omega_{\Lambda}$=0.7 and H$_{0}$=70 km s$^{-1}$ Mpc$^{-1}$, assume a (universal) \citet{Kroupa01} initial mass function (IMF) and solar metallicity of 12 + log$_{10}$(O/H)$_{\odot}$=$Z_{O/H\odot}$=8.83 and $\Zsun=0.017$ \citep{GrevesseSauval98}.

\section{An observation based $\lowercase{f}_{SFR}(Z,\lowercase{z})$ determination}\label{sec: framework intro}
We construct f$_{SFR}$(Z,z) building on the framework detailed in \citet{ChruslinskaNelemans19}.
In essence, their method is based on a combination of three key ingredients: the distribution of galaxy stellar masses (galaxy stellar mass function - GSMF), and the distributions describing the star formation rates (SFR) and metallicities of galaxies at fixed stellar mass $M_{*}$. All distributions are redshift-dependent.
The GSMF is used to obtain the number density of galaxies of different masses. The contribution of galaxies of different masses to the total SFRD can then be obtained by weighing the number density of galaxies in a given mass range by their SFR. Finally, f$_{SFR}$(Z,z) is obtained by assigning a metallicity to each $M_{*}$.
\\
The SFR and metallicity distributions are modelled as log-normal distributions centered on the empirical star formation-mass relation (SFMR) and the mass - (gas-phase) metallicity relation (MZR) respectively, with dispersions $\sigma_{\rm SFMR}$ and $\sigma_{\rm MZR}$ (representing the intrinsic scatter around the two relations).
Note that the observational gas-phase metallicity estimates probe the oxygen abundance ($Z_{O/H}$=12+log$_{10}$(O/H)\footnote{In contrast to the total metal mass fraction $Z$ or iron abundance commonly used in the simulations/theoretical studies. $Z_{O/H}$ is often used as a proxy for the total $Z$, but note that this requires assuming a particular (typically solar-like) abundance pattern.}), and that is the metallicity measure used in our study.
Additional scatter in $Z_{O/H}$ is introduced to model
the spread in the metallicity at which the stars are forming within the galaxies.
To determine all the necessary ingredients, the authors assemble a compilation of observational results describing the MZR, SFMR and GSMF, combined over a wide range of redshifts ($z$) and $M_{*}$.
Several variations of the base relations (SFMR, MZR, GSMF)
are explored in order to discuss the impact of the uncertain
absolute metallicity scale (steming from the differences between the estimates obtained with different metallicity determination methods), the shape of the high mass end of the SFMR and the redshift evolution of the low mass end of the GSMF (see Sec. 2 in \citealt{ChruslinskaNelemans19} and references therein)
\\
In this study we explore an alternative method to obtain the redshift-dependent metallicity distribution of star forming galaxies, based on the fundamental metallicity relation.
We then modify the SFR distribution to better account for the contribution of star forming galaxies that are strong outliers to the general SFMR (starbursts).
Our motivation for those modifications and the details of our approach are laid out in Sec. \ref{sec: FMR} and Sec. \ref{sec: starbursts} respectively.

\section{The fundamental metallicity relation}\label{sec: FMR}

The fundamental metallicity relation (FMR) is a three parameter dependence linking M$_{*}$, SFR and gas phase metallicity of star forming galaxies \citep{Ellison08,Mannucci10}. 
The relation implies the mass-metallicity correlation, as observed in the MZR, 
but it also introduces an anti-correlation between metallicty and SFR, such that galaxies
of  the  same  stellar  mass  showing  higher  than  average 
SFR  also  have  lower  metallicities (i.e. the galaxy's offset from the average mass-metallicity relation -MZR-
is anti-correlated with its offset from the average SFR-mass relation). 
\\
The existence of such SFR-metallicity anti-correlation has been reported in numerous observational studies
\citep[e.g.][]{Mannucci11,Yates12,AndrewsMartini13,Bothwell13,Lara-Lopez13,Salim14,Zahid14b,Yabe15,Hunt16_obs,Sanders18,Cresci19,Curti20,Sanders20} and established up to redshift $z\sim3.5$.\\
Such correlation is also expected from theoretical models of galaxy evolution and is thought to reflect the changes in the SFR-fuelling gas fraction present within the galaxy \citep[e.g.][]{Ellison08,Dave11,Yates12,DeLucia20}.
Lower SFR for a given M$_{*}$ then implies that the galaxy has already used up most of its cold gas reservoir
(fuelling its SFR in the past), and therefore (in the absence of strong outflows)
its interstellar medium is more metal rich than that of the galaxy of the same mass that is still highly star forming.
At low SFR there are also fewer supernovae that can remove (especially at low M$_{*}$) metal rich gas from the galaxy.
In turn, inflowing metal poor material lowers  the  metallicity  and 
provides additional fuel for the SFR, increasing the latter.
As long as the time-scales on which the SFR and metallicity evolve are of the same order, 
the anti-correlation between the galaxy offsets from the MZR and SFMR is expected to hold \citep[e.g.][]{Torrey18}.
Those timescales are similar for the feedback/SFH implementations used in the large-scale cosmological simulations such as EAGLE and Illustris-TNG, which warrants the existence of the FMR up to high redshifts in the simulations \citep{Lagos16,DeRossi17,Torrey18}.
However, \citet{Torrey18} point out that the strength of the correlation - especially for low mass galaxies at high redshift - may be reduced if models with particularly strong and/or bursty feedback are used.
\citet{Lagos16} find that the shape of the FMR in the EAGLE simulations depends on the adopted model of star formation.
Therefore, observational confirmation of the existence or breakdown of the FMR at high redshifts will help to discriminate between the different feedback and star formation prescriptions.
\\
Observationally, the FMR is found to show little to no evolution with redshift up to z$\sim$3 within the uncertainty of the current data and range of galaxy properties probed (see \citealt{Cresci19} and \citealt{Sanders20} for recent discussion).
Since both the mass and SFR distributions of star forming galaxies change over time, the apparent lack of FMR evolution implies that galaxies at different redshifts probe different parts of the locally established relation.
At fixed  M$_{*}$, a decrease in the average galaxy metallicity as a function of redshift is still 
expected, as the typical SFR is higher at earlier cosmic times.\\
\newline
 At $z\gtrsim2$, the rate of decrease in metallicity implied by a non-evolving FMR is much weaker than that reported
by various studies discussing the redshift evolution of the MZR \citep[e.g.][]{Maiolino08,Mannucci09}.
Similarly to the FMR, the MZR (and its evolution with redshift) is  virtually unconstrained at redshifts $z\gtrsim 3.5$ \citep{MaiolinoMannucci19}. 
This uncertainty in the rate of metallicity evolution is one of the key factors affecting the f$_{\rm SFR}$(Z,z).
In particular, different high redshift extrapolations of empirical relations used in the literature lead to drastically different conclusions about the birth metallicities of stars forming beyond the peak of the cosmic SFH \citep[e.g.][]{ChruslinskaNelemansBelczynski19,ChruslinskaNelemans19,Boco21}.
\\
Recent observational studies find support for rapid early metal enrichment in high redshift galaxies, pointing towards a weak MZR evolution that is more in line with
the apparently invariant FMR
(see Sec. 3 in \citealt{Boco21} and references therein for a recent discussion).
Such weaker metallicity evolution is also in agreement with the results of cosmological simulations and semi analytical models of galaxy evolution \citep[e.g.][]{Yates12,Torrey19}.
Furthermore, \citet{Sanders20} show that 
when the differences in the properties of the ionized gas within HII regions (responsible for the emission lines used to estimate the metallicity; in particular lower iron to oxygen ratio for the same O/H found in high redshift galaxies) of galaxies at different redshifts are taken into account, the inferred MZR evolution is milder (i.e. metallicities of high redshift galaxies are biased low if those differences are neglected). In fact, the MZR evolution found by \citet{Sanders20} is consistent with the non- or weakly evolving FMR within the redshift and mass range probed in their study.
The existence of a FMR also means that 
if the galaxy sample is biased towards high SFRs, the inferred average metallicity is underestimated. Such biases can be expected in high redshift and low stellar mass galaxy samples, affecting the MZR shape (the low mass end slope) and the rate of evolution with redshift (a decrease in the normalisation).
\\
\newline
In light of the above discussion, we conclude that the assumptions about the metallicity evolution made in \citet{ChruslinskaNelemans19}, which rely on the MZR obtained by \citet{Mannucci09} and extrapolate its redshift evolution based on their two highest redshift bins, are likely to overestimate the rate of metallicity decrease at $z\gtrsim2$ and require revision: either assuming a weaker MZR evolution, or using the FMR to assign metallicity to galaxies.\\
The use of the FMR instead of a MZR in f$_{\rm SFR}$(Z,z) determination appears advantageous for two reasons:
(i) the redshift invariance of the FMR allows to circumvent the problem of the uncertain MZR evolution with redshift - the metallicity distribution at each $z$ is then defined by the local FMR and the SFMR and GSMF (whose $z$ dependence is generally better constrained than that of the MZR).
We stress that with the current data there is no guarantee that at z$\gtrsim$3 the FRM remains (close to) redshift invariant. By assuming a non-evolving FMR throughout the entire cosmic history we explore another extreme assumption (with respect to \citet{ChruslinskaNelemans19}) about the rate of metallicity evolution at high redshift.
(ii) if all star forming galaxies follow the FMR (at present, there is no clear evidence to the contrary),
it can be used to describe metallicity of galaxies belonging to populations for which there are little examples of metallicity determinations (e.g. strong outliers of the star forming main sequence that are not described by the MZR, such as starburst galaxies, see Sec. \ref{sec: starbursts}).
\\
\newline
However, the major problem with this approach is that 
there is no agreement on the exact form of this three parameter
mass-SFR-metallicity dependence.
The f$_{\rm SFR}$(Z,z) derived with the FMR will necessarily
strongly depend on the choice of this relation, just as it strongly depends on the choice of MZR and its extrapolated evolution.
While the discussion of the systematic effects on the f$_{\rm SFR}$(Z,z) introduced by the choice of a particular form of the MZR is relatively straightforward \citep{ChruslinskaNelemans19} and to a large extent boils down to the discussion of differences caused by the use of different metallicity determination methods (which are well documented in the literature, e.g. \citealt{KewleyEllison08}, \citealt{MaiolinoMannucci19}), an analogous discussion for the FMR is not.
The observationally inferred FMR is known to non-trivially depend on the metallicity determination method, the SFR determination method and the galaxy sample selection criteria \citep[e.g.][]{Yates12,Hunt16_obs,Kashino16,Telford16,Cresci19}.
We note that it is not clear how the FMR depends on each of those factors.
Therefore, one needs to be cautious when using different literature results concerning the properties of galaxy populations (e.g. the SFMR) in combination with a particular FMR estimate - if those results are based on different techniques, this would represent an internally inconsistent approach.
\\
Instead of using a particular example from the literature, in our analysis we resort to a phenomenological description of the FMR.
We aim to capture the robust observational/theoretical features of the mass-SFR-metallicity dependence and describe the variations in the shape of this dependence caused by different choices of the local MZR and SFMR (that reflect the differences in the metallicity and SFR determination methods used in the literature).
This allows us to ensure the consistency of the method and discuss the realistic extremes of the f$_{\rm SFR}$(Z,z).

\subsection{Constructing the FMR from $z\sim0$ scaling relations}\label{Sec: FMR - model}

\begin{figure*}
\vspace*{-0.5cm}
\includegraphics[width=1.8\columnwidth]{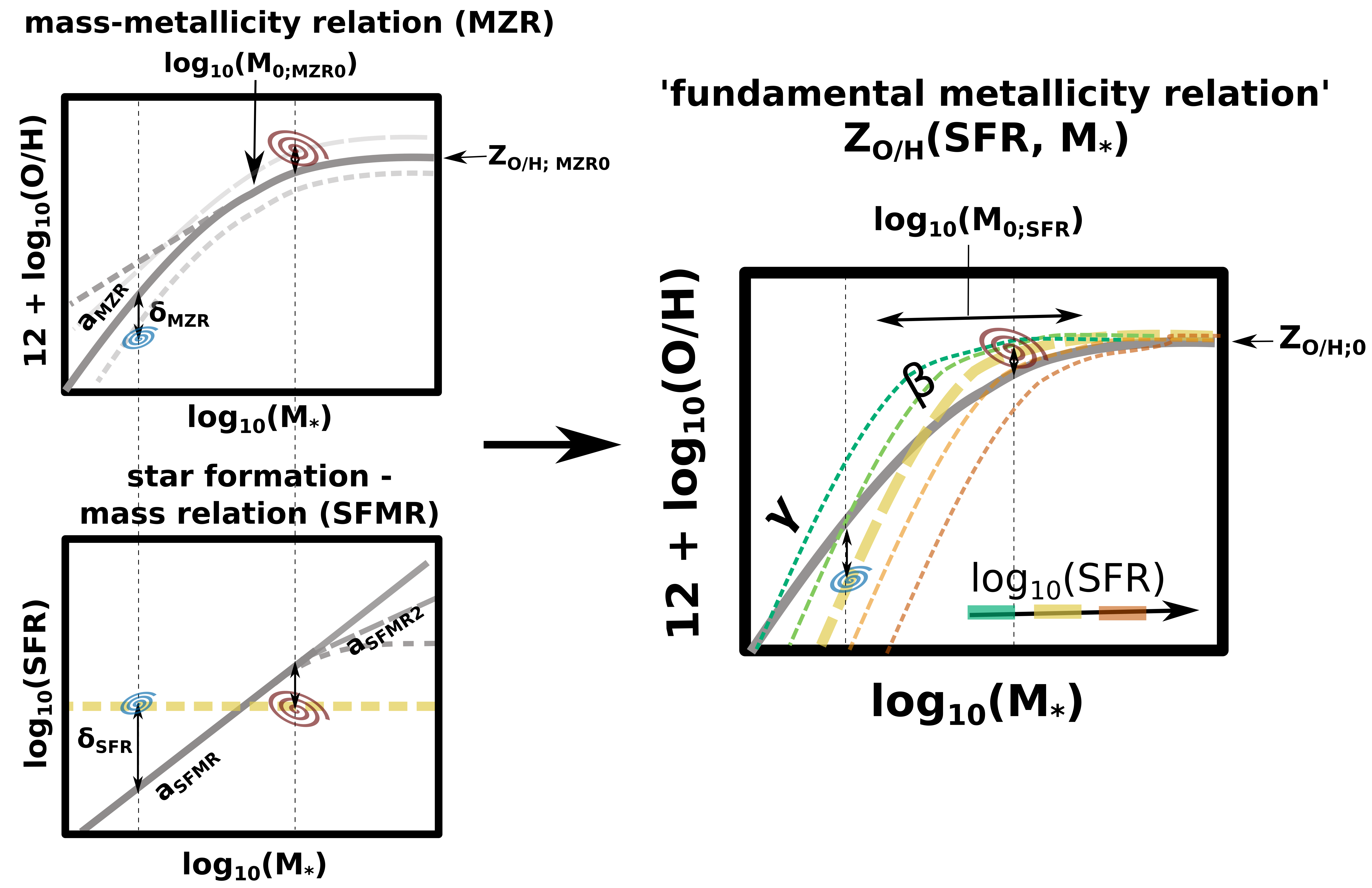}
\vspace*{-0.1cm}
\caption{
Sketch illustrating the local MZR (top left), SFMR (bottom left) and the three parameter mass-metallicity-SFR dependence Z$_{O/H}$(SFR, M$_{*}$) (right), indicating the key parameters of those relations and the observed anti-correlation between the galaxy's offsets from the MZR ($\delta_{MZR}$) and SFMR ($\delta_{SFMR}$).
We consider different assumptions about the MZR and SFMR relations (i.e. normalization $\rm Z_{O/H; MZR0}$ and slope $a_{\rm MZR}$ of the MZR, high mass end of the SFMR $a_{SFMR2}$) to cover the range of possibilities present in the literature, as indicated with the thick gray lines in the left panels.
In Sec. \ref{Sec: FMR - model} we relate the parameters of the Z$_{O/H}$(SFR, M$_{*}$) to the parameters of the local 2D relations.
}
\label{fig: FMR construction sketch}
\end{figure*}

 To construct our model FMR, we describe its 2D projection on the  Z$_{O/H}$-log$_{10}$(M$_{*}$) plane.
 This projection is the most commonly shown in observational studies.
 We refer to it as $Z_{O/H}(SFR, M_{*})$ (see Fig. \ref{fig: FMR construction sketch} for the illustration).
 \\
 At fixed SFR, the shape of Z$_{O/H}$(SFR,M$_{*}$) is found to show the same characteristic features as the MZR: the low mass part of the relation is almost linear (Z$_{O/H}$ $\propto$ log$_{10}$(M$_{*}$), it bends around a certain turnover mass and flattens at high masses, approaching a constant metallicity value 
 (see e.g. \citealt{Mannucci10,Mannucci11,Yates12,AndrewsMartini13,Hunt16_obs,Telford16,Cresci19,Curti20,Sanders20}).
 To describe this dependence, we  use the parametrisation from \citet{Curti20}:
 \\
 \begin{equation}\label{eq: Curti parametrisation}
  \rm Z_{O/H}(SFR, M_{*}) = Z_{O/H;0} - \frac{\gamma}{\beta} log\left( 1 + (\frac{M_{*}}{M_{0;SFR}})^{-\beta} \right)
 \end{equation}
 where $Z_{O/H;0}$ is the asymptotic metallicity of the high-mass end of the relation, $M_{0;SFR}$ is the turnover mass, $\gamma$ is the slope of $Z_{O/H}(SFR, M_{*})$ at $M_{*}<<M_{0;\rm SFR}$ and $\beta$ regulates the width of the knee of the relation. 
 The $Z_{O/H}(SFR, M_{*})$ dependence on SFR is expected primarily in the turnover mass $M_{0;\rm SFR}$, as illustrated in the right panel of Figure \ref{fig: FMR construction sketch}.
The left panel of Figure \ref{fig: FMR construction sketch} illustrates the average $z\sim0$ MZR and SFMR.
We relate the $Z_{O/H}(SFR, M_{*})$ parameters to those of the average
 $z\sim0$ scaling relations - each of the parameters of eq. \ref{eq: Curti parametrisation} is discussed in the relevant subsection below. 
Discussion of the observational properties of the $Z_{O/H}(SFR, M_{*})$ in the reminder of this section
and our choices are motivated by the results shown in the references given at the beginning of this section - we do not list them again in each of the subsections, unless we refer to a result from a specific paper(s).
\newline
\\
Equation \ref{eq: Curti parametrisation} describes the shape of the relation,
but the main observable property that characterizes the FMR is 
the quantity that relates the galaxy's offset from the average MZR and SFMR
(i.e. quantifies the strength of the SFR-metallicity correlation).
To describe the strength of the SFR-metallicity correlation
we introduce the coefficient $\nabla_{\rm FMR}$\footnote{
This is similar to e.g. \citet{Salim14}, \citet{Sanders18} and \citet{Sanders20}.
Note that many observational studies use the term 'strength of the SFR-metallicity correlation' when referring to the spread between the Z$_{O/H}$(M$_{*}$,SFR) curves at different fixed SFR (which is effectively done by comparing the fitted values of the '$\alpha$ parameter' introduced by \citealt{Mannucci10}). See also footnote \ref{ftn: alpha vs nabla FMR}}, defined as follows:
\begin{equation}\label{eq: FMR offsets}
\delta_{MZR}=-\nabla_{FMR} \cdot \delta_{SFMR}
\end{equation}
where $\delta_{MZR}$ = Z$_{O/H, gal.}$ - Z$_{O/H, at \ MZR}$
and $\delta_{SFMR}$ = log$_{10}$(SFR$_{gal.}$) - log$_{10}$(SFR$_{at \ SFMR}$)
are the galaxy's offsets from the average MZR and SFMR .
Higher values of $\nabla_{\rm FMR}$ imply a stronger correlation.

\subsubsection{The asymptotic metallicity}

A clear feature present among the observational $Z_{O/H}(SFR, M_{*})$ determinations is the high mass flattening. 
We note that some authors report a reversal of the (anti-)correlation at high M$_{*}$ rather than a simple flattening, e.g. \citealt{Yates12,Kashino16}).
\citet{Telford16} point out that certain sample selection criteria (e.g. applying signal to noise ratio cuts on oxygen lines used to estimate the metallicity) may lead to biases against the massive, low SFR galaxies and therefore affect the inferred low SFR and high mass part of the FMR. They argue for such biases as the cause of the reversal of the correlation at high masses as seen in \citet{Kashino16}. The effects of dust (and the applied dust SFR corrections) may also induce biases against massive, metal rich galaxies \citep{Telford16}.
\\
Change in the relation at high masses is also supported theoretically, and could be attributed to the increased importance of AGN feedback that can rapidly influence the SFR while the metallicity continues to evolve much more gradually \citep[e.g.][]{DeRossi17,Torrey18}.
We therefore assume that the high mass flattening is a robust feature of the $Z_{O/H}(SFR, M_{*})$ and include it within our model.
Observationally, the value of the asymptotic metallicity $Z_{O/H;0}$ at which the relation flattens appears to be roughly independent of the SFR and coincide with that of asymptotic metallicity of the average z$\sim$0 MZR ($Z_{O/H;MZR0}$).
As such, it is affected by the choice of the metallicity determination method, which leads to systematic offsets in the normalisation of the MZR
(compare examples shown in \citealt{Telford16} and in Fig. 2 in \citealt{Cresci19}).\\
We fix $Z_{O/H;0}=Z_{O/H;MZR0}$, where $Z_{O/H;MZR0}$ is defined by the choice of the $z\sim0$ MZR.

\subsubsection{The slope at low masses and high SFR}

At the low mass end, the
Z$_{O/H}(SFR, M_{*})$ for fixed log$_{10}$(SFR) seems to be well approximated with a linear dependence.
Accordingly, in this regime ($M_{*}<< M_{0;SFR}$), eq. \ref{eq: Curti parametrisation} reduces to:
 \begin{equation}\label{eq: FMR linear regime}
  \rm Z_{O/H}(SFR, M_{*}) = Z_{O/H;0} + \gamma log_{10}\left( M_{*} \right) - \gamma log_{10}\left( M_{0;SFR} \right)
 \end{equation}
The slope of this dependence is generally steeper than that of the average $z\sim0$ MZR.
$\gamma$ seems to be affected by the choice of the metallicity determination technique \citep[see e.g. examples shown in][]{Yates12,Hunt16_obs,Telford16,Cresci19}, but it is difficult to read off any systematic trend based on the examples shown in the literature (especially keeping in mind the presence of other differences in the methods that also affect the relation).
Z$_{O/H}(SFR, M_{*})$ obtained for different log$_{10}$(SFR),
 show no clear SFR dependence and appear nearly parallel to each other.
 For simplicity, here we assume that $\gamma$ is independent of SFR.
However, we note that in their recent study, \citet{Curti20} find that $\gamma$ decreases with decreasing SFR to values similar to that of their average z$\sim$0 MZR slope (see Fig. 9 therein).
This can lead to significantly different metallicities of low M$_{*}$, low SFR galaxies when the FMR is extrapolated down to low masses ($\lesssim 10^{8}\Msun$).
In appendix \ref{app: SFR dependent nabla} we introduce a additional variation of our model that accounts for such a dependence and discuss its influence on our results.
\\
We can recover the characteristics summarised above by applying eq. \ref{eq: FMR offsets}
 in the low-mass regime and assuming that $\nabla_{\rm FMR}$=$\nabla_{\rm FMR0}$ is fixed in this regime. 
 The slope $\gamma$ then follows directly from the slopes of the MZR and SFMR and the correlation coefficient $\nabla_{\rm FMR0}$. 
 For Z$_{O/H} = \rm a_{MZR} log_{10}(M_{*}) + b_{MZR}$
 and $\rm log_{10}(SFR) = a_{SFMR} log_{10}(M_{*}) + b_{SFMR}$ (appropriate at  $M_{*}<< M_{0;\rm SFR}$), we thus obtain:
\begin{equation}\label{eq: FMR full correlation}
  \begin{aligned}
 \rm Z_{O/H}(SFR,M_{*}) & = -\nabla_{\rm FMR0} \ \rm log_{10}(SFR) +\\ &(\nabla_{\rm FMR0} \times a_{SFMR} + a_{MZR}) \ \rm log_{10}(M_{*}) +\\ &
 \nabla_{\rm FMR0} \times b_{SFMR} + b_{MZR}
 \end{aligned}
\end{equation}
 The slope of the $Z_{O/H}(SFR, M_{*})$ relation at fixed SFR is then related to the slopes of the low mass parts of the MZR ($a_{\rm MZR}$) and SFMR ($a_{SFMR}$) 
 and the parameter describing the strength of the SFR-metallicity correlation $\nabla_{\rm FMR0}$:
 \begin{equation} \label{eq: gamma}
 \gamma=\nabla_{\rm FMR0} \ a_{SFMR} + a_{MZR}
 \end{equation} 
 The values $a_{\rm MZR}$ and $a_{SFMR}$ are set by the choice of the local scaling relations. 
 The choice of $\nabla_{\rm FMR0}$ is discussed in Sec. \ref{sec: FMR - nabla}.
 The dependence of $\gamma$ on those parameters in the relevant range of values is shown in the right top panel of Fig. \ref{fig: FMR parameters vs local relations}.
 $b_{MZR}$ can be expressed with the parameters of the z$\sim$0 average MZR as: $b_{MZR}=Z_{O/H;MZR0} - a_{MZR} log\left( M_{0;MZR0} \right)$\footnote{See eq. \ref{eq: FMR linear regime} - the functional form of the MZR is commonly described by eq. \ref{eq: Curti parametrisation}, although $\beta$=$\gamma$ is often used. That is also the parametrisation used in \citealt{ChruslinskaNelemans19}}.
 
 \subsubsection{Turnover mass as a function of SFR}

Observational studies 
suggest that $M_{0;\rm SFR}$ increases with SFR
(i.e. for higher SFR, $Z_{O/H}(SFR, M_{*})$ flattens at higher masses).
The steepness of this dependence governs the spacing between the Z$_{O/H}$(M$_{*}$,SFR) curves obtained for different SFR (see Fig. \ref{fig: FMR construction sketch}).
Examples shown in \citet{Telford16} and in Fig. 2 of \citet{Cresci19} suggest that both the metallicity and SFR determination methods have impact on the distance between the different SFR lines (and so on $M_{0;SFR}$).
\\
\citet{Curti20} find that a linear dependence
(log$_{10}$($M_{0;SFR}$)$\propto$ log$_{10}$(SFR)) can well describe the trend seen for their z$\sim$0 galaxy sample.
 \\
 We note that the linear dependence between log($M_{0;\rm SFR}$) and log$_{10}$(SFR) is a natural consequence of the assumptions listed earlier in this section, that $\gamma$ and $Z_{O/H;0}$ are independent of SFR.
 In the $M_{*}<< M_{0;SFR}$ regime eq. \ref{eq: Curti parametrisation} reduces to eq. \ref{eq: FMR linear regime},
 which in our description is equal to eq. \ref{eq: FMR full correlation}.
 From this equality\footnote{
Within our framework the value of the slope of this dependence is 
also the value of the $\alpha$ parameter that minimizes the scatter in the 2D projection of the FMR that is conventionally used to discuss the strength of the three parameter dependence: $Z_{O/H}-\mu_{\alpha}$ where $\mu_{\alpha}=log(M_{*})-\alpha \ log(SFR)$, as originally proposed by Mannucci et al. (2010).
Although, as can be seen in eq. \ref{eq: FMR M0SFR}, $\alpha$ 
is not a good measure of the strength of the
correlation between the SFR and metallicity, as it depends strongly on the shape of the two relations,
in particular on the slope of the low mass end of the MZR.
What really describes the strength of the correlation is $\nabla_{\rm FMR}$.
\label{ftn: alpha vs nabla FMR}
}:
\begin{equation}\label{eq: FMR M0SFR}
  \begin{aligned}
  \rm log_{10}\left( M_{0;SFR} \right) & =
  \frac{\nabla_{\rm FMR0}}{\gamma} log_{10}\left( SFR \right) +\\
  & \frac{a_{MZR}log_{10}\left( M_{0;MZR0} \right) -\nabla_{\rm FMR0} b_{SFMR}}{\gamma} \\
  & = \alpha \ log_{10}\left( SFR \right) + m_{0}
  \end{aligned}
\end{equation}
The dependence of $\alpha$ on $a_{\rm MZR}$, $a_{\rm SFMR}$ 
and $\nabla_{\rm FMR0}$ is shown in Fig. \ref{fig: FMR parameters vs local relations}.

\subsubsection{The value of $\beta$}\label{sec: model: beta}

$\beta$ describes the bending of the $Z_{O/H}(SFR, M_{*})$ relation. 
 The higher the value of $\beta$, the sharper the transition from the 
 linear to flat regime and the smaller the range of log$_{10}$(M$_{*}$) over which the transition happens (see Fig. \ref{fig: beta dependence} for illustration).\\
$\beta$ could in principle depend on the SFR.
 \citet{Curti20} let $\beta$ as a free parameter when fitting their $Z_{O/H}(SFR, M_{*})$ 
 relation for different SFR bins. While their best fit relations give different $\beta$ values
 in different SFR bins, the authors note that there is no clear dependence on SFR.
 Ultimately, they quote a single value of $\beta\sim2$ to describe the FMR.
 \\
 In principle, we can numerically solve for $\beta$ using the condition that galaxies with zero offset from the z$\sim$0 SFMR 
 lie on the z$\sim$0 MZR.
 This condition allows us to properly recover the z$\sim$0 MZR when sampling the galaxy properties
 from the z$\sim$0 GSMF and SFMR and assigning their metallicities with $Z_{O/H}(SFR, M_{*})$.
This condition is only applicable in the SFR range that is covered 
 by the z$\sim$0 SFMR ($\gtrsim10^{-3}\Msun$/yr and $\lesssim 10^{2}\Msun$/yr).
 Numerous solutions are allowed - especially at low SFR$\lesssim 10^{-1} \Msun/yr$,  where the intersection with the z$\sim$0 MZR is reached at log($M_{*}$)$<$log($M_{0;\rm SFR}$) i.e. where $Z_{O/H}(SFR, M_{*})$ is virtually insensitive to the choice of $\beta$.
 We use the numerical solutions as a guide and adopt a simplified 
 $\beta$ dependence on log$_{10}$(SFR), verifying that $Z_{O/H}(SFR, M_{*})$ 
 for galaxies at z$\sim$0 SFMR leads to $Z_{O/H}$ that agree with the corresponding values from the z$\sim$0 MZR to within 0.01 dex (note that this is smaller than the typically found residual scatter about the  $Z_{O/H}(SFR, M_{*})$ of $\sim$0.05 dex).
 Where necessary, we extrapolate the obtained dependence to lower and higher SFR values than probed by z$\sim$0 SFMR. The resulting dependence is shown in Fig. \ref{fig: beta summary} in Appendix \ref{app: FMR variations}.

\subsubsection{The strength of the SFR-metallicity correlation}\label{sec: FMR - nabla}

Observationally, $\nabla_{\rm FMR}$ appears to be a function of $M_{*}$ and SFR. Such dependence is evident in the analysis presented by \citet{Salim14} (see Table 1 therein, where $\kappa$ corresponds to our $\nabla_{\rm FMR}$), where the authors compare the strength of the correlation for galaxies split in different mass and SFR bins.
This is one of the key observed characteristics of the FMR that we aim to reproduce within our description.
\\
The fact that $Z_{O/H;0}$ appears to be roughly independent of the SFR means that the observed $Z_{O/H}(SFR, M_{*})$ flattening does not simply reflect the presence of the high mass flattening of the MZR (and potentially SFMR), but also indicates a weakening/disappearance of the SFR-Z$_{O/H}$ correlation in the high $M_{*}$ regime (i.e. small/zero $\nabla_{\rm FMR}$).
Note that, by construction (fixing the value of $Z_{O/H;0}$), we recover such high $M_{*}$ weakening of the correlation in our method.
\\
A similar behaviour is also seen in the Z$_{O/H}$ - log$_{10}$(SFR) projection of the FMR, where at fixed log$_{10}$(M$_{*}$) the flattening appears at relatively low SFR values (compared to the SFMR).
In other words, the correlation between the SFR and metallicity is found to weaken/disappear at low SFR/low specific SFRs (sSFR=SFR/M$_{*} \lesssim 10^{-10} - 10^{-10.5}$/yr). 
When projected onto the  Z$_{O/H}$-log$_{10}$(M$_{*}$) plane, this means that the correlation is weaker 'above' than 'below' the average $z\sim$0 MZR. That is also the case within our description: for the same absolute value of $\delta_{SFMR}$, $\nabla_{\rm FMR}$ is smaller above than below the $z\sim0$ MZR due to the SFR-dependent location of the flattening and -in certain model variations- the $\beta$ dependence on SFR.
\\
While observational studies seem to agree on the existence of those high $M_{*}$ and low SFR/low sSFR weaker correlation regimes, there is no agreement on the precise region of the FMR (in terms of the Z$_{O/H}$, log$_{10}$(M$_{*}$) and log$_{10}$(SFR) values) in which they appear.
Within our description we can qualitatively reproduce those trends. The transition between the strong correlation regime and the part of the FMR where the correlation weakens happens at different log$_{10}$(M$_{*}$) and sSFR depending on the parameters of the $z\sim0$ MZR and SFMR.
We stress that we only assume that $\nabla_{\rm FMR}$ is fixed (i.e. $\nabla_{\rm FMR}$=$\nabla_{\rm FMR0}$, independent of $M_{*}$ and SFR) in the low mass and high SFR part of the FMR - as guided by the observed characteristics of $Z_{O/H}(SFR, M_{*})$ (but see Appendix \ref{app: SFR dependent nabla}, where we further relax this assumption and introduce a SFR-dependent $\nabla_{\rm FMR}$ in this region of the FMR). 
We do not explicitly assume any value or dependence for $\nabla_{\rm FMR}$ in the remaining part of the relation.
\\
\newline
$\nabla_{\rm FMR0}$ is the only parameter used in our description of the $Z_{O/H}(SFR, M_{*})$ that is not defined by the choice of the local MZR and SFMR, as a potential dependence of this parameter on the metalicity/SFR derivation method is unclear and only few determinations of $\nabla_{\rm FMR}$ are given in the literature.
Using the H$\alpha$ based SFR and metallicity derivation method of \citet{Mannucci10}, \citet{Salim14} find $\nabla_{FMR}\sim$0.3 in the strong correlation regime.
Applying a infrared based SFR derivation method, they find somewhat lower values of $\nabla_{FMR}\sim$0.2.
In their study focusing on $z\sim2.3$, \citet{Sanders18} use several example metallicity calibrations (both strong line-based/theoretical and 'direct'/empirical) and find $\nabla_{FMR}\sim$0.11 - 0.27, hinting at potential dependence on the choice of  metallicity  indicator  and  calibration.
However, as argued by \citet{Sanders20}, different metallicity calibrations (or redshift-dependent adjustments to z$\sim$0 based calibrations) may be needed to correctly derive metallicities of star forming galaxies at different redshifts.
This is taken into account in \citet{Sanders20}, where different empirical metallicity calibrations are used at z$\sim$0 and at higher redshifts. 
They find best-fit $\nabla_{\rm FMR}$=0.27 at z$\sim$0 and somewhat shallower dependence at z$\sim$2.3 ($\nabla_{FMR}\sim$0.19), although consistent within the uncertainty with z$\sim$0 determination (see their Fig.10).
\\
\citet{Torrey18} show the strength of the SFR-metallicity correlation split in several log$_{10}$(M$_{*}$) and redshift bins as found in the IllustrisTNG simulations (see Table 1 therein).
They report values in the range $\nabla_{FMR}\sim$0.25 - 0.34 (except for the lowest log$_{10}$(M$_{*}$)$\sim$9 and highest mass bin log$_{10}$(M$_{*}$)$>$10.5 at z$\sim$0 , where noticeably weaker $\nabla_{\rm FMR}$ is found: 0.19 and 0.1 respectively), with no clear mass or redshift dependence.
\\
\newline
We use $\nabla_{\rm FMR0}$=0.27 as recently found by \citet{Sanders20} as our fiducial choice.
Given the uncertainty of this parameter, we consider values between 0.17 - 0.3 dex to discuss the sensitivity of our results to this choice
\footnote{
Lower values of $\nabla_{\rm FMR}$ were also reported, but we find that $\nabla_{\rm FMR0}\lesssim$0.17 underpredict the redshift evolution of the MZR at $z\lesssim$1.5 (where the impact of biases discussed in Sec. \ref{sec: FMR} is not so severe, and so its evolution is reasonably constrained). 
Much lower $\nabla_{\rm FMR0}$ also make it difficult to reproduce the width of the $z\sim0$ MZR of $\sigma_{MZR}\sim$0.1 dex (unless the scatter around the FMR is in reality higher than the typically reported $\sigma_{FMR}\sim$0.05 dex).
}.
\begin{figure*}
\vspace*{-0.5cm}
\includegraphics[width=1.9\columnwidth]{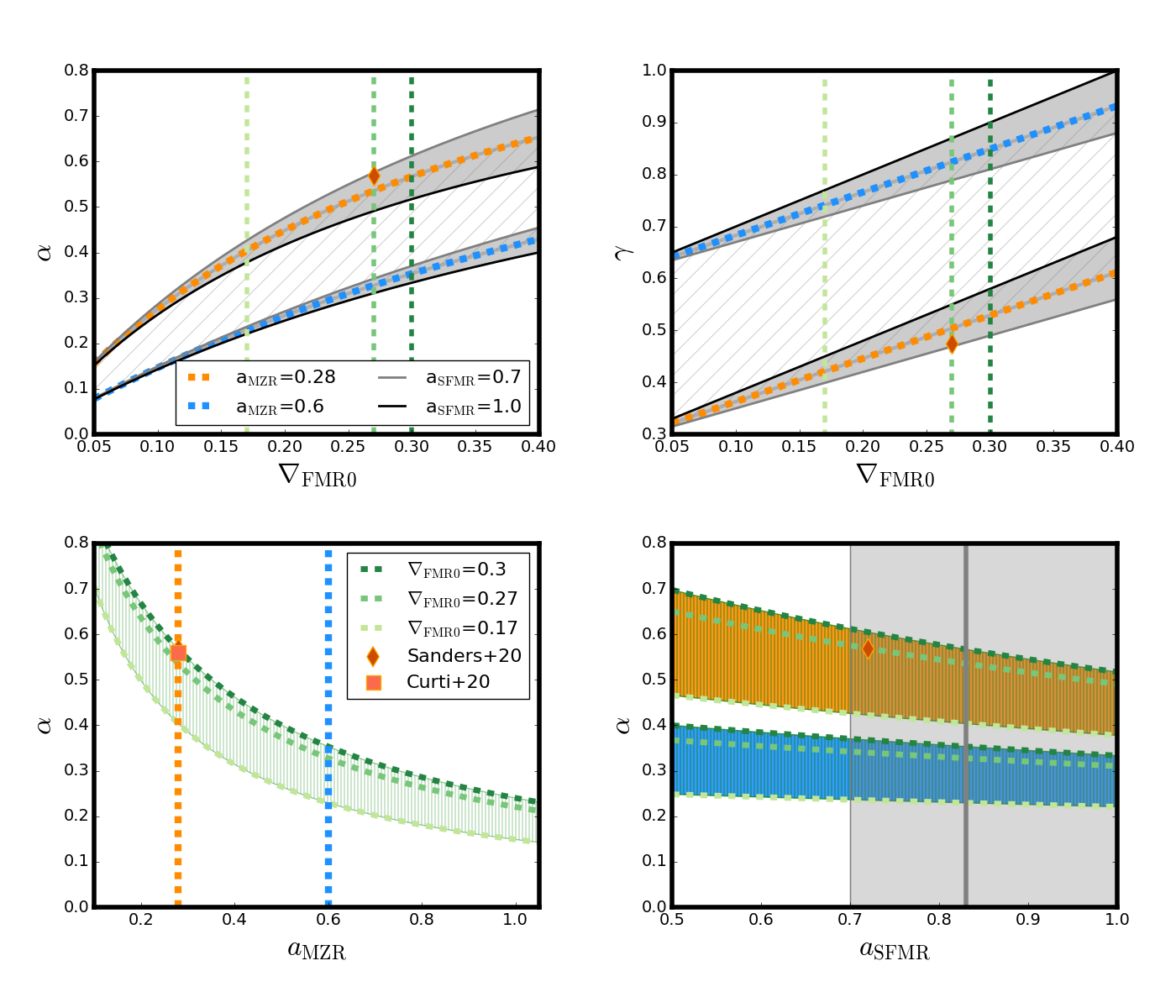}
\vspace*{-0.3cm}
\caption{
Dependence of the parameters $\gamma$ 
( $Z_{O/H}(SFR, M_{*})$ slope at M$_{*}<<$M$_{0;SFR}$, see eq. \ref{eq: Curti parametrisation})
and $\alpha$ (slope of the log$_{10}$(M$_{0;SFR}$) dependence on log$_{10}$(SFR), see eq. \ref{eq: FMR M0SFR}) on the slopes of the $z\sim0$ MZR and SFMR and the parameter $\nabla_{\rm FMR0}$.
Hatched areas in the top panels span between the values obtained for a$_{MZR}\sim$0.3 (orange lines/areas) and 0.6 (blue lines/ares) and indicate the variation caused by the different MZR slopes considered in this study. 
Gray areas in the top panels and in the bottom right panel 
span between a$_{SFMR}$=0.7-1 to illustrate the variation caused by the change in the SFMR slope.
The hatched green area in the bottom left panel shows the variation caused by varying $\nabla_{\rm FMR0}$ in the range 0.17-0.3, assuming a$_{SFMR}$=0.83 (our default low/intermediate mass slope of the SFMR).
The dashed green line falling within that range indicates our fiducial choice of $\nabla_{\rm FMR0}$=0.27.
Orange (blue) ranges in the bottom right panel show the range of $\alpha$ obtained with a$_{MZR}\sim$0.3 (a$_{MZR}\sim$0.6) and $\nabla_{\rm FMR0}$ in the range 0.17-0.3 for each a$_{SFMR}$.
The orange square shows $\alpha$ and a$_{MZR}$ as found by \citet{Curti20}.
The brown diamond shows $\alpha$/$\gamma$ calculated with $a_{\rm MZR}$, $a_{SFMR}$ and $\nabla_{\rm FMR0}$ from \citet{Sanders20} (note that it overlaps with the estimate from \citealt{Curti20} in the bottom left panel).
}
\label{fig: FMR parameters vs local relations}
\end{figure*}

\subsubsection{Calculating metallicities of galaxies at different $z$ with $Z_{O/H}(SFR, M_{*})$}
$Z_{O/H}(SFR, M_{*})$ constructed as described in this Section 
is fully determined by the choice of $\nabla_{\rm FMR0}$ and the parameters of the $z\sim0$ MZR and SFMR.
We introduce the variations of those local relations that we explore in this study in Sec. \ref{sec: model variations}.
We further assume that $Z_{O/H}(SFR, M_{*})$ does not evolve with redshift - metallicity evolution with redshift is then a result of an evolving GSMF and SFMR.
We sample galaxy masses from the redshift-dependent GSMF and their SFR from the distribution centered around the redshift-dependent SFMR, as discussed in Sec. \ref{sec: framework intro}.
We then use $Z_{O/H}(SFR, M_{*})$ to assign the metallicity.
Furthermore, observational studies indicate the residual scatter around the FMR $\sigma_{FMR}\sim$0.05 dex. To account for that, we add a normally distributed scatter $\sigma_{FMR}$=0.05 dex to metallicities assigned with our $Z_{O/H}(SFR, M_{*})$.

\subsection{Differences with respect to Chruslinska \& Nelemans (2019)}\label{Sec: FMR ChN19}

Even though the metallicity distribution of galaxies used in \citet{ChruslinskaNelemans19} relies on the redshift dependent MZR,
the authors include the simplified form of the mass-metallicity-SFR dependence within their framework, assuming that the galaxy offsets from the average MZR and SFMR are fully anti-correlated,
i.e. setting the coefficient from eq. \ref{eq: FMR offsets} to $\nabla_{FMR}=\frac{\sigma_{\rm MZR}}{\sigma_{\rm SFMR}}\approx 0.33$,
where $\sigma_{\rm MZR}$=0.1 dex and $\sigma_{\rm SFMR}$=0.3 dex
describe the scatter around the average MZR and SFMR respectively.
The same $\nabla_{\rm FMR}$ is used independent of $M_{*}$, SFR or redshift.
In the strong correlation regime discussed in Sec. \ref{sec: FMR - nabla} (at low/intermediate $M_{*}$ and high SFR) and at $z\sim$0 their description of the  mass-metallicity-SFR dependence is effectively the same as implemented in this study (except for the lower value of $\nabla_{\rm FMR}$ used in this work).
However, the strength of the observed SFR-metallicity correlation appears to weaken at high $M_{*}$ and low SFR/sSFR.
This behaviour is not captured with the simple description given by eq. \ref{eq: FMR offsets} with fixed $\nabla_{\rm FMR}$.
Another important difference is that, in this study the Z$_{O/H}$(M$_{*}$,SFR)
(once defined with $z\sim0$ relations) is assumed to be redshift-invariant. 
\citet{ChruslinskaNelemans19} use eq. \ref{eq: FMR offsets} to calculate offsets relative to SFMR and MZR as found at any given redshift.
This means that, due to strong MZR evolution at $z\gtrsim2$, their Z$_{O/H}$(M$_{*}$,SFR) is redshift-dependent.

\subsection{FMR example: comparison with Sanders et al. (2020)}

\begin{figure*}
\vspace*{-0.5cm}
\includegraphics[scale=0.5]{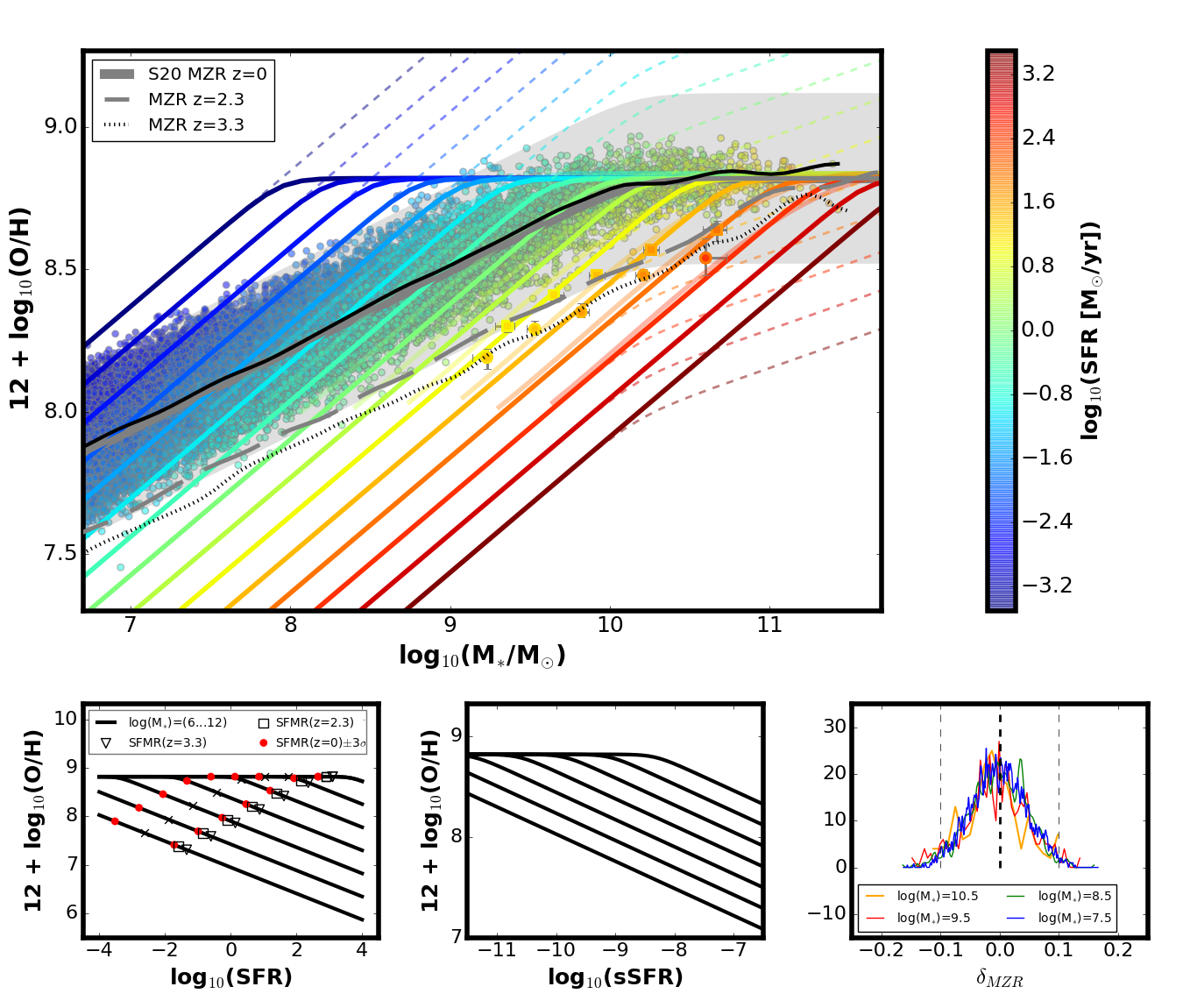}
\vspace*{-0.3cm}
\caption{
Example $Z_{O/H}(SFR, M_{*})$ obtained with our method (coloured lines - different log$_{10}$(SFR) values) and using the $z\sim0$ MZR, SFMR and $\nabla_{\rm FMR0}$ from \citet{Sanders20} (S20). 
Colored points are sampled from the GSMF and SFMR at z$\sim$0.
Black line is plotted to show that z$\sim$0 MZR (shown as thick solid gray line and shaded area - 3 $\sigma_{\rm MZR}$ region) is well reproduced with our $Z_{O/H}(SFR, M_{*})$ and z$\sim$0 GSMF and SFMR.
The gray dashed lines show fits to the maximum density of galaxies sampled from GSMF and SFMR at $z$=2.3 and 3.3 to indicate the projected MZR evolution.
Data points are z$\sim$2.3 and 3.3 stacks from S20.
Faint lines show FMR fitted by S20.
Coloured dashed lines show $Z_{O/H}(SFR, M_{*})$ as would be obtained with eq. \ref{eq: FMR offsets} and $\nabla_{\rm FMR0}$=0.27 used at all masses and SFRs.
Projections onto $Z_{O/H}$-log$_{10}$(SFR) 
and $Z_{O/H}$-log$_{10}$(sSFR) planes are shown in the bottom panels for several log$_{10}$(M$_{*}$) values (black lines - log$_{10}$(M$_{*}$) from 6 to 12).
In the bottom left panel we indicate the SFR corresponding to SFMR value for a given M$_{*}$ at $z=0$ (cross), 2.3 (square) and 3.3 (triangle). Red dots indicate the SFR of a galaxy $\pm 3\sigma_{SFMR}$ away from the SFMR at $z\sim0$.
The rightmost bottom panel shows the offsets of $z\sim0$ model galaxies from the $z\sim0$ MZR, plotted for several mass bins.
}
\label{fig: S20 S20 FMR270 example}
\end{figure*}

In this Section we present an example $Z_{O/H}(SFR, M_{*})$ obtained with our phenomenological model, where we keep all the input assumptions as similar as possible to \citet{Sanders20}.
We choose this study, as the authors provide estimates of all the ingredients that are necessary to construct our $Z_{O/H}(SFR, M_{*})$ (the $z\sim0$ MZR, SFMR and $\nabla_{\rm FMR0}$)\footnote{
Note that while the MZR is often shown in observational studies discussing the FMR, the SFMR that describes the galaxy sample used in that study is typically not and $\nabla_{\rm FMR}$ is rarely estimated. This makes it difficult to directly compare our results with other FMR estimates given in the literature.
}, which allows for a direct comparison with their results.
The resulting $Z_{O/H}(SFR, M_{*})$ is shown in Fig. \ref{fig: S20 S20 FMR270 example}. 
In the bottom panels we include two other common 2D projection of the FMR: the $Z_{O/H}$-log$_{10}$(SFR) plane (bottom left panel) and the $Z_{O/H}$-log$_{10}$(sSFR) plane (bottom middle panel).
The thick gray line in the main panel of Fig. \ref{fig: S20 S20 FMR270 example} shows the $z\sim0$ MZR as given in \citet{Sanders20}. The coloured points around that line represent the $z\sim0$ population of star forming galaxies from our model
\footnote{$M_{*}$ are sampled from our $z\sim0$ GSMF, SFR are sampled from SFMR from \citet{Sanders20} with Gaussian scatter $\sigma_{SFMR}$ and the metallicity is assigned using our $Z_{O/H}(SFR, M_{*})$ with scatter $\sigma_{FMR}$}.
The black line indicates a fit to the maximum density region occupied by the $z\sim0$ galaxy sample described above, to ensue that the $z\sim0$ MZR is reproduced.
The right bottom panel shows the $Z_{O/H}$ residuals around the MZR obtained that way for several example log$_{10}$(M$_{*}$),
which shows that the typically indicated intrinsic width of the relation $\sigma_{MZR}\sim$0.1 dex is reasonably reproduced (note that this quantity is not an input in our model).
The thick solid coloured lines in the main panel show $Z_{O/H}(SFR, M_{*})$ for various fixed log$_{10}$(SFR) values.
The faint coloured solid lines in the background were obtained with the best-fit FMR given by \citet{Sanders20} (see eq. 10 therein), plotted roughly in the range of $M_{*}$ and SFR probed by their galaxy sample.
We also plot their $z\sim2.3$ (squares) and $z\sim$3.3 (circles) data points (obtained for stacked spectra of the observed galaxies, see Table 1 therein).
The inner (outer) colours in all symbols correspond to upper (lower) bound on the log$_{10}$(SFR) within the uncertainties provided by the authors for each of the stacks.
The long dashed and dotted gray lines indicate the projected $z\sim2.3$ and $z\sim3.3$ MZR, calculated in the same way as the black line, but using the $z\sim2.3$ and $z\sim3.3$ SFMR from Sanders et al. (2020) respectively, and our GSMF estimated at the corresponding redshifts. The overall agreement is remarkable.
\\
We also show $Z_{O/H}(SFR, M_{*})$ for various fixed log$_{10}$(SFR) values as would have been obtained with
the approach used in \citet{ChruslinskaNelemans19} i.e., using eq. \ref{eq: FMR offsets} with $\nabla_{\rm FMR}$=$\nabla_{\rm FMR0}$=0.27 fixed (i.e. the same at all SFRs and masses, see colored dashed lines in the main panel of Fig. \ref{fig: S20 S20 FMR270 example}).
It can be seen that the $Z_{O/H}(SFR, M_{*})$ constructed that way is identical with our model at low/intermediate $M_{*}$ and high SFRs (the strong correlation regime), and starts to deviate at high $M_{*}$ and low SFR (above the $z\sim$0 MZR), where $\nabla_{FMR}<\nabla_{FMR0}$.
This demonstrates that the mass and SFR dependence of the SFR-metallicity correlation  discussed in Sec. \ref{sec: FMR - nabla} is present in our description.

 \begin{figure*}
\vspace*{-0.5cm}
\includegraphics[scale=0.48]{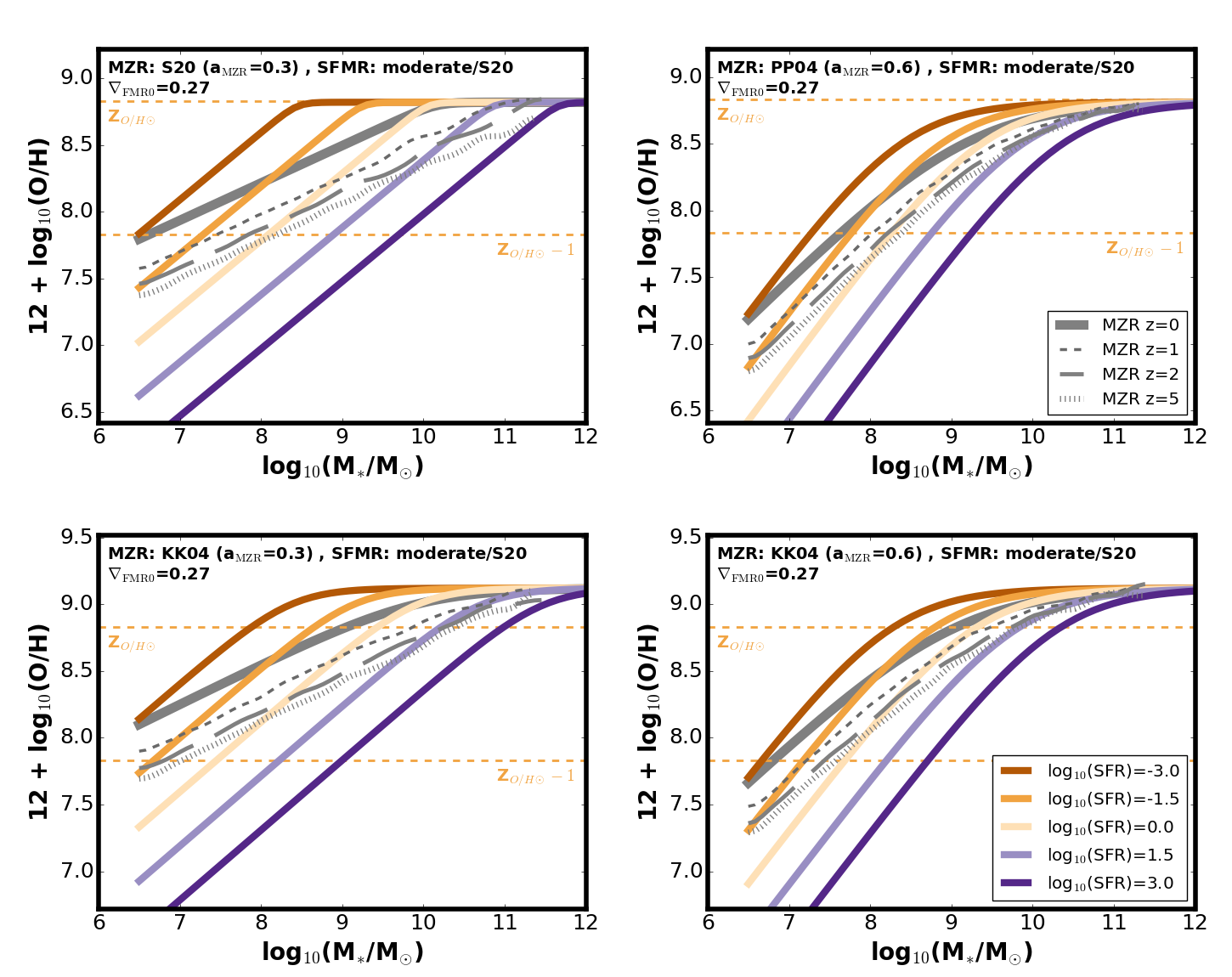}
\vspace*{-0.3cm}
\caption{
$Z_{O/H}(SFR, M_{*})$ obtained as described in Sec. \ref{Sec: FMR - model} - different colours show $Z_{O/H}(SFR, M_{*})$ at fixed log$_{10}$(SFR) (increasing from left to right, see legend).
Different panels correspond to different choices of the $z\sim0$ MZR (indicated by the thick gray line).
All panels assume SFMR with a moderate flattening ($z\sim0$ SFMR with a$_{SFMR}$=0.72 at high masses and a$_{SFMR}$=0.83 at low masses) and $\nabla_{\rm FMR0}$=0.27.
Gray dashed lines indicate the projected MZR evolution as obtained by performing fits to maximum density of galaxies in 12+log$_{10}$(O/H) - log$_{10}$(M$_{*}$) plane, where the galaxies were sampled from GSMF (assuming $\alpha_{GSMF}$=$\alpha_{fix}$) and SFMR at several higher redshifts and their Z$_{O/H}$ was assigned using the $Z_{O/H}(SFR, M_{*})$ shown in each of the panels.
Horizontal orange lines indicate solar and 10\% solar metallicity, assuming Z$_{O/H \odot}$=8.83 \citep{GrevesseSauval98}.
}
\label{fig: FMR MZR choice}
\end{figure*}

\subsection{Model variations}\label{sec: model variations}

 \begin{table}
\centering
\small
\caption{Parameters of the different versions of the $z\sim0$ MZR and SFMR used in this study. The first column gives the name used to reference the variation throughout this study.
\newline
$^a$ different SFMR parametrisation (see footnote \ref{ftn: Tomczak16})
}
\begin{tabular}{c c c c c }
\hline
 &\multicolumn{4}{c}{ \textbf{MZR $z\sim0$} }\\
 & $a_{\rm MZR}$ & log$_{10}\left( \rm M_{0;MZR0} \right)$ & Z$_{O/H;MZR0}$ & $\beta_{MZR}$ \\ \hline
 PP04   & 0.6 & 9.19 &  8.81  & 0.6   \\
 S20 & 0.28 & 10.16 &  8.82  & 3.43   \\
 KK04a & 0.57 & 9.03 & 9.12  & 0.57    \\
 KK04b & 0.3 & 9.9 & 9.12  & 0.74   \\ \hline
 &\multicolumn{4}{c}{ \textbf{SFRM $z\sim0$} } \\
  & $a_{SFMR}$ & log$_{10}\left( \rm M_{0;SFMR0} \right)$ & $a_{SFMR2}$ & $b_{SFMR}$ \\ \hline
 no flattening   & 0.83 & - &  0.83  & -8.241  \\
 sharp flattening$^{a}$ & 0.83 & 9.89 &  \multicolumn{2}{c}{s$_{0}$=-0.033} \\ \hline
 moderate/S20 & 0.83 & 9.73 & 0.72  & -8.241 \\
 moderate/S14 & 0.83 & 9.73 & 0.49  & -8.241 \\ \hline

\end{tabular}
\label{tab: MZR and SFMR params}
\end{table}

In this study we consider several variations of the base $z\sim0$ MZR and SFMR relations,
representing extreme choices of the shapes of the two relations and the MZR normalisation $Z_{O/H;MZR0}$ reported in the literature.
We also consider two variations of the GSMF: either with fixed low mass end slope ($\alpha_{GSMF}$=-1.45), or with $\alpha_{GSMF}$ steepening with redshift (see Sec. 3.1. in \citealt{ChruslinskaNelemans19} for the details).\\
This allows us to explore the extremes of the f$_{SFR}$(Z,z) distribution. 
We briefly introduce each of the MZR and SFMR variations used to construct our Z$_{O/H}$(SFR,M$_{*}$) below.

\subsubsection{MZR variations}
The MZR is parametrised in the same way as Z$_{O/H}$(M$_{*}$,SFR), i.e.:
 \begin{equation}\label{eq: MZR parametrisation}
  \rm Z_{O/H}(M_{*}) = Z_{O/H;MZR0} - \frac{a_{MZR}}{\beta_{MZR}} log\left( 1 + (\frac{M_{*}}{M_{0;MZR0}})^{-\beta_{MZR}} \right)
 \end{equation}
MZR parameters for the variations considered in this study are given 
in the top rows of Table \ref{tab: MZR and SFMR params}.
\\
Variations PP04 and KK04a are identical as in \citet{ChruslinskaNelemans19} 
(where KK04a was labelled KK04) and represent MZR estimates based on the \citet{PettiniPagel04} O3N2 and \citet{KobulnickyKeweley04} metallicity calibrations respectively.
Both MZRs have similar slopes $a_{MZR}\sim0.6$, but differ in normalisation by $\sim$0.3 dex.
This difference represents a well known systematic offset between the metallicities
derived with the theoretical calibrations \citep[e.g.][]{KobulnickyKeweley04} and the so-called direct method or empirical calibrations \citep[e.g.][]{PettiniPagel04,Sanders20}, where the latter yield estimates that are $\sim$2 times lower than the former.
The choice of KK04 and PP04 calibrations maximises the difference in $Z_{O/H;MZR0}$ (and in the low/high metallicity tails of the f$_{SFR}$(Z,z), see \citealt{ChruslinskaNelemans19}), i.e. metallicity estimates obtained with other methods fall in between  \citep[see e.g. Fig 15 in][]{MaiolinoMannucci19}\footnote{Note that the \citet{Zahid14a} estimate shown in Fig. 15 in \citet{MaiolinoMannucci19} is based on theoretical metallicity calibration from \citet{KobulnickyKeweley04}}.
\\
Furthermore, we consider two additional variations: S20 and KK04b,
that have $Z_{O/H;MZR0}$ as in the PP04 and KK04a variations respectively 
but a shallower low mass slope $a_{MZR}\sim0.3$.
The S20 variation is based on the recent direct-method based determination from \citet{Sanders20}.
The parameters of the KK04b variation were chosen in such a way that the MZR is identical with the one in KK04a variation at log$_{10}$(M$_{*}/\Msun)\gtrsim$8.7 and differs at lower masses (where the original determination was not constrained by the data).
Our motivation for including those variations is the following: 
the shape of the z$\sim$0 MZR, in particular its low mass end slope and log$_{10}\left( M_{0;MZR0} \right)$ is what sets $\gamma$ and $\rm log_{10}\left( M_{0;SFR} \right)$ within our framework, as the dependence on the SFMR parameters is reduced by multiplication by $\nabla_{FMR}\lesssim 0.3$.
However, the MZR slope and turnover mass are not well constrained with the current data (the purely linear regime is often not probed) and depend on the adopted parametrisation, for the same galaxy sample one can fit different a$_{MZR}$ and turnover masses \citep[e.g.][for the parametrisation used in eq. \ref{eq: Curti parametrisation}, $\beta$ is also somewhat degenerate with the MZR turnover mass and slope]{Curti20}.
Recent MZR determinations by \citet{Curti20} and \citet{Sanders20} both indicate shallower a$_{MZR}\sim$0.3 than typically found in earlier studies \citep[a$_{MZR}\sim$0.6, e.g.][]{AndrewsMartini13,Zahid14b}.
\citet{Curti20} and \citet{Sanders20} discuss several potential biases in earlier studies that may be causing this difference (e.g. relatively bright, high SFR galaxies may dominate the low mass end of the sample and therefore induce a bias towards lower metallicity).
Alternatively, massive star-based metallicity determination methods (yielding $Z_{O/H;MZR0}$ consistent with empirical/direct methods) lead to a$_{MZR}\sim$0.6, similar to earlier studies \citep[see discussion in][]{Sanders20}.
Given those differing results and until better observational constraints on 
the low mass end of the MZR are available, we consider 0.3$\lesssim$a$_{MZR}\lesssim$0.6 
as realistic.
Note that this uncertainty in the MZR slope at M$_{*}\lesssim10^{8.5}\Msun$ leads to significant differences in metallicity when the relation is extrapolated down to low stellar masses  (e.g. at log$_{10}$(M$_{*}$/$\Msun$)=6 there is $\approx$0.56 dex difference between the metallicity estimated with KK04a and KK04b z$\sim$0 MZR variations and $\approx$0.76 dex difference if S20 and PP04 z$\sim$0 MZR variations are compared).
By considering a range of slopes we explore the uncertainty associated with the low mass MZR extrapolation on the (low metallicity part of) f$_{\rm SFR}$(Z,z) estimate.
In this study, the MZR is only used to construct $Z_{O/H}(SFR, M_{*})$ and we do not need to describe its evolution with redshift.

\subsubsection{SFMR variations}

To describe the $z\sim0$ SFMR we follow \citet{ChruslinskaNelemans19} 
and assume the low/intermediate mass slope $a_{SFMR}$ and normalisation $b_{SFMR}$ 
based on \citet{Boogaard18}.
As discussed in \citet{ChruslinskaNelemans19}, the shape of the high mass end
is debated: while many authors report a varying degree of flattening above a certain mass \citep[e.g.][]{Schreiber15,Tomczak16,Bisigello18}, others see no evidence for the change of slope \citep[e.g.][]{RenziniPeng15,Pearson18}.
We consider the same extreme variations as described in \citet{ChruslinskaNelemans19}:
'no flattening' - a linear relation between log$_{10}$(SFR) and log$_{10}$(M$_{*}$)
with a single slope at all masses and
'sharp flattening' - with a$_{SFMR}\sim0$ at high masses (using the parametrisation from \citet{Tomczak16}\footnote{Parametrised as follows: $\rm log_{10}(SFR)=s_{0} - log_{10}\left(1 + (\frac{M_*}{ M_{0;SFMR0} })^{-a_{SFMR}}\right)$\label{ftn: Tomczak16}}).
We focus on those two extremes to discuss our results, but use additional
'moderate' variations in some of the figures introducing our FMR model.
Those assume that the slope of the SFMR changes from $a_{SFMR}$ to $a_{SFMR2}<a_{SFMR}$ 
at log$_{10}$(M$_{*}$)=log$_{10}\left( \rm M_{0;SFMR0} \right)$.
Variation moderate/S20 assumes a high mass slope as in the $z\sim0$ SFMR shown in \citet{Sanders20}, while moderate/S14 follows the prescription of \citet{Speagle14}.
The relevant parameters for all the SFMR variations are given in the bottom rows of Table \ref{tab: MZR and SFMR params}.\\
The SFMR evolution with redshift is described as in \citet{ChruslinskaNelemans19}.

\subsubsection{The FMR for the considered model variations}
In this section we summarize the key differences between
$Z_{O/H}(SFR, M_{*})$ obtained for the different
variations of the $z\sim$0 MZR, SFMR and $\nabla_{\rm FMR0}$ 
considered in this study.
\\
It can be seen from eq. \ref{eq: FMR M0SFR} that 
the impact of the SFMR parameters on the SFR-dependent turnover mass $\rm log_{10}(M_{0;SFR})$ (and on $\alpha$)
is reduced by the multiplication by $\nabla_{FMR}\lesssim0.3$.
Considering a range of $a_{SFMR}\sim$0.7-1 spanned by
different determinations of the low mass slope of the SFMR present in the literature \citep[see e.g. Fig. 10 in][]{Boogaard18} instead of using a fixed value $a_{SFMR}$=0.83 would only affect $\alpha$ by $\lesssim$0.08.
This is illustrated by the gray bands in the top panels in 
Fig. \ref{fig: FMR parameters vs local relations}.
At the same time, for the range of $a_{\rm MZR}$ considered in this study
$\alpha$ varies between $\sim$0.2 and 0.55 (compare the orange and blue lines in Fig. \ref{fig: FMR parameters vs local relations},
where smaller values correspond to steeper MZR slopes).
Variation in $\nabla_{\rm FMR}$ between 0.17 and 0.3 affects $\alpha$ by $\sim$0.15 (see green lines and bottom left panel in Fig. \ref{fig: FMR parameters vs local relations}).
\\
The choice of MZR has a decisive role in setting the $Z_{O/H}(SFR, M_{*})$. $\nabla_{\rm FMR0}$ also visibly affects the relation, while SFMR has a relatively mild impact on its shape within our framework.
$Z_{O/H}(SFR, M_{*})$ obtained for the different 
variations of the $z\sim 0$ MZR considered in this study and for our fiducial $\nabla_{\rm FMR0}$=0.27 is shown in Fig. \ref{fig: FMR MZR choice}. The shift in normalisation between the PP04/S20 and KK04a/b variations (compare top and bottom panels) as well as the difference in the spacing between the different log$_{10}$(SFR) lines depending on $a_{\rm MZR}$ (compare right and left panels) are evident.
The analogous figures showing the impact of different SFMR (Fig. \ref{fig: FMR SFMR choice})
and $\nabla_{\rm FMR0}$ (Fig. \ref{fig: FMR FMR slope choice}) choices are shown in Appendix \ref{app: FMR variations}.
The additional $Z_{O/H}(SFR, M_{*})$ variation with a SFR-dependent $\nabla_{\rm FMR}$ is discussed in Appendix \ref{app: SFR dependent nabla} and illustrated in Fig. \ref{fig: FMR varN}.
\\
The choice of the SFMR variation (its high mass end) has the strongest effect on the $\beta$ parameter ( see Fig. \ref{fig: beta summary} in the Appendix \ref{app: FMR variations}, which shows the $\beta$ - log$_{10}$(SFR) relation for different choices of the local scaling relations).
 In practice, cases with a SFMR with no flattening/only mild change of slope are well described with a single value of $\beta$. 
 The dependence on SFR becomes apparent in cases with SFMR showing a significant deviation from a single power law.
 The obtained values of $\beta$ are generally smaller for steeper MZR. $\beta$ is only weakly affected by the choice of $\nabla_{\rm FMR0}$ (shifting towards smaller values with decreasing $\nabla_{\rm FMR0}$).
 We note that using the  $z\sim0$ MZR from Curti et al. (2020),
 we can recover their best fit $\beta \sim 2$ (when we assume single power law SFMR).
 \\
 Before showing the results, we first discuss the treatment of starburst galaxies in our models.

\section{Starburst galaxies}\label{sec: starbursts}

A common approach to describe the SFR distribution of galaxies
is to use a gaussian distribution centered around the redshift-dependent SFMR \footnote{But see \citealt{Boco19,Boco21} for alternative method based on the galaxy SFR functions (number  density  of  galaxies  per  logarithmic bin of SFR).}.
In reality, the SFR of star forming galaxies at fixed stellar mass seems to follow a bimodal, double-gaussian shape. The secondary peak of this distribution is attributed to starburst (SB) galaxies - strong SFMR outliers that feature SFR a few times higher \footnote{Note that various criteria are used in the literature to distinguish starbursts and regular star forming galaxies.
Most commonly the criteria are based on the SFR and require that
the starburst SFR is at least a factor of a few (factors between 3 and 10 are used in the literature) higher than that of an average galaxy of the same mass and at the same redshift \citep[e.g.][]{Orlitova20}
} than those of the regular star forming galaxies of the same mass and redshift.
Several authors estimate the fraction of starburst galaxies (f$_{SB}$; the ratio between the number of galaxies associated with the starburst component of the SFR distribution and the total number of star forming galaxies in the considered mass and redshift range)
and report values f$_{SB}\sim$2-3\% \citep[e.g.][]{Rodighiero11,Sargent12,Bethermin12,Ilbert15,Schreiber15}. Despite the relatively low f$_{SB}$ values, the starburst's contribution to the total cosmic SFRD could still amount to $\sim10\%$ at z$\sim$2 due to their high SFR.
\citet{Boco21} use the double gaussian distribution of galaxy SFRs from \citet{Sargent12} and suggest that accounting for the starburst component can improve the consistency between the cosmic SFRD at z$\gtrsim$2 determined
with the use of galaxy stellar mass functions paired with SFMR \citep[as used in][]{ChruslinskaNelemans19} and that estimated with the use of SFR functions (which better account for the SFR of dusty galaxies at high redshifts; as used in \citealt{Boco19}).
\\
\newline
Crucially, the above mentioned f$_{SB}$ determinations are based on galaxy samples limited to relatively massive objects (log$_{10}$(M$_{*}$/$\Msun$)$>$10) and $z\lesssim$2.
Studies of \citet{Caputi17} and \citet{Bisigello18} extend the analysis of the distribution of star forming galaxies in the log$_{10}$(SFR) - log$_{10}$(M$_{*}$) plane to much lower masses and higher redshifts.
While their results are consistent with the previous determinations at high log$_{10}$(M$_{*}$),
they show that f$_{SB}$ is a strong function of stellar mass (increasing towards lower log$_{10}$(M$_{*}$)) and $z$ (increasing with redshift).
Moreover, they indicate that starburst galaxies follow a distinct sequence in the log$_{10}$(SFR) - log$_{10}$(M$_{*}$) plane that is located $\sim$1 dex above the SFMR. This offset of the starburst sequence relative to SFMR is considerably higher than previously reported \citep[e.g.][]{Sargent12,Bethermin12}.
This suggests that the contribution of starburst galaxies to the total SFRD budget can be much higher than previously estimated. 
If starburst galaxies follow the general FMR (as suggested by the results of \citealt{Hunt12}; to our knowledge, there is no evidence to the contrary),
they would contribute to the star formation at relatively low metallicities compared to galaxies on the SFMR.
Therefore, they affect the low metallicity tail of the f$_{SFR}$(Z,z) - crucial for the discussion of the origin of transients as long gamma ray bursts and double black hole mergers.
We aim to discuss the possible impact of starbursts on the f$_{SFR}$(Z,z) in view of the results reported in 
\citet{Caputi17} and \citet{Bisigello18}.
In the following section \ref{Sec: starburst implementation} we describe the 
method used to include the contribution of starburst galaxies within our framework.

\subsection{Method and considered variations}\label{Sec: starburst implementation}
To account for the contribution of starburst galaxies in our calculations, we follow the procedure outlined below:
\begin{itemize}
\item At each redshift, we sample log$_{10}$(M$_{*}$)
of star forming galaxies from the galaxy stellar mass function as described in \citet{ChruslinskaNelemans19}.
\item We use f$_{SB}$ to describe the fractions of starburst galaxies and regular star forming galaxies at each  log$_{10}$(M$_{*}$) and $z$. Our choice is outlined further in this section.
\item The SFR of regular galaxies is given by the SFMR from \citet{ChruslinskaNelemans19}.
The SFR of starburst galaxies at each log$_{10}$(M$_{*}$) follows a normal distribution with scatter $\sigma_{SB}$.
The peak of the distribution can be related to the galaxy's log$_{10}$(M$_{*}$) with a linear relation:
$\rm log_{10}(SFR_{SB}) = a_{SB} log_{10}(M_{*}) + b_{SB}$,
to which we refer as the starburst sequence
We set the value of $b_{SB}$ by defining the offset of the
starburst sequence from the SFMR, i.e. $b_{SB} = b_{SFMR} + \Delta_{SB-SFMR}$.
The assumed parameters are given further in this section.
\item To describe the metallicity at which the starburst galaxies produce stars, we assume that they follow the same FMR as regular star forming galaxies (described in Sec. \ref{Sec: FMR - model}).
\end{itemize}
We consider two sets of parameters describing the properties of starbursts.
Firstly, we follow the same implementation as used in the recent study by \citet{Boco21}, which is based on the works of \citet{Sargent12} and \citet{Bethermin12}. This implementation assumes a constant f$_{SB}$=0.03 (independent of mass and redshift), a starburst sequence with scatter $\sigma_{SB}$=0.24, located $\Delta_{SB-SFMR}$=0.59 dex above the SFMR and parallel to the SFMR (specifically, we assume the starburst sequence slope of a$_{SB}$=0.83, corresponding to the low mass slope of the SFMR from \citealt{ChruslinskaNelemans19}
\footnote{Note that this is different than in \citealt{Boco21}, 
who assume that the slope of the starburst sequence is parallel to that of the SFMR, where the SFMR is described as a single power law with the parameters given by \citet{Speagle14}.})
In light of the results of \citet{Caputi17} and \citet{Bisigello18} this implementation severely underestimates both f$_{SB}$ and the SFR of galaxies on the starburst sequence.
Therefore, it likely provides the absolute lower limit on the contribution of starbursts.
\\
Secondly, we follow the results of \citet{Caputi17} and \citet{Bisigello18}
to model the mass and redshift dependence of f$_{SB}$ and the properties of the starburst sequence. We provide the details of this implementation in Sec. \ref{Sec: fSB implementation B18/C17} and \ref{Sec: SB sequence implementation B18/C17} and refer to it as the B18/C17 implementation in the reminder of this paper.

\subsubsection{The fraction of starbursts}\label{Sec: fSB implementation B18/C17}

\begin{figure}
\vspace*{-0.5cm}
\includegraphics[width=1.\columnwidth]{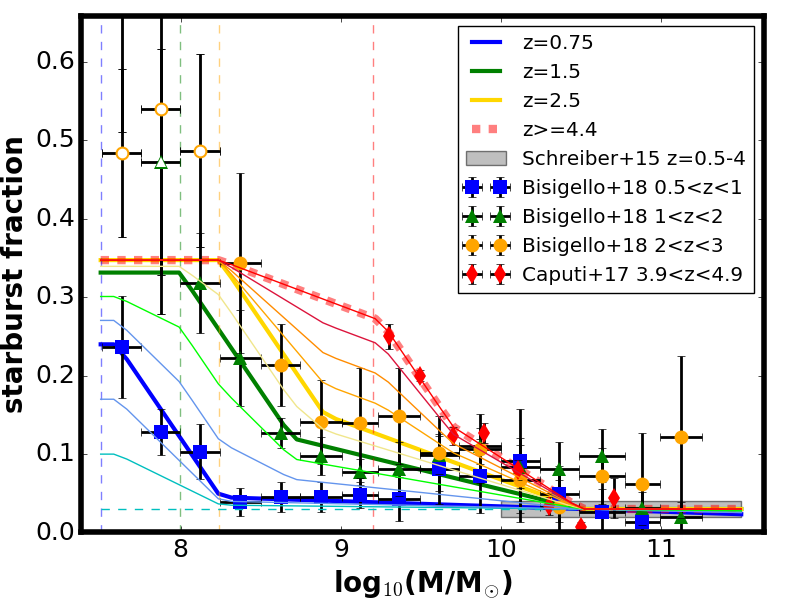}
\vspace*{-0.3cm}
\caption{
Fraction of starbursts as a function of galaxy stellar mass.
Thick lines show the assumed dependence at z$\sim$0.75, 1.5 and 2.5
based on the data from Bisigello et al. (2018)
and at z$\sim$4.4 based on Caputi et al. (2017).
Thin lines show the interpolation between those redshifts
and extrapolation to z=0 (thin horizontal dashed line), plotted with 0.25 step in redshift at z$<$1.5
and with 0.5 step in redshift above.
Vertical dashed lines mark log$_{10}$(M$_{fix}$) (see Table. \ref{tab: SB fraction}) 
for different redshift bins.
Open symbols mark estimates where the galaxy sample from Bisigello et al. (2018) is below
90\% mass completeness.
}
\label{fig: SB fraction}
\end{figure}

To describe the mass and redshift dependence of $f_{SB}$, 
we use the observational estimates obtained by \citet{Bisigello18} 
(provied for three redshift bins: $0.5<z<1$, $1<z<2$, $2<z<3$; see Fig. 9 therein)
and \citet{Caputi17} ($3.9<z<4.9$, see their Fig. 7).
The assumed fraction of starbursts versus stellar mass and redshift is shown in Fig. \ref{fig: SB fraction}.
At each redshift bin covered by the data, 
we assume the following relation between $f_{SB}$ and log$_{10}$(M$_{*}$):\\

%   \[
  \begin{equation}\label{eq: SB fraction}
  f_{SB} = \begin{cases}
	\rm a_{1} log_{10}(M_{*}) + b_{1} & \text{if } \rm   log_{10}(M_{\rm fix}) \leq log_{10}(M_{*}) \leq log_{10}(M_{0;SB}) \\
    \rm a_{2}\ log_{10}(M_{*}) + b_{2} & \text{if } \rm log_{10}(M_{0;SB}) < log_{10}(M_{*}) \leq 10.5 \\
    \rm \ 0.03 & \text{if } \rm log_{10}(M_{*}/\Msun) > 10.5 \\
          \end{cases}
\end{equation}
%  \]

 All stellar masses in eq. \ref{eq: SB fraction} are in solar units $\Msun$.
 The adopted coefficients are given in Table \ref{tab: SB fraction}.
 We assume that the above relation holds strictly in the middle of each redshift bin
 and interpolate between them to describe  $f_{SB}$ at $0.5<z<4.4$.
 At z$>$4.4 we use the relation from z=4.4. That way we obtain a conservative estimate of the starburst contribution at higher redshifts.
 We extrapolate the constructed dependence to lower redshifts,
 setting a constant starburst fraction of 3\% at all masses at z=0.
\\
 At z=0.75,1.5 and 2.5, log$_{10}$(M$_{\rm fix}$) is the lower edge of the lowest mass bin 
 where the galaxy sample from \citet{Bisigello18} is within 90\% stellar-mass completeness.
 We make a conservative assumption and use a fixed $f_{SB}$ value below that mass (i.e., 
  $f_{SB} =\rm a_{1} log_{10}(M_{\rm fix}) + b_{1}$). 
  At $z=4.4$, log$_{10}$(M$_{\rm fix}$)=9.25 corresponds to the lowest mass for which the fraction of starbursts has been estimated in \citet{Caputi17}.
At this redshift, $f_{SB}$ is fixed at the same mass as at z=2.5 (i.e. log$_{10}$(M$_{*}$)=8.24) and we assume a linear relation to describe the dependence between log$_{10}$(M$_{*}$)=8.24 and log$_{10}$(M$_{*}$)=9.25 .
 This exception is necessary to avoid a decrease in the fraction of starbursts
 at log$_{10}$(M$_{*}$)$<$9.25 at z$>$2.5, 
 which would break the trend seen in the data (see Figure \ref{fig: SB fraction}).
 At log$_{10}$(M$_{*}$)$>$10.5 there is a lot of scatter in the data and 
 the trend is inconclusive. In this mass range we use a fixed value $f_{SB}$=0.03 at all redshifts, as found in earlier studies that focused on the most massive galaxies \citep[e.g.][]{Rodighiero11,Schreiber15}.
 The steeper, lower mass part of $f_{SB}$ - log$_{10}$(M$_{*}$) dependence is well described with the same slope across all four redshift bins. Therefore, for simplicity we assume $a_{1}=-0.3$.  The slope of the relation in the higher mass part ($a_{2}$) 
 and the mass log$_{10}$(M$_{0;SB}$) separating the high and low mass parts of the relation
 increase with redshift.

\begin{table}
\centering
\small
\caption{
Coefficients used in eq. \ref{eq: SB fraction},
describing f$_{SB}$ as a function of log$_{10}$(M$_{*}$)
at redshift $z_{bin}$. The remaining coefficient $a_{1}=-0.3$
at all $z_{bin}$
}
\begin{tabular}{c c c c c c}
\hline
 $z_{bin}$ & b$_{1}$ & a$_{2}$ & b$_{2}$ & log(M$_{\rm 0;SB}$/$\Msun$) & log(M$_{\rm fix}$/$\Msun$) \\ \hline
  \hline
  0.75 & 2.52 & -0.0067 & 0.1 & 8.25 & 7.6 \\ \hline
  1.5  & 2.73 & -0.05 & 0.555 & 8.7  & 7.99 \\ \hline
  2.5  & 2.82 & -0.075 & 0.817 & 8.9  & 8.24 \\ \hline
  4.4  & 3.045 & -0.131 & 1.408 & 9.7  & 9.25 \\ \hline  
\end{tabular}
\label{tab: SB fraction}
\end{table}
\subsubsection{The starburst sequence}\label{Sec: SB sequence implementation B18/C17}

\begin{figure}
\vspace*{-0.5cm}
\includegraphics[width=1.\columnwidth]{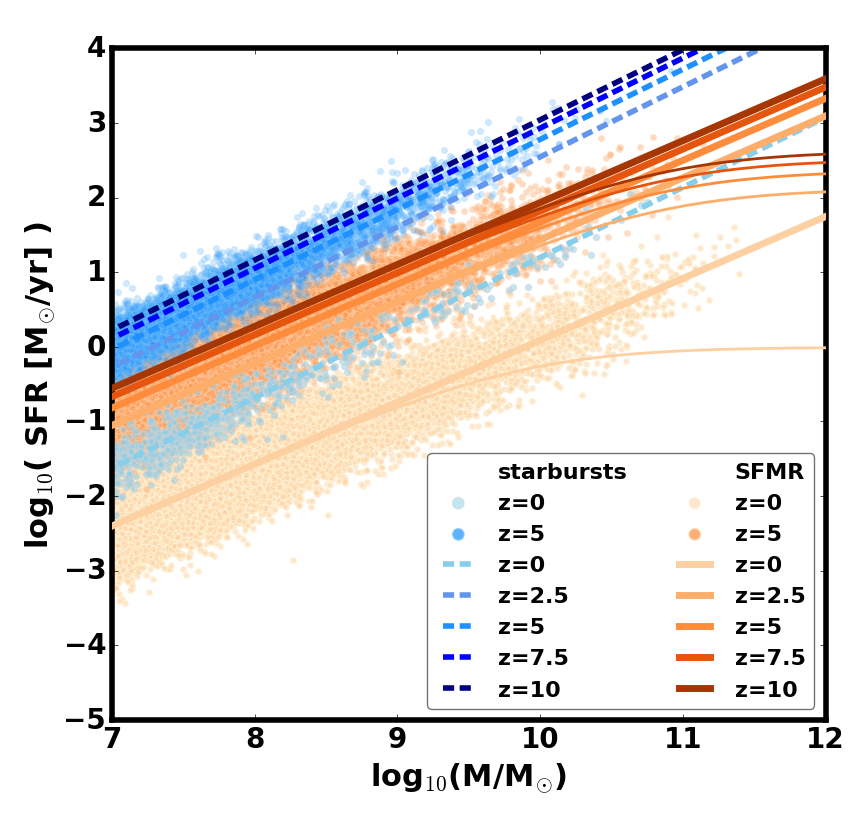}
\vspace*{-0.3cm}
\caption{
Starburst sequence (SB - dashed lines) compared with the SFMR for extreme assumptions about the high mass slope of the SFMR: no flattening (thick solid lines) and sharp flattening (thin solid lines).
Colours correspond to different redshifts.
Points sampled from the GSMF and assigned to the SFMR or starburst sequence using starburst fraction shown in Fig. \ref{fig: SB fraction} are plotted for z=0 and z=5 to illustrate the increasing fraction of star forming galaxies that populate the starburst sequence at higher redshifts and lower stellar masses.
The properties of starbursts are based on the results from \citet{Caputi17} and \citet{Bisigello18}.
%starbursts.py
}
\label{fig: SB sequence}
\end{figure}

We guide our description of the SFR distribution of starburst galaxies with the results shown in Fig. 3 and Fig. 7 from \citet{Caputi17} and Fig. 6 and Fig. 7 from \citet{Bisigello18}.
The resulting starburst sequence is shown in Fig. \ref{fig: SB sequence}.
The width of the starburst sequence in \citet{Caputi17} and in \citet{Bisigello18} is smaller that that of the SFMR, although the values are not given. 
Guided by the results shown in the figures, we assume $\sigma_{SB}$=0.2 dex.
Both \citet{Caputi17} and \citet{Bisigello18} 
find a starburst sequence that is steeper than the SFMR.
There is no clear evidence for $a_{SB}$ evolution with redshift. For simplicity, we assume $a_{SB}$=0.94
(average between the best fit values in 3 redshift
bins from Bisigello et al. and high redshift estimate from Caputi et al.).
To set the offset of the starburst sequence from the SFMR $\Delta_{SB-SFMR}$ we focus on the results for the intermediate masses log$_{10}$(M$_{10}/\Msun$)$\sim$9-9.5
(where the sample is complete and the SFMR is not affected strongly by the potential flattening at the high mass end).
We find that $\Delta_{SB-SFMR}$ in both \citet{Bisigello18} and \citet{Caputi17} is about 1 dex and we assume this value in our calculations.
Note that the SFMR shown in \citet{Caputi17} is likely an upper limit on the SFMR location at z$\sim$4.4, which suggests that at those high redshifts the offset might be even larger.

\section{Results: the distribution of the cosmic SFRD over metallicities and redshift}\label{sec: results - fsfr(Z,z)}

In this section we discuss the distributions of the cosmic SFRD over metallicities and redshift (f$_{SFR}$(Z,z)) obtained for different variations of our observation-based model (including different choices of the local MZR, SFMR, GSMF, $\nabla_{\rm FMR0}$ and prescriptions to account for starburst galaxies).\\
Ultimately, we aim to explore the extreme f$_{SFR}$(Z,z) cases in terms of the amount of SFRD occurring at low (below 10\% solar metallicity; Z$_{O/H}\leqslant$Z$_{O/H\odot}$-1) and high (above solar Z$_{O/H}\geqslant$Z$_{O/H\odot}$) metallicity.
\\
In the remainder of this paper we distinguish between two sets of models based on the choice of the local scaling relations (MZR and SFMR): the 'low metallicity'
f$_{SFR}$(Z,z) cases - obtained with $z\sim0$ MZR with low normalisation (PP04 or S20) and SFMR with sharp flattening at high masses, and the 'high metallicity' 
f$_{SFR}$(Z,z) cases - obtained with $z\sim0$ MZR with high normalisation (KK04a or KK04b) and SFMR with no flattening.
Other combinations of the local MZR and SFMR lead to more moderate metallicity distributions.
% \\
% In Sec. \ref{sec: results FMR effect} we compare the 
% f$_{SFR}$(Z,z) obtained with the non-evolving FMR and with the redshift-dependent MZR based approach used in \citet{ChruslinskaNelemans19}.
% We discuss the impact of $\nabla_{\rm FMR0}$ and the MZR slope (distinguishing PP04/KK04a and S20/KK04b MZR variations) in Sec. \ref{sec: results - MZR slope and FMR slope}.
% In Sec. \ref{sec: results - starbursts} we discuss the f$_{SFR}$(Z,z) under different assumptions about the contribution of starburst galaxies.
\\
\newline
We note that in order to describe the f$_{SFR}$(Z,z) at high redshifts and low metallicities (as shown in this Section and needed, for instance, in applications to gravitational wave astrophysics) one needs to extrapolate the FMR well beyond the regions where it is constrained by current observations (in particular to z$\gtrsim$3 and $M_{*}\lesssim10^{8}\Msun$).
The importance of the assumed z$\gtrsim$3 extrapolation is discussed in Sec. \ref{sec: results FMR effect}, where we compare the 
 f$_{SFR}$(Z,z) obtained with the non-evolving FMR and with the redshift-dependent MZR based approach used in \citet{ChruslinskaNelemans19}.
Metallicity of $M_{*}\lesssim10^{8}\Msun$ galaxies assigned with our FMR is primarily sensitive to the MZR slope and normalisation and the strength of the SFR-metallicity anti-correlation at low masses/SFRs. Those factors are discussed in Sec. \ref{sec: results - MZR slope and FMR slope} and further in  Appendix \ref{app: SFR dependent nabla}.
In Sec. \ref{sec: results - starbursts} we discuss the f$_{SFR}$(Z,z) under different assumptions about the contribution of starburst galaxies.
\subsection{f$_{SFR}$(Z,z) with redshift-invariant FMR}\label{sec: results FMR effect}
\begin{figure*}
\vspace*{-0.5cm}
\includegraphics[width=1.9\columnwidth]{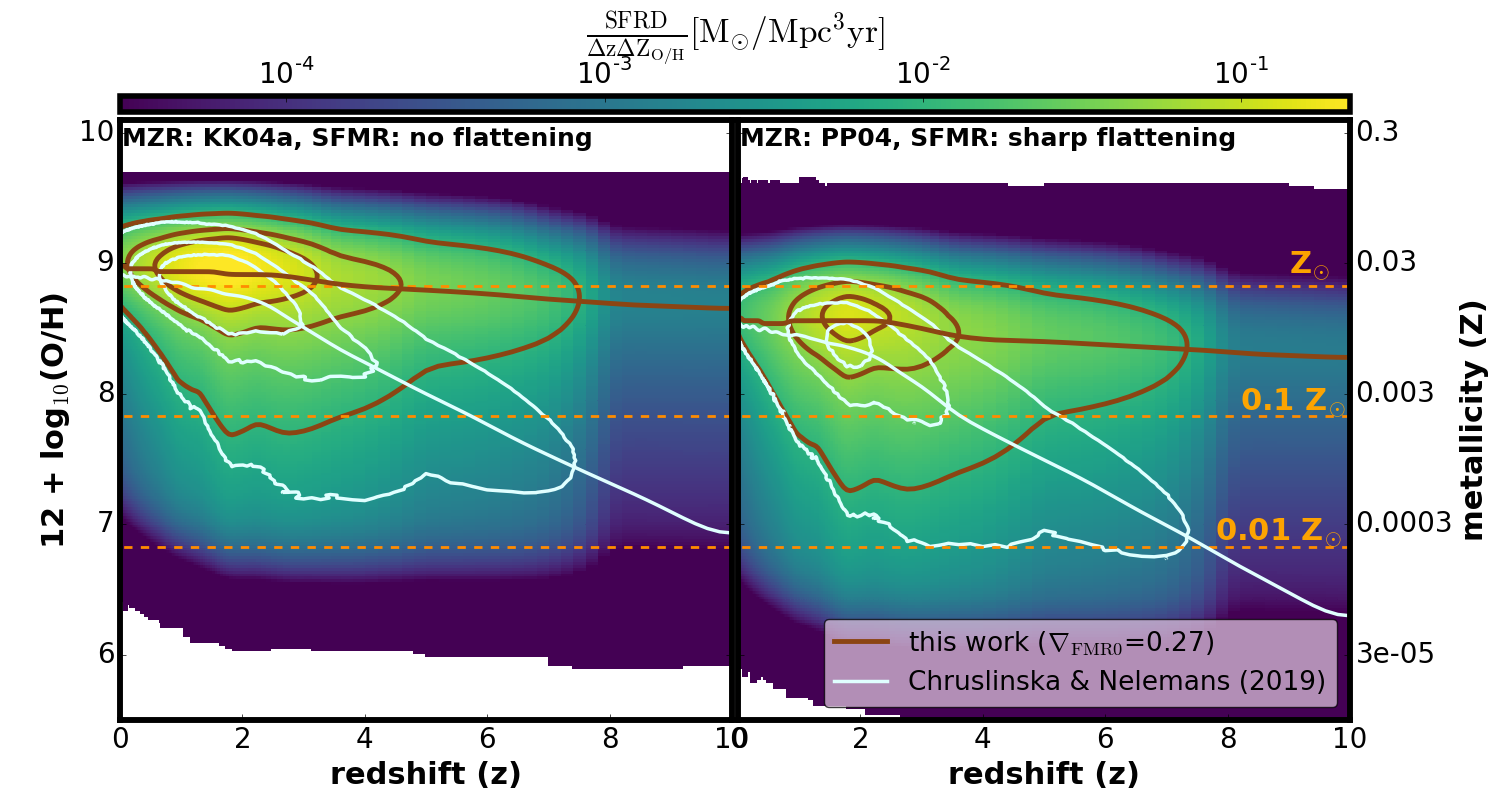}
\vspace*{-0.3cm}
\caption{
Distribution of the star formation rate density (SFRD) at different metallicities and redshift (z)
in this paper (background colours, brown contours; assuming non-evolving FMR and $\nabla_{\rm FMR0}=0.27$)
compared to \citet{ChruslinskaNelemans19} (white contours, assuming redshift-dependent MZR).
Colour indicates the amount of SFRD happening in different redshift and metallicity bins, contours indicate constant SFRD and are plotted for 0.01,0.05 and 0.1 [$\Msun$/Mpc$^{3}$yr] (with the highest value corresponding to the most inward contour).
Left – model variation with KK04a $z\sim0$ MZR and SFMR with no flattening at high masses compared with the high metallicity extreme from \citet{ChruslinskaNelemans19}.
Right – model variation with PP04 $z\sim0$ MZR and SFMR with sharp flattening at high masses compared with the low metallicity extreme from \citet{ChruslinskaNelemans19}.
All models assume GSMF with non-evolving low mass end.
Orange horizontal dashed lines indicate solar, 10\% solar and 1\% solar metallicity (assuming \citealt{GrevesseSauval98} solar metallicity scale).
The right metallicity axis was obtained from $Z_{O/H}$ assuming solar abundance ratios.
Note that beyond z$\gtrsim$3 the distribution relies on extrapolation of the MZR/FMR evolution with redshift - in this respect, the plotted contours contrast somewhat extreme assumptions.
}
\label{fig: SFRD(Z,z) - FMR effect}
\end{figure*}

\begin{figure}
\vspace*{-0.5cm}
\includegraphics[width=1.\columnwidth]{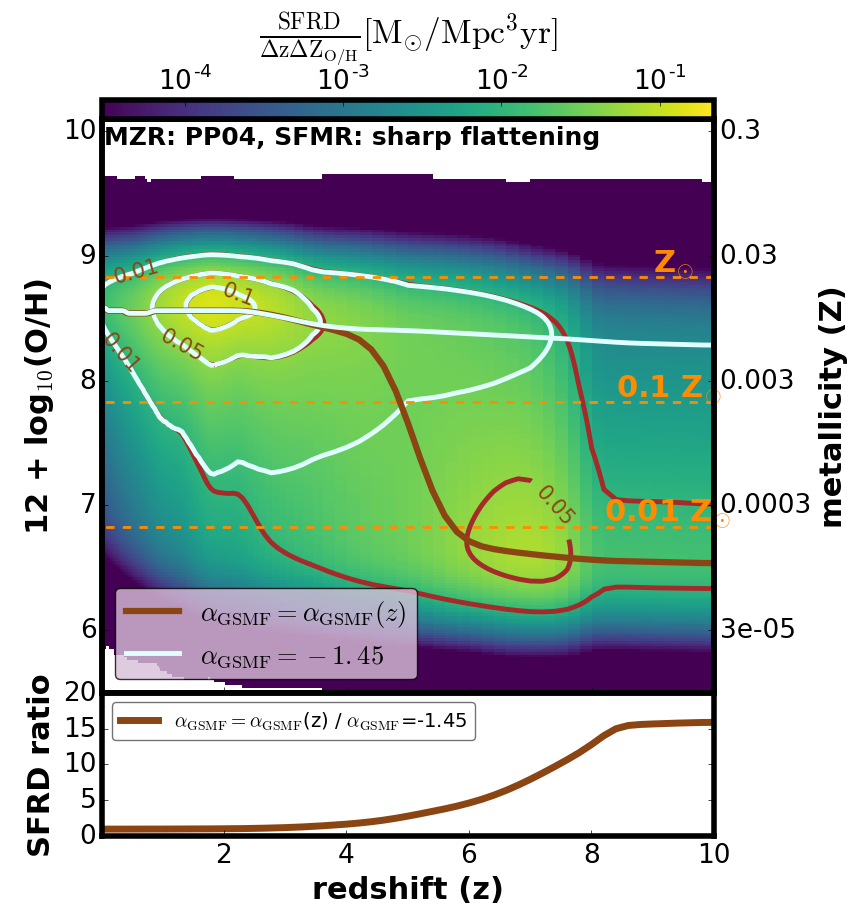}
\vspace*{-0.3cm}
\caption{
Example illustrating the effect of different assumptions about the low mass end slope of the GSMF ($\alpha_{\rm GSMF}$) on the f$_{\rm SFR}$(Z,z).
Top panel: brown constant SFRD contours and background colours - variation with $\alpha_{GSMF}$ steepening with redshift (i.e. predicting increasing number density of low mass galaxies), white contours - fixed $\alpha_{\rm GSMF}$=-1.45.
Contours  are plotted for 0.01,0.05 and 0.1 [$\Msun$/Mpc$^{3}$yr].
Both model variations assume $\nabla_{\rm FMR0}$=0.27 and the same 
SFMR and $z\sim0$ MZR (indicated in the figure).
f$_{\rm SFR}$(Z,z) significantly deviate at $z\gtrsim3$: a redshift-dependent $\alpha_{\rm GSMF}$ leads to most of the SFRD happening at low metallicity and to a higher total SFRD(z).
Bottom panel: ratio of the total SFRD at each redshift as obtained in the two variations.
}
\label{fig: SFRD(Z,z) - GSMF example}
\end{figure}

The comparison between the f$_{SFR}$(Z,z) constructed with the use of redshift-invariant Z$_{O/H}$(SFR,M$_{*}$) and the corresponding distributions from \citet{ChruslinskaNelemans19} is shown in Fig. \ref{fig: SFRD(Z,z) - FMR effect}.
We compare the variations with $z\sim0$ MZR and SFMR choices as in the high (left) and low (right) metallicity extremes from \citet{ChruslinskaNelemans19} (see Sec. 4.2 and Fig. 8 therein).
In this section we do not explicitly include starburst galaxies,
so that all assumptions that are not related to the metallicity distribution of galaxies are the same as in \citet{ChruslinskaNelemans19}.
\\
\newline
Fig. \ref{fig: SFRD(Z,z) - FMR effect} shows that the metallicity distributions start to deviate around $z\gtrsim$1.5 (compare brown and white contours), where the redshift-dependent MZR predicts a steeper decrease in metallicity than what results from the non-evolving FMR.
The difference becomes striking at $z\gtrsim$3, but
we stress that neither the MZR nor the FMR is currently constrained in this redshift regime. In particular, there is no guarantee that the FMR holds or continues to show weak/no redshift evolution beyond that redshift.
On the other hand, as discussed in Sec. \ref{sec: FMR}, the extrapolated MZR evolution as assumed by \citet{ChruslinskaNelemans19} is likely to overestimate the rate of decrease in metallicity.
Therefore, in Fig. \ref{fig: SFRD(Z,z) - FMR effect} we contrast two seemingly extreme assumptions.
Until better observational constraints are available, 
this comparison can serve to illustrate (likely a conservative) range of uncertainty of the high redshift part of f$_{SFR}$(Z,z) resulting from the extrapolated evolution of the galaxy metallicity distribution with redshift.
The extrapolated MZR evolution leads to a $z\sim10$ peak metallicity that is almost 2 dex lower than what results from the non-evolving FMR assumption.
The latter leads to SFRD concentrated at higher metallicities (irrespective of the model variation), but the extended low metallicity tail is still present at all redshifts.
\\
We note that the difference between the  f$_{SFR}$(Z,z) obtained with a redshift dependent MZR and with the non-evolving FMR was recently discussed in  \citet{Boco21} - qualitatively our results are the same as discussed therein.
However, rather than discussing a example f$_{SFR}$(Z,z) obtained with a particular FMR or MZR taken from the literature, here we model the FMR consistently with the choice of the local SFMR and MZR and can explore the uncertainties of the final result.
\\
 All variations shown in Fig. \ref{fig: SFRD(Z,z) - FMR effect}
 assume a GSMF with non-evolving low mass slope.
As discussed in \citet{ChruslinskaNelemans19} (see Sec. 4.1 therein), this assumption 
has relatively little effect on f$_{SFR}$(Z,z) at $z\lesssim$3,
but strongly affects the result at higher redshifts (both its low metallicity tail and the total SFRD).
This is illustrated in Fig. \ref{fig: SFRD(Z,z) - GSMF example} -  the effect of the steepening low mass end of the GSMF on f$_{SFR}$(Z,z) is analogous for other model variations.

\subsubsection{The effect of the $z\sim0$ MZR slope and $\nabla_{\rm FMR0}$} \label{sec: results - MZR slope and FMR slope}
The effect of the choice of the  $z\sim0$ MZR slope ($a_{\rm MZR}$) and $\nabla_{\rm FMR0}$
on f$_{SFR}$(Z,z) is shown in Fig. \ref{fig: SFRD(Z,z) - MZR and FMR slope}.
Similarly to Fig. \ref{fig: SFRD(Z,z) - FMR effect}, the left and right panels show high and low metallicity model variations respectively.
Top (bottom) panels show variations with $a_{MZR}\sim$0.3 ($a_{MZR}\sim$0.6).
Different contours illustrate the impact of $\nabla_{\rm FMR0}$ 
\\
\newline
As can be expected, f$_{SFR}$(Z,z) for variations with steeper $z\sim0$ MZR (higher $a_{\rm MZR}$ values) have a more extended low metallicity tail at each redshift.
However, variations with lower $a_{\rm MZR}$ values feature a much steeper metallicity evolution at $z\lesssim 4$.
At higher redshifts the metallicity peak of f$_{SFR}$(Z,z) decreases at a similar rate for the variations with the same $\nabla_{\rm FMR0}$ and same total SFRD (i.e. the same SFMR and GSMF assumptions), irrespective of $a_{\rm MZR}$.
\footnote{
For instance, for the $\nabla_{FMR0}=0.27$ cases shown in Fig. \ref{fig: SFRD(Z,z) - MZR and FMR slope}, the peak metallicity decreases by $\sim$0.55 dex ($\sim$0.4 dex) between z=0 and z=4 in the model with S20 (KK04b) $z\sim0$ MZR.
This decrease is only $\sim0.1$ dex ($\sim$0.15dex) for the models with steeper PP04 (KK04a) $z\sim0$ MZR. Between z=4 and z=10, the peak metallicity decreases by $\sim0.1$ dex ($\sim$0.18 dex) for the cases with SFMR with no (sharp) flattening (irrespective of the $a_{\rm MZR}$).
}
This can be understood by looking at the comparison of Z$_{O/H}$(SFR,M$_{*}$) obtained for different $z\sim0$ MZR, shown in Fig. \ref{fig: FMR MZR choice}.
In this plane, galaxies with the same log$_{10}$(M$_{*}$) at higher redshifts shift to higher log$_{10}$(SFR) lines.
The rate of this shifting is dictated by the redshift-dependent SFMR.
The offset in Z$_{O/H}$ between the different log$_{10}$(SFR) curves at fixed log$_{10}$(M$_{*}$) is bigger for variations with lower $a_{\rm MZR}$ (compare left and right panels in Fig \ref{fig: FMR MZR choice}) and similarly for higher $\nabla_{\rm FMR0}$ (see Fig. \ref{fig: FMR FMR slope choice} in the Appendix \ref{app: FMR variations}).
This translates into steeper decrease of the f$_{SFR}$(Z,z) metallicity peak with redshift in the variations with lower $a_{\rm MZR}$ and higher $\nabla_{\rm FMR0}$.
At sufficiently high redshifts (where almost all galaxies occupy the high SFR, linear regime of Z$_{O/H}$(SFR,M$_{*}$)) the spacing in Z$_{O/H}$ for different log$_{10}$(SFR) curves is set by the choice of $\nabla_{\rm FMR0}$.
Lower $\nabla_{\rm FMR0}$ result in weaker metallicity evolution (compare white and brown contours in Fig. \ref{fig: SFRD(Z,z) - MZR and FMR slope}).
Therefore, for fixed $\nabla_{\rm FMR0}$ and
the same assumptions about the SFMR, the rate of metallicity evolution at $z\gtrsim 4$ and fixed log$_{10}$(M$_{*}$) is essentially the same.
However, the full f$_{SFR}$(Z,z) distribution is also affected by the GSMF. 
If the low mass end of the GSMF is allowed to steepen with redshift, the contribution of the low mass (and so - low metallicity) galaxies
to the total SFRD at high redshifts is considerably higher than if the low mass end of the GSFM is fixed (see example shown in Fig. \ref{fig: SFRD(Z,z) - GSMF example}). 
\\
The FMR is not constrained for low mass, low SFR galaxies.
Results discussed above assume that those galaxies are well described by the same relation as the galaxies with higher M$_{*}$ and SFR, and that the strength of the SFR-metallicity correlation does not change in this part of the parameter space (i.e. $\nabla_{\rm FMR}$=$\nabla_{\rm FMR0}$).
In appendix \ref{app: SFR dependent nabla} we additionally discuss the variation 
with a SFR-dependent $\nabla_{\rm FMR}$ in which the SFR-metallicity correlation is assumed to disappear (i.e. the FMR breaks down) at SFR corresponding to $z\sim0$ SFMR galaxies with $M_{*}\lesssim 10^{8}\Msun$. Overall, this assumption has minor impact on the estimated f$_{\rm SFR}$(Z,z) distribution (see Fig. \ref{fig: fSFRD varN}).

% allowing
% for a conservative estimate of the low mass galaxies to the low metallicity SFRD budget.

\begin{figure*}
\vspace*{-0.5cm}
\includegraphics[width=1.9\columnwidth]{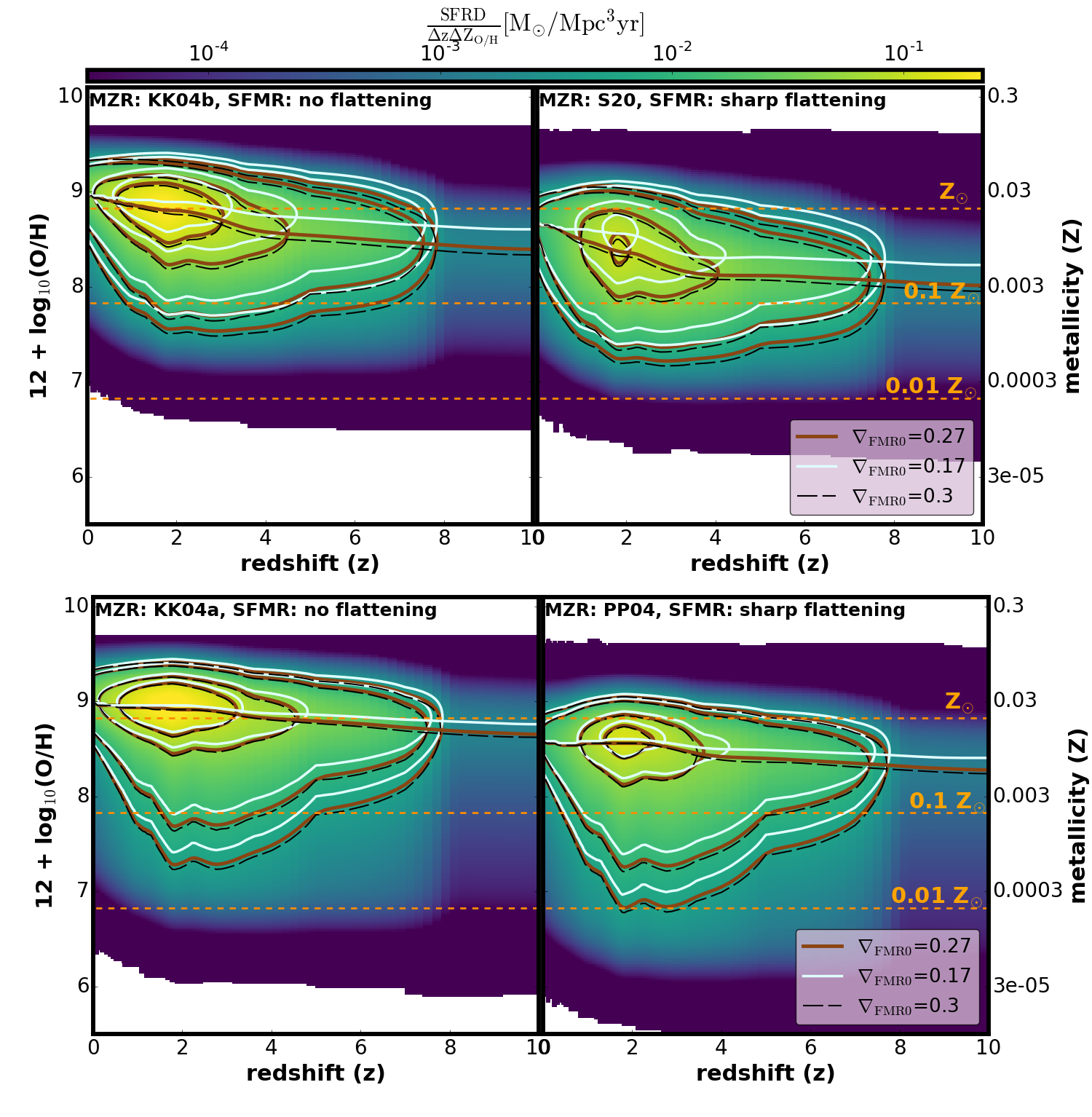}
\vspace*{-0.3cm}
\caption{
Distribution of the star formation rate density (SFRD) at different metallicities and redshift (z) for different assumptions about the $z\sim0$ MZR slope (top panels: variations with $a_{MZR}\sim$0.3, bottom panels: $a_{MZR}\sim$0.6) and $\nabla_{\rm FMR0}$ (compare different contours; background colours are plotted for variations with $\nabla_{\rm FMR0}$=0.27).
Left (right) panels show model variations with $z\sim0$ MZR and SFMR that maximise the SFRD at high (low) metallicity.
Constant SFRD contours are plotted for 0.005,0.01,0.05 and 0.1 [$\Msun$/Mpc$^{3}$yr] (the highest value corresponds to the innermost contour).
All model variations assume a GSMF with non-evolving low mass end.
Variations with flatter $z\sim$0 MZR lead to a more compact metallicity distribution, but their peak decreases with redshift faster than in variations with a steeper $z\sim$0 MZR.
}
\label{fig: SFRD(Z,z) - MZR and FMR slope}
\end{figure*}

\subsection{f$_{SFR}$(Z,z): the impact of starbursts}\label{sec: results - starbursts}

\begin{figure*}
\vspace*{-0.5cm}
\includegraphics[width=1.9\columnwidth]{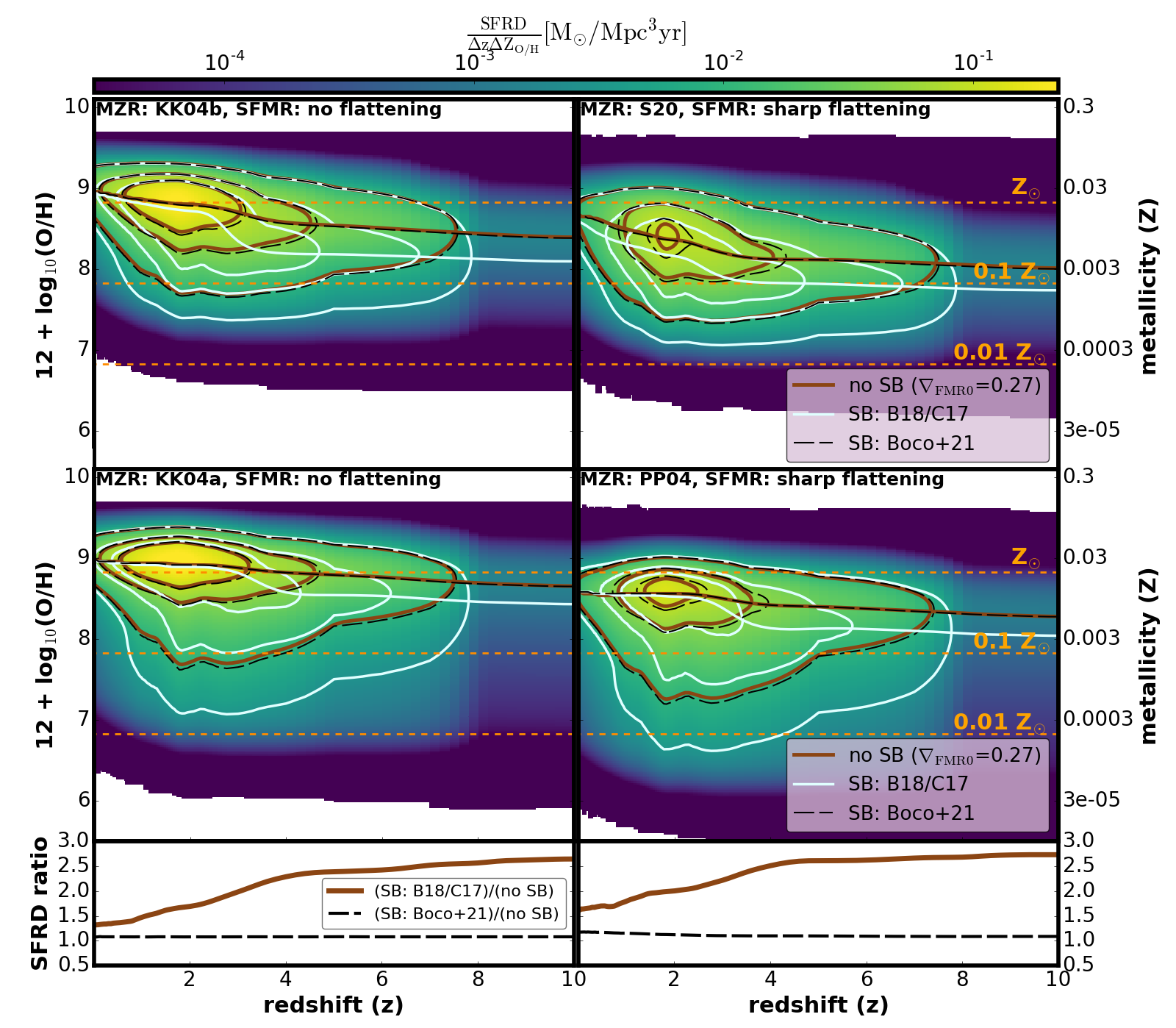}
\vspace*{-0.3cm}
\caption{
Distribution of the star formation rate density (SFRD) at different metallicities and redshift (z) for different assumptions about starburst galaxies:
no starbursts (background colours, brown contours),
starburst sequence and f$_{SB}$ model based on the results of Bisigello et al. (2018) and Caputi et al. (2017)  (B18/C17, white contours) and as in Boco et al. (2021) (black dashed contours).
Top and middle panels correspond to different  $z\sim0$ MZR and SFMR variations (same as in Fig. \ref{fig: SFRD(Z,z) - MZR and FMR slope}). Contours indicate constant SFRD and are plotted for 0.01,0.05 and 0.1 [$\Msun$/Mpc$^{3}$yr] (with the highest value corresponding to the innermost contour).
Note that different assumptions about starburst galaxies lead to different total SFRD at each redshift - bottom panels show the ratio of the SFRD integrated over metallicities at each $z$ in the model with starbursts (brown - B18/C17 implementation, black - Boco et al 2021 implementation) to the corresponding model without starbursts (the choice of MZR does not affect the total SFRD).
All models assume GSMF with non-evolving low mass end and $\nabla_{\rm FMR0}$=0.27.
}
\label{fig: SFRD(Z,z) - SB effect}
\end{figure*}
The comparison of f$_{SFR}$(Z,z) obtained with and without including starbursts is shown Fig. \ref{fig: SFRD(Z,z) - SB effect}.
Top and middle panels show the same $z\sim0$ MZR and SFMR variations as in Fig. \ref{fig: SFRD(Z,z) - MZR and FMR slope}. Different contours correspond to different starburst implementations (introduced in Sec. \ref{Sec: starburst implementation}).
The effect of including starbursts in our calculations is twofold: 
first, a fraction $f_{SB}$ of star forming galaxies is now assigned a higher SFR than what would result from the SFMR.
This means that the total SFRD at each redshift is higher than in the model that does not explicitly account for starbursts (see bottom panel in Fig. \ref{fig: SFRD(Z,z) - SB effect}).
Second, assuming that starbursts follow the FMR, they contribute to star formation happening at relatively low (for their mass and redshift) metallicity. Therefore, including starbursts shifts the peak of the f$_{SFR}$(Z,z) to lower metallicities and broadens the low metallicity part of the distribution.
\\
\newline
Both effects are clearly seen when the variations including the B18/C17 starburst implementation
(white contours in Fig. \ref{fig: SFRD(Z,z) - SB effect}) are compared with the corresponding variations that do not include starbursts (brown contours).
The lower edges of white contours extend to lower metallicities with respect to no starburst case at all redshifts, while the upper, high metallicity edges remain the same. This difference increases with redshift due to increasing f$_{SB}$.
The total SFRD is about 2.5 times higher at $z\gtrsim$4 in the case with starbursts (brown line, bottom panel in Fig. \ref{fig: SFRD(Z,z) - SB effect}). Note that we fix f$_{SB}$ beyond $z=4.4$ - the highest redshift bin covered by the data in \citet{Caputi17}.
If the trend seen at lower redshift continues, the difference at high redshifts would be even higher.
\\
If instead we follow the starburst implementation as used in the recent study by \citet{Boco21} (black dashed contours), the difference with respect to cases without starbursts is negligible. 
This is expected, given the low fixed f$_{SB}$=3\% and the fact that in this prescription the starburst sequence is within the 2$\sigma_{SFMR}$ scatter of the SFMR.
The difference is slightly more pronounced if SFMR with a sharp flattening at high masses is used (right panels in Fig. \ref{fig: SFRD(Z,z) - SB effect}), as in those cases there is a larger difference in SFR between massive galaxies on the SFMR and on the starburst sequence.
The inclusion of starbursts as in \citet{Boco21} also barely affects the total SFRD (see black dashed line in the bottom panel in Fig. \ref{fig: SFRD(Z,z) - SB effect}).

\section{Metallicity-dependent cosmic SFH}

In this section we discuss the cosmic SFH - SFRD integrated over all metallicities as a function of redshift/cosmic time, as well as the high and low metallicity cuts of the cosmic SFH (i.e. SFRD occurring above and below certain metallicity thresholds) obtained for different model variations.
Note that this is essentially a different way to present our results, which provides less detailed information than the full f$_{SFR}$(Z,z) distributions shown in Sec. \ref{sec: results - fsfr(Z,z)}.
It allows us to zoom into the interesting parts of the 
distribution and demonstrate the uncertainty of its various parts more clearly.
However, we stress that the discussed cuts of the cosmic SFH are sensitive to the considered low/high metallicity thresholds.
In some cases, factors that have minor impact on the overall f$_{SFR}$(Z,z)
distribution may lead to considerable uncertainty in the SFH cut for a particular choice of metallicity threshold.
The 10\% solar/solar Z$_{O/H}$ thresholds used here were chosen to zoom into the low/high metallicity tails of the distribution.
In general, the relevant thresholds may vary depending on the considered problem.
\\
When showing the cosmic SFH and its various metallicity cuts,
in each case we indicate the range of possible outcomes spanned by the model variations with extreme assumptions about the considered factors.
We indicate which of the considered factors drives the uncertainty of the f$_{SFR}$(Z,z) in different regimes (i.e. low/high metallicity, low/high redshift).
As discussed in Sec. \ref{sec: results - fsfr(Z,z)}, the difference between the f$_{SFR}$(Z,z) obtained for variations with starburst implementation as in \citet{Boco21} (i.e. with fixed, small fraction of starbursts) and the corresponding variations with no starbursts is negligible. Therefore, in this section we only discuss the former.

\subsection{Cosmic SFH}\label{sec: results - total SFH}
 \begin{figure*}
\vspace*{-0.5cm}
\includegraphics[width=1.7\columnwidth]{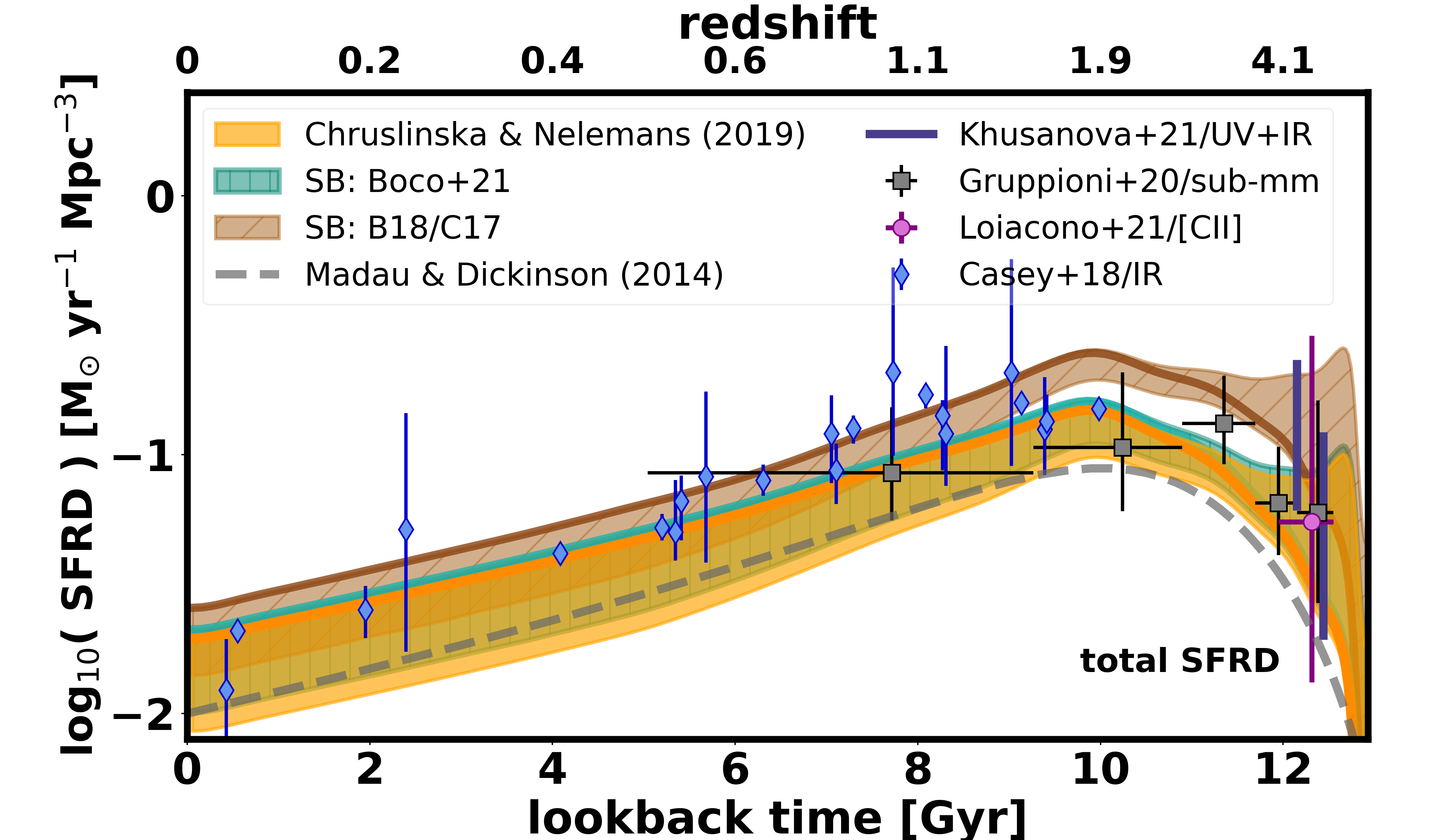}
\vspace*{-0.3cm}
\caption{
Star formation rate density as a function of lookback time/redshift.
The plotted ranges span between the estimates obtained for different variations of model assumptions about the SFMR and GSMF.
The ranges in different colours correspond to different assumptions about the starburst galaxies (turquoise: as in \citet{Boco21}, brown: based on \citet{Caputi17}-C17 and \citet{Bisigello18}-B18).
The orange region shows the corresponding range from \citet{ChruslinskaNelemans19} (with no starburst contribution).
The thick solid orange and brown lines show how the upper edge of the corresponding range would look like under the assumption of a non-evolving low mass end of the GSMF (the effect on the turquise range follows closely the one that can be seen for \citealt{ChruslinskaNelemans19}, the lower edge is not affected by this assumption).
The dashed gray line shows the estimate from \citet{MadauDickinson14} for reference.
We overplot the z$\lesssim$2 infrared-based estimates complied by \citet{Casey18} and the recent observational high redshift SFRD estimates by \citet{Loiacono21}, \citet{Khusanova20} and \citet{Gruppioni20}, that account for the contribution of heavily dust obscured galaxies. Note that the latter two provide a lower limit. See text for the discussion of the comparison with observations.
Where necessary, we correct for the difference in SFR due to varying IMF assumptions.
}
\label{fig: total SFRD}
\end{figure*}

The cosmic SFH obtained for different model variations considered in this study
 is shown in Figure \ref{fig: total SFRD}.
 Coloured ranges distinguish different assumptions about the starburst galaxies. 
 They span between the extreme cases resulting from different assumptions about the high mass end of the SFMR and the low mass end of the GSMF.
  Note that the assumptions about the metallicity ($z\sim0$ MZR, redshift evolution of MZR/non-evolving FMR, $\nabla_{\rm FMR}$) have no impact on the (total) cosmic SFH, but affect its different metallicity cuts.
  For the convenience of use and discussion, we approximate the results shown in Figure \ref{fig: total SFRD} with a broken power law: SFRD = $A$ (1+z)$^{\kappa}$, fitting $A$ and $\kappa$ in several redshift ranges for the upper and lower edge of each variation.
  The resulting coefficients are given in Table \ref{tab: total SFRD}.
  Coefficients obtained when only variations with non-evolving low mass end of the GSMF are considered are provided in Table \ref{tab: total SFRD, alpha GSMF fixed} in the Appendix \ref{app: cosmic SFH appendix}.
  \\
  Considering variations with fixed assumptions about the starburst galaxies, we note the following:
  \begin{itemize}
      \item The uncertainty in the cosmic SFH at z$<$2 is dominated by the assumptions about high mass end of the SFMR: the lower edges of all ranges correspond to variations with SFMR with sharp flattening at high masses (as in low metallicity variations), while the upper edges correspond to variations with SFMR with no flattening (as in high metallicity variations). Assumptions about the GSMF have secondary role in this redshift range.
      \item The evolution of the low mass end of the GSMF dominates the uncertainty at higher redshifts:  the lower (upper) edges of all ranges correspond to variations with non-evolving (steepening) low mass end of the GSMF. At the same time, the importance of the high mass end of the SFMR is reduced due to decreasing number density of the most massive galaxies (see e.g. Fig. 3 in \citealt{ChruslinskaNelemans19}).  
      \item The upper edges of all ranges feature an upturn in the SFRD evolution at $z\gtrsim$4, before the sharp decrease at $z>7$.
       This is due to strongly increasing number density of low mass galaxies in the variations in which the low mass end of the GSMF steepens with redshift. 
       We note that such upturn only appears when the
       SFMR and GSMF are extrapolated to log$_{10}$(M$_{*})<8$ (see discussion in \citealt{ChruslinskaNelemans19}).
      \item All variations show a sharp decrease in the SFRD at $z\sim 7$.
            This is due to the rapid evolution of the GSMF normalisation
            between $z\sim7$ and 8, seen when observational GSMF estimates from different redshifts 
            are combined (see Fig. 3 in \citealt{ChruslinskaNelemans19}).
  \end{itemize}
The first three effects are best seen by comparing the thick solid lines in Fig. \ref{fig: total SFRD} (corresponding to variations with SFMR with no flattening and GSMF with non-evolving low mass end) with the upper edges of the relevant ranges.\\
Comparing the variations including starbursts (turquoise/brown ranges) with those from \citet{ChruslinskaNelemans19} (that do no explicitly account for their contribution; orange range), one can see that:
\begin{itemize}
    \item The SFRD is increased in the variations including starbursts; starbursts implementation used in Boco et al. (2021) (turquoise) leads to negligible difference, while the B18/C17 implementation (brown) leads to a factor of $\sim$2.5 increase at high redshifts
    (see Sec. \ref{sec: results - starbursts})
    \item variations including starbursts span a narrower range in cosmic SFH
    (the lower edge is lifted with respect to no starbursts case). 
    This is due to the fact that assigning high SFR (compared to SFMR values) to a fraction of galaxies (starbursts) effectively reduces the difference between the SFRD in model variations with no/sharp SFMR flattening.
    The effect is stronger at high redshifts for the B18/C17 prescription, as f$_{SB}$ increases with redshift.
    \item the increasing fraction of starbursts in the B18/C17 variations leads to a  broader peak of the cosmic SFH and a shallower SFRD decrease at high redshifts.
\end{itemize}

In Figure \ref{fig: total SFRD} we contrast the total SFRD that results from various realisations of our observation-based model with several observational determinations of this quantity.
At z$<$2, there is about a factor of 2 (2.5) offset between the SFRD obtained with model variations that assume SFMR with no flattening at high masses (and C17/B18 starburst implementation) and the commonly used, best-fit estimate from \citet{MadauDickinson14} (gray dashed line).
However, we note that this offset is within the scatter and uncertainty of the z$\lesssim$2 observational estimates - in particular, infrared based SFRD determinations suggest higher SFRD than the estimate obtained by \citet{MadauDickinson14}, more in agreement with our SFRD determination based on the GSMF, SFMR, and starburst sequence (see also Fig. 1 in \citealt{Casey18} - the infrared-based data complied by those authors are also included in Fig. \ref{fig: total SFRD}; and Fig. 2 and discussion in \citealt{Boco21}). 
% Therefore, we cannot constrain our models based on the comparison with z$\lesssim$2 data.
\\
At higher redshifts ($2<z<7$), our model variations (even those not including starbursts) in general show shallower SFRD decrease than the estimate provided by \citet{MadauDickinson14} (SFRD$\propto (1+z)^{-2.9}$). 
This is striking for the upper edges of our estimates (see also Tab. \ref{tab: total SFRD}).
Shallower SFRD evolution than estimated by \citet{MadauDickinson14} is in line with the most recent high redshift observational estimates \citep[e.g.][although the uncertainties in those measurements are still substantial]{Khusanova20,Gruppioni20,Loiacono21},
indicating that a significant fraction of the SFRD at $z\gtrsim 2$ still occurs in dust obscured galaxies (and hence was missed by the earlier - mostly UV light-based surveys).
% However, the slow SFRD decrease in our models may happen (at least partially) for a different reason. As discussed above, we obtain much shallower SFRD evolution (or even a reversal of the decreasing trend) (i) if the abundance of low mass galaxies increases with redshift ($\alpha_{\rm GSMF}=\alpha_{\rm GSMF}(z)$) and/or (ii) if we allow for a significant contribution of starburst galaxies (f$_{SB}$ increasing with redshift).
% While massive starburst galaxies in our model may effectively belong to the population of galaxies that significantly contributes to the dust-obscured SFRD uncovered by the recent surveys, the faint (low mass) end of the galaxy population is not directly probed with those observations.
Those estimates\footnote{We plot the total SFRD estimate from \citet{Khusanova20} in which the IR contribution was obtained with GSMF rather than UV luminosity function used in the calculations, as that is closer to our approach.} are also included in Fig. \ref{fig: total SFRD}.
\\
However, the comparison of our results with those estimates is not straightforward for two main reasons:
a) observations may be incomplete: 
the estimate provided by \citet{Khusanova20} is based on a UV selected sample and may not fully account for extremely dusty galaxies, while the estimate provided by \citet{Gruppioni20} does not account for the contribution of UV sources
b)  the faint (low mass) end of the galaxy population is not directly probed by observations. All estimates shown in Fig. \ref{fig: total SFRD} include (different) extrapolations to account for their contribution to the total SFRD.
In our models we extrapolate all of the empirical relations down to log$_{10}$(M$_{*}/\Msun)=6$ - as discussed above, the result of this extrapolation at high $z$ is very sensitive to the GSMF slope.
Extrapolations used in the high redshift observational estimates give less weight to low-mass galaxies than our models with steepening low mass end of the GSMF.
The estimate by \citet{Khusanova20} includes extrapolations down to M$_{*}$/luminosity limit comparable to the extrapolation limit used in our study, but is averaged over the results obtained with different GSMF/UV luminosity functions (with different low mass end slopes) from the literature.
\citet{Gruppioni20} extrapolate the infrared luminosity function down to low luminosities.
However, such galaxies are expected to be relatively unobscured and the faint-end slope of the infrared luminosity function is much flatter than that of the UV luminosity function (and the GSMF used in our study) at high redshifts - in that sense, this estimate underestimates the contribution of low mass galaxies to the total SFRD.
Therefore, the observational estimates are likely lower limits when compared to our models.
\\
We note that all model variations with fixed low mass end of the GSMF and no/small fraction of starbursts fall below the recent $z\sim3$ limit from \citet{Gruppioni20}. 
Given the fact that this determination underestimates the contribution of low mass galaxies to the total SFRD with respect to our models, this offset may hint at the need for higher contribution of massive galaxies than included in those model variations (e.g. higher fraction of starburst galaxies or higher overall normalisation of the SFMR at those redshifts).
We also note that the z$\sim$4.5 estimate by \citet{Khusanova20} falls above our estimates obtained with fixed low mass end of the GSMF and no/small fraction of starbursts, but it is unclear how to properly compare this averaged estimate with our results. All model variations are consistent with the limits obtained by \citet{Gruppioni20} and \citet{Loiacono21} (although the latter is not particularly constraining due to large uncertainty that overlaps with all of the considered model variations) at similar redshifts.

\begin{table*}
\centering
\small
\caption{
Coefficients of the power law fits to the cosmic SFH (lower and upper edges of the ranges shown in  Fig. \ref{fig: total SFRD}) obtained for different variations of our observation-based model.
Within each redshift range indicated in the first column total SFRD can be approximated with the dependence: SFRD=A$\cdot(1+z)^{\kappa} [\Msun/yr]$.
Note that the cosmic SFH in 'no SB' variation is identical with the cosmic SFH from \citet{ChruslinskaNelemans19}.
}
\begin{tabular}{c c c c c c c }
\hline
 \multicolumn{7}{c}{ \textbf{cosmic SFH - lower edge} }\\
 variation:  & \multicolumn{2}{c}{no SB} & \multicolumn{2}{c}{SB: Boco+21} & \multicolumn{2}{c}{ SB: B18/C17} \\ 
$z$ range & $\kappa$ & A & $\kappa$ & A & $\kappa$ & A  \\ \hline
0$<z\leqslant$1 & 2.6546 & 0.0076 & 2.6093 & 0.0091 & 2.7828 & 0.0123 \\ \hline
1$<z\leqslant$1.8 & 2.4050 & 0.0091 & 2.3416 & 0.0109 & 2.7119 & 0.0129 \\ \hline
1.8$<z\leqslant$3 & -1.0535 & 0.3195 & -1.1284 & 0.3891 & -0.7558 & 0.4585 \\ \hline
3$<z\leqslant$7 & -2.2771 & 1.7427 & -2.2909& 1.9499 & -1.9628 & 2.4435\\ \hline
7$<z\leqslant$8.8 & -12.5425 & 3.249$\cdot10^{9}$ & -12.5526 & 3.607$\cdot10^{9}$ & -12.4895 & 7.845$\cdot10^{9}$\\ \hline
8.8$<z\leqslant$10. & 0 & 0.0012 & 0 & 0.0013 & 0 & 0.0033 \\ \hline
\hline
 \multicolumn{7}{c}{ \textbf{cosmic SFH - upper edge} }\\
  variation & \multicolumn{2}{c}{no SB} & \multicolumn{2}{c}{SB: Boco+21} & \multicolumn{2}{c}{ SB: B18/C17} \\ 
$z$ range & $\kappa$ & A & $\kappa$ & A & $\kappa$ & A  \\ \hline
0$<z\leqslant$1 & 2.2214 & 0.0194 & 2.2222 & 0.0209 & 2.3850 & 0.0248 \\ \hline
1$<z\leqslant$1.8 & 1.8708 & 0.0247 & 1.8708 & 0.0266 & 2.2515 & 0.0272 \\ \hline
1.8$<z\leqslant$4 & -1.2436 & 0.6097 & -1.2445 & 0.6585 & -0.5342 & 0.4793 \\ \hline
4$<z\leqslant$7 & 0.6591 & 0.0285 & 0.6585 & 0.0307 & 0.7477 & 0.0609 \\ \hline
7$<z\leqslant$8.8 & -9.3760 & 3.29$\cdot10^{7}$ & -9.3864 & 3.624$\cdot10^{7}$ & -9.4740 & 1.037$\cdot10^{8}$ \\ \hline
8.8$<z\leqslant$10. & 0 & 0.0168 & 0 & 0.018 & 0 & 0.0421 \\ \hline
\hline
\end{tabular}
\label{tab: total SFRD}
\end{table*}

\subsection{A low metallicity cut of the cosmic SFH}\label{sec: results - low metallicity SFH}

\subsubsection{Note on the low metallicity threshold}
When referring to low metallicity, throughout this paper we use a limit of $<$10\% solar oxygen abundance (with $Z_{O/H\odot}$ following \citealt{GrevesseSauval98}) and the numbers quoted later in this Section necessarily depend on this choice.
This threshold is chosen arbitrarily and serves for illustrative purposes.
In general, the relevant definition of low metallicity varies depending on the discussed problem and application.
For instance, the abrupt drop commonly seen in the metallicity dependence of the formation efficiency of black hole - black hole mergers formed in the course of isolated binary evolution appears at metallicities 10\% - 50\% solar (depending on the considered model of binary/stellar evolution, e.g. \citealt{ChruslinskaNelemansBelczynski19}, Broekgaarden et al. (2021) in prep.). 
In that sense, depending on the model, the interesting low metallicity threshold could be as high as 50\% solar metallicity.
Furthermore, from the point of view of evolution of massive stars and the related transients, the most relevant metallicity tracer is iron rather than oxygen (which serves as a metallicity proxy in this study; the oxygen abundance is the most easily obtained from emission line measurements and so it is the most commonly used metallicity proxy in studies of interstellar medium of galaxies).
That is primarily due to high sensitivity of the radiation-driven mass loss rates of hot, massive stars to the abundance of this element \citep[e.g.][]{Vink01, VinkdeKoter05, GrafenerHamann08, Sanders20}.
In general, interstellar medium (ISM) of galaxies is enriched in oxygen and iron on different timescales due to the fact that different sources dominate the production of those elements (core-collapse supernovae with $\lesssim$10 Myr delay with respect to star formation for oxygen and SNIa supernovae with $\sim$0.1-1 Gyr delay for iron; e.g. \citealt{Wheeler89}).
This means that especially young, highly star-forming galaxies (i.e. typical at high redshifts) are expected to show overabundance of oxygen relative to iron with respect to solar ratios. 
Recent studies indeed find evidence for super-solar O/Fe$\gtrsim$2 (O/Fe)$|_{\odot}$ in the ISM of $z\sim 2$ star forming galaxies \citep[e.g.][]{Steidel16,Topping20,Cullen21}.
Consequently, oxygen (or, more generally, $\alpha$-element)-enhanced abundance ratios are also found in local stars that formed at earlier cosmic epochs \citep[e.g.][]{Bensby04,Tolstoy09}.
There is no universal way to translate metallicity measure based on oxygen abundance to iron abundance, in particular the typical oxygen to iron ratio in the ISM (and the interesting low metallicity threshold - if oxygen is used as a proxy) can be expected to vary with redshift and possibly galaxy characteristics. A more thorough discussion of this issue is beyond the scope of this work and is left to future studies.

\subsubsection{Results}

\begin{figure*}
\vspace*{-0.5cm}
\includegraphics[width=1.7\columnwidth]{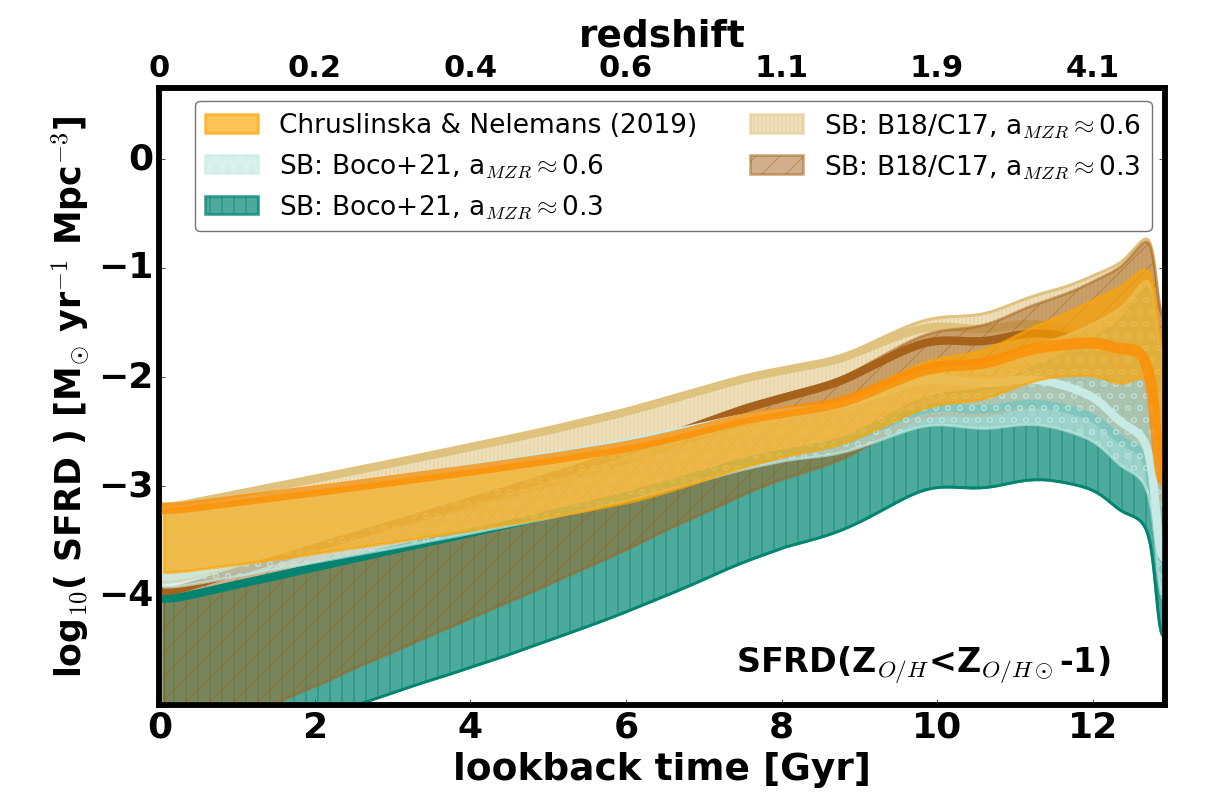}
\vspace*{-0.3cm}
\caption{
Star formation rate density at low metallicity (i.e. below 10\% solar, assuming solar oxygen abundance Z$_{O/H\odot}$=8.83 \citet{GrevesseSauval98}) as a function of lookback time/redshift.
The plotted ranges show estimates obtained for different variations of assumptions about the starburst  (SB) galaxies  (blue/turquoise: as in \citet{Boco21}, brown/beige: based on \citet{Caputi17}-C17 and \citet{Bisigello18}-B18) and low mass slope of the redshift $z\sim0$ MZR ($a_{\rm MZR}$). 
The orange region shows the range of the low metallicity SFHs obtained by \citet{ChruslinskaNelemans19}.
Thick solid lines in different colours indicate the location of the upper edge of the corresponding ranges obtained with the assumption of non-evolving low mass end of the GSMF, demonstrating the importance of this assumption beyond the peak of the cosmic SFH.
All variations from this study assume $\nabla_{\rm FMR0}=0.27$.
}
\label{fig: low Z SFRD}
\end{figure*}

SFRD that occurs at Z$_{O/H}<$Z$_{O/H\odot}$-1 as a function of redshift/time for different variations of our model is shown in Figure \ref{fig: low Z SFRD}.
% \textbf{We stress that the numbers quoted in this Section are sensitive to the considered low metallicity threshold - especially in the high metallicity variations at z$<$2, as by considering 10\% solar metallicity cut we zoom into the very tail of f$_{\rm SFR}$(Z,z).
% }
Note that the assumptions about the high mass end of the SFMR have no impact on the low metallicity tail of f$_{SFR}$(Z,z) and the low metallicity SFH.
Similarly as for the cosmic SFH, we provide the power law fits to the upper and lower edges of the ranges shown in Figure \ref{fig: low Z SFRD} fitted in several $z$ ranges. Those are listed in Tab. \ref{tab: low Z SFH GSMF fixed} and Tab. \ref{tab: low Z SFH} in the Appendix \ref{app: cosmic SFH appendix}.
The following features are common to all model variations:
\begin{itemize}
\item At  $z\lesssim 2$ the SFRD occurring at low metallicity increases faster than the total SFRD. The exact slope depends on the variation.
 In this redshift range the width of the ranges is set by the 
assumptions about the $z\sim0$ MZR:
 lower/upper edges of all ranges correspond to variations with high/low MZR normalisation (as in the high/low metallicity variations). 
 At those redshifts, assumptions about
 the low mass end of the GSMF have negligible impact 
 on the amount of star formation happening at low metallicity.
\item Beyond the peak of the cosmic SFH assumptions about the low mass end of the GSMF set the width of the ranges and the behaviour of the upper edges.
The lower/upper edges correspond to variations with high/low MZR normalisation and non-evolving/steepening GSMF low mass end.
\item If $\alpha_{GSMF}$ is fixed, the low metallicity cut of the SFH shows a broad peak between z=2-4 (depending on the variation)
 and a relatively mild evolution up to z$\sim$7.
 \item If $\alpha_{GSMF}$ evolves with redshift, the increasing abundance of low mass galaxies (forming stars at low metallicity)
 causes a continuous increase in the low metallicity cut of the SFH up to z$\sim$7.5-8.
\item All variations show a sharp decrease in the low metallicity SFRD at z$\gtrsim$8, reflecting the sharp drop in the total SFRD at those redshifts.   
\end{itemize}
The impact of the assumed $z\sim0$ MZR slope can be seen by comparing
pale blue and turquoise or beige and brown ranges in Figure \ref{fig: low Z SFRD}.
At low redshifts, variations with steeper MZR slope lead to $\sim$10 times higher SFRD at low metallicity.
This is due to the presence of a more extended low metallicity tail of the f$_{SFR}$(Z,z) (see Sec. \ref{sec: results - MZR slope and FMR slope}). 
Lower $a_{\rm MZR}$ lead to more compact f$_{SFR}$(Z,z) - in those cases, for the high metallicity variations 
even the low mass galaxies contribute to SFRD at Z$_{O/H}$ above the considered threshold of 10\% solar metallicity.
This difference is reduced at higher redshifts due to steeper metallicity evolution of variations with lower $a_{\rm MZR}$ values - the slope of the low metallicity SFH redshift dependence at $z\lesssim2$ depends on the $z\sim0$ $a_{\rm MZR}$. 
\\
The slope of the dependence is also affected by the contribution of starbursts (compare pale blue and beige or turquoise and brown ranges).
As expected, if the fraction of starbursts increases with redshift (B18/C17 implementation), the low metallicity cut of the SFH increases faster towards high redshifts.
At $z\lesssim$0.2 (within the last $\sim$2 Gyr) there is little difference in the amount of SFRD occurring at low metallicity  between the variations with no/low fraction of starbursts and the B18/C17 implementation. The contribution of starbursts only becomes evident at higher redshifts. At the peak of the cosmic SFH variations with the B18/C17 starbursts implementation lead to $\gtrsim$3 times higher SFRD at Z$_{O/H}<$Z$_{O/H\odot}$-1 than the corresponding variations with no/low fraction of starbursts
\\
As discussed above, at $z\gtrsim2$ the slope of the redshift dependence of the low metallicity SFRD and the location of the peak of this dependence is dominated by the assumptions about $\alpha_{GSMF}$.
Observational tracers of star formation happening at low metallicity available at high redshifts (e.g. the volumetric rate of long gamma ray bursts or, potentially, that of black hole -- black hole mergers as a function of redshift) could help to discriminate between the different variations of our models.
In particular, it could provide information about the contribution of low mass galaxies to the total SFRD at high redshifts (thereby constraining jointly the low mass SFMR and GSMF) - difficult to constrain with regular galaxy surveys.
\\
\newline
Considering all model variations discussed in this section,
the overall spread between the estimates of the 
SFRD happening below Z$_{O/H}<$Z$_{O/H\odot}$-1 ranges between $\sim$1.6 dex around the peak of the cosmic SFH and $\sim$3 dex at $z>$8.
At $z\sim$0 there is $\sim$2.3 dex variation between the different estimates.
The uncertainty is dominated by the assumptions about $a_{\rm MZR}$.
At the peak of the cosmic SFH 
the spread is somewhat reduced due to steeper metallicity evolution in variations with lower $a_{\rm MZR}$.
At $z\sim$8 the spread increases to $\sim$2.5 dex if the low mass end of the GSMF is allowed to steepen with redshift, and remains close to $\sim$1.6 dex
if only variations with fixed $\alpha_{\rm GSMF}$ are considered.
\\
Note that the overall spread between the low metallicity SFH estimates discussed in this study is significantly larger than $\sim$0.5-1.2 dex found in \citet{ChruslinskaNelemans19} for the same low metallicity threshold.
In their study, effectively only the variations due to the uncertain MZR normalisation and the redshift evolution of the low mass end of the GSMF (which are also included in our analysis) were explored.
This difference highlights the importance of the additional factors discussed in this study for the low metallicity cut of the cosmic SFH.\\
Note that the spread in the low metallicity SFH estimates as discussed above is not only due to differences in the metallicity distribution of the SFRD at different redshifts found for the considered model variations, but also due to their different total SFRD.
The latter differences are effectively eliminated in Fig. \ref{fig: low Z SFRD - fraction} in Appendix \ref{app: cosmic SFH appendix}, where we additionally show the fraction of the total SFRD that happens at $Z_{O/H}<Z_{O/H\odot}$-1 as a function of redshift for the discussed model variations.
\\
\newline
Finally, we note that the low metallicity part of the SFH is very sensitive to the adopted log$_{10}$(M$_{*}$) limit down to which all the empirical relations are extrapolated. This is especially important at $z\lesssim$1, where the low metallicity star formation is limited to low mass galaxies, and at z$\gtrsim$4, due to the potential steepening of the low mass end of the GSMF.
With the current data there is no guarantee that those relations can be safely extrapolated to galaxy masses below log$_{10}$(M$_{*}/\Msun$)$\lesssim$8.
However, given their abundance, completely neglecting the contribution of those galaxies is not a satisfactory solution either.
In variations favoring high metallicity (i.e. with high MZR normalisation, shallow MZR slope, no/low contribution of starburst galaxies)
ignoring the contribution of galaxies with log$_{10}$(M$_{*}/\Msun$)$<$8
would lead to negligible amount of SFRD happening below 10\% solar metallicity even beyond the peak of the cosmic SFH.
If the SFR-metallicity correlation significantly weakens/disappears at low specific SFR (see additional model variation discussed in Appendix \ref{app: SFR dependent nabla} and Fig. \ref{fig: lowZ cut varN}), this may also significantly reduce the estimated amount of SFRD happening below 10\% solar metallicity in variations with relatively flat MZR. In those cases, the low/intermediate redshift star formation can only take place below 10\% solar metallicity in low mass galaxies that are MZR outliers (see e.g. left bottom panel in Fig. \ref{fig: FMR MZR choice} and in Fig. \ref{fig: FMR varN}), and so the result strongly depends on the adopted FMR extrapolation.
Until better constraints on the properties of the low mass end the star forming galaxy population are available, those extrapolations are important source of uncertainty for both the total cosmic SFH, as well as for its low metallicity cut.

\subsection{A high metallicity cut of the cosmic SFH}\label{sec: results - high metallicity SFH}

\begin{figure*}
\vspace*{-0.5cm}
\includegraphics[width=1.7\columnwidth]{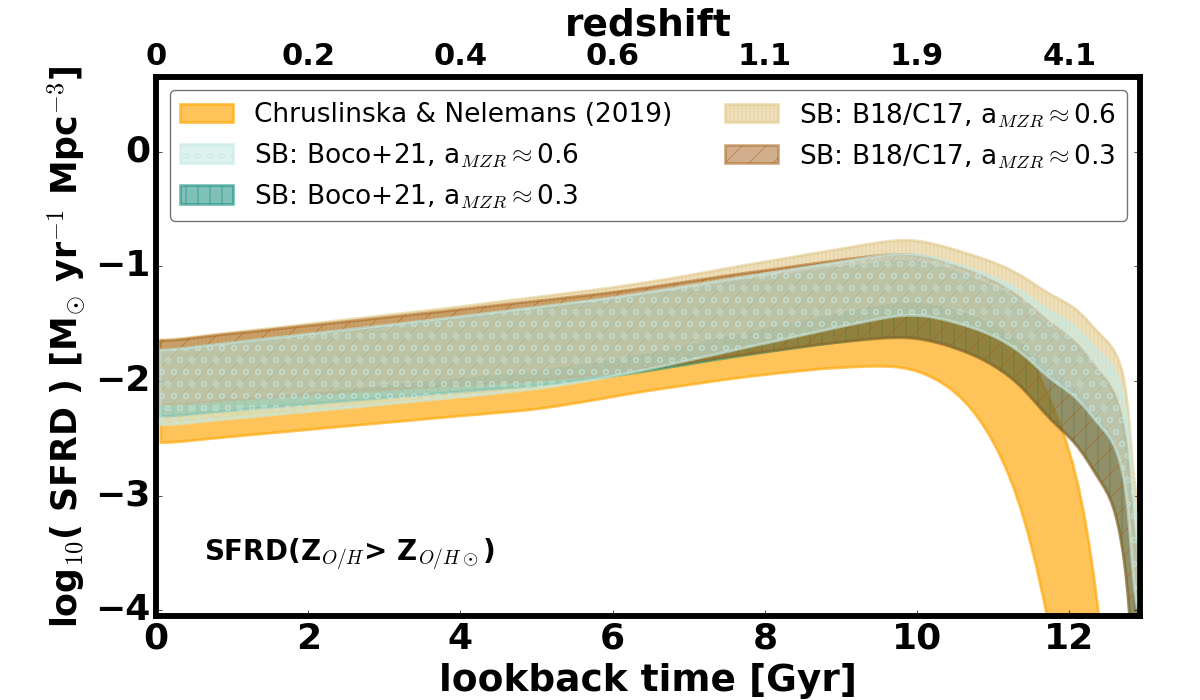}
\vspace*{-0.3cm}
\caption{
Star formation rate density at high metallicity (above solar, assuming Z$_{O/H\odot}$=8.83 \citet{GrevesseSauval98}) as a function of lookback time/redshift.
The plotted ranges show estimates obtained for different variations of assumptions about the starburst galaxies (blue/turquoise: as in \citet{Boco21}, brown/beige: based on \citet{Caputi17}-C17 and \citet{Bisigello18}-B18) and low mass slope of the redshift $z\sim0$ MZR ($a_{\rm MZR}$). The ranges span between the low and high metallicity extremes as defined in the text
and assume $\nabla_{\rm FMR0}=0.27$.
The orange region shows the corresponding range from \citet{ChruslinskaNelemans19}.
Note that starburst galaxies contribute to SFRD at relatively low metallicity and have little impact on the high metallicity cut of the SFRD.
}
\label{fig: high Z SFRD}
\end{figure*}

The high metallicity cut of the cosmic SFH - i.e. SFRD that occurs at Z$_{O/H}>$Z$_{O/H\odot}$ as a function of redshift/time for different variations of our model is shown in Fig. \ref{fig: high Z SFRD}.
This quantity is relatively well constrained in our models when compared to the low metallicity cut of the SFH discussed in the previous section:
the overall spread between the estimates obtained with different model variations from this study ranges from $\sim$0.8 dex at $z\sim0$ to $\sim$1.8 dex at $z\gtrsim$7.
The biggest difference appears between the estimate from \citealt{ChruslinskaNelemans19} (orange range)
and all the variations from this study, which lead to considerably higher SFRD occurring at high metallicity beyond the peak of the cosmic SFH.
As discussed in Sec. \ref{sec: results FMR effect}, this is due to the fact that the non-evolving FMR assumption 
leads to much slower metallicity evolution when compared to that implied by the extrapolated redshift dependent MZR.
The general features that can be seen for all  ranges shown in Fig. \ref{fig: high Z SFRD} can be summarised as follows:
\begin{itemize}
    \item The high metallicity cut of the SFH shows a gradual increase towards the peak of the SFH ($z\sim2$) and a sharp decrease at high redshifts. This reflects the evolution of the total SFRD.
   However, the increase towards the peak is shallower and the later decrease is sharper -  both due to the overall shifting of the SFRD towards lower metallicities at earler epochs.
   \item The width of the ranges is set by the assumptions about the SFMR and $z\sim0$ MZR: upper edges of all ranges correspond to high metallicity variations, lower edges of all ranges correspond to low metallicity variations.
   Assumptions about the low mass end of the GSMF have negligible effect on the high metallicity cut of the cosmic SFH.
\end{itemize}
Comparing the ranges obtained for different assumptions about the MZR slope (brown and beige or pale blue and turquoise), one can see that the results start to deviate at $z\gtrsim 1$, where variations with steeper MZR lead to higher SFRD at high metallicity. 
This is due to slower evolution of the metallicity distribution at $z\lesssim 4$ for those variations, as discussed in Sec. \ref{sec: results - MZR slope and FMR slope}.
Assumptions about starbursts have little effect on the Z$_{O/H}>$Z$_{O/H\odot}$ metallicity cut of the cosmic SFH (compare light blue/turquoise and beige/brown ranges).
This is due to the fact that (given the FMR), those galaxies only contribute to star formation at relatively low metallicities.
\\
The fraction of the total SFRD that happens at $Z_{O/H}>Z_{O/H\odot}$ as a function of redshift for the discussed model variations is shown
in Fig. \ref{fig: high Z SFRD - fraction} in Appendix \ref{app: cosmic SFH appendix}.

\subsection{The impact of $\nabla_{\rm FMR0}$ on the low/high metallicity cut of the cosmic SFH}

\begin{figure*}
\vspace*{-0.5cm}
\includegraphics[width=1.7\columnwidth]{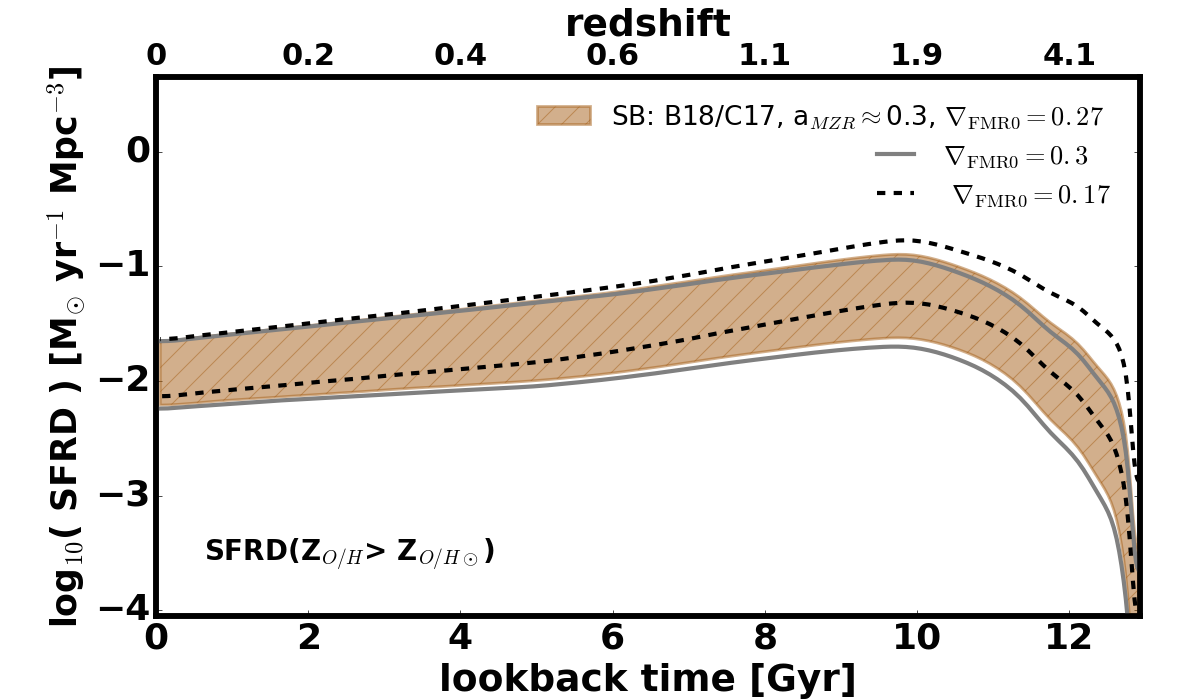}
\vspace*{-0.3cm}
\caption{
Example illustrating the effect of $\nabla_{\rm FMR0}$ on the high metallicity cut of the cosmic SFH.
The plotted range shows estimate obtained with B18/C17 starbursts implementation and $a_{\rm MZR}\sim$0.3, assuming $\nabla_{\rm FMR0}$=0.27 (same as in Fig. \ref{fig: high Z SFRD}).
Gray solid (black dashed) lines mark the location of the upper and lower edges
of that range obtained with  $\nabla_{\rm FMR0}$=0.3 ( $\nabla_{\rm FMR0}$=0.17).
}
\label{fig: high Z SFRD - nabla effect}
\end{figure*}

\begin{figure*}
\vspace*{-0.5cm}
\includegraphics[width=1.7\columnwidth]{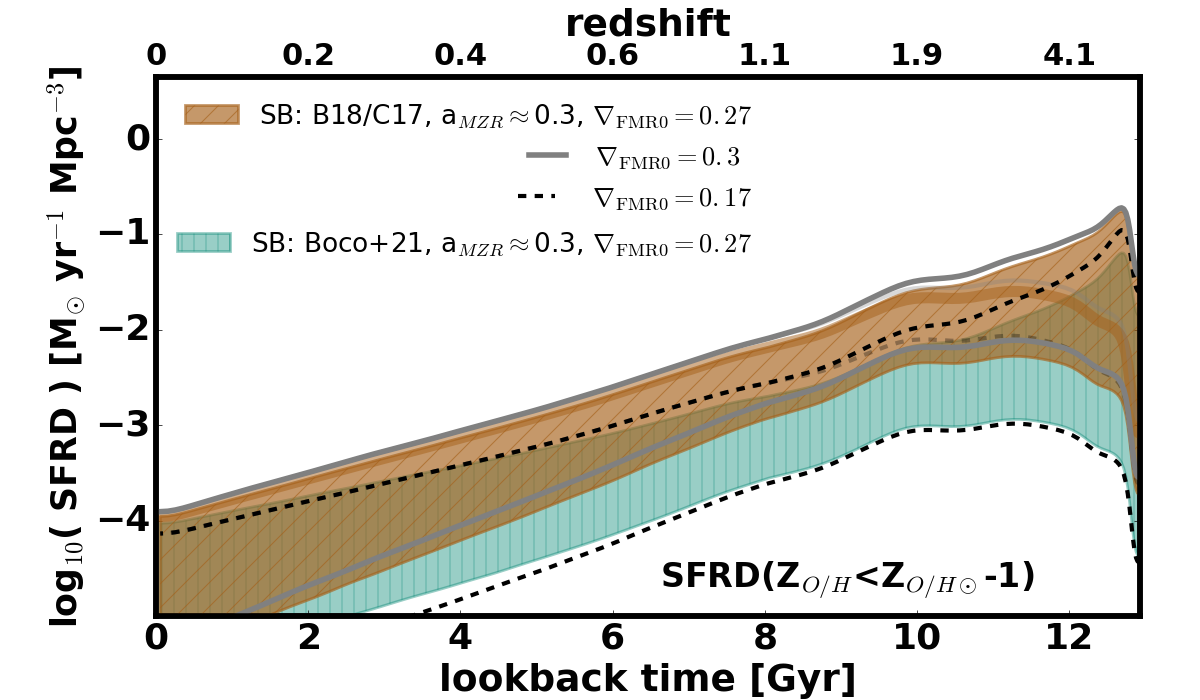}
\vspace*{-0.3cm}
\caption{
Example illustrating the effect of $\nabla_{\rm FMR0}$ on the low metallicity cut of the cosmic SFH.
The brown range shows estimate obtained with B18/C17 starbursts implementation and $a_{\rm MZR}\sim$0.3, assuming $\nabla_{\rm FMR0}$=0.27 (same as in Fig. \ref{fig: low Z SFRD}).
Gray solid (black dashed) lines mark the location of the upper and lower edges
of that range obtained with  $\nabla_{\rm FMR0}$=0.3 ( $\nabla_{\rm FMR0}$=0.17).
Faint lines in the corresponding colours show the evolution of the upper edge under the assumption of  non-evolving low mass end of the GSMF.
For comparison, we also show the range obtained with Boco et al. (2021) starbursts implementation, $a_{\rm MZR}\sim$0.3 and $\nabla_{\rm FMR0}$=0.27 (turquoise, same as in Fig. \ref{fig: low Z SFRD}). If the SFR-metallicity correlation is much weaker than implied by our fiducial $\nabla_{\rm FMR0}$=0.27, the impact of starbursts on the low metallicity cut of the SFH is substantially reduced.
}
\label{fig: low Z SFRD - nabla effect}
\end{figure*}

As discussed in Sec. \ref{sec: FMR - nabla}, the range of extreme reasonable assumptions about $\nabla_{\rm FMR0}$ and the potential dependence of this parameter for instance, on the metallicity derivation technique, cannot be easily identified with the currently available observational results. We therefore discuss the sensitivity of our result to the choice of this factor.
The impact of  $\nabla_{\rm FMR0}$ on the high and low metallicity cut of the cosmic SFH for the example model variations is shown in Fig. \ref{fig: high Z SFRD - nabla effect} and Fig. \ref{fig: low Z SFRD - nabla effect}, respectively.
In essence, higher (lower) $\nabla_{\rm FMR0}$ increase (decrease) the amount of SFRD that happens at low metallicity (and vice versa for the SFRD at high metallicity).
Lower $\nabla_{\rm FMR0}$ means that the difference between the metallicities of highly star-forming galaxies and their quiescent counterparts of the same mass is reduced.
As a consequence, if the anti-correlation between the SFR and metallicity is much weaker than with our fiducial assumptions, 
the impact of starbursts on the low metallicity cut of the SFH is substancially reduced. 
This can be seen in Fig. \ref{fig: low Z SFRD - nabla effect}, where the lower and upper edges of the range obtained with B18/C17 starburst implementation and $\nabla_{\rm FMR0}$=0.17 (dashed lines)
roughly match the range obtained for the corresponding model variations assuming low fraction of starbursts (turquoise range).
This also means that the impact of the choice of $\nabla_{\rm FMR0}$ on the variations with lower fraction  of starbursts and/or starburst sequence with smaller offset from the SFMR than in B18/C17 implementation, is smaller than in the examples shown in Fig. \ref{fig: low Z SFRD - nabla effect} and \ref{fig: high Z SFRD - nabla effect}.
As discussed in Sec. \ref{sec: results - MZR slope and FMR slope},  $\nabla_{\rm FMR0}$ also affects the rate of the overall decrease in metallicity - and so the slope of the different metallicity cuts of the cosmic SFH as a function of redshift.
However, while many different factors affect the slope of the low metallicity cut of the SFH below the peak of the cosmic SFH (see Sec. \ref{sec: results - low metallicity SFH}), at high redshifts assumptions about the low mass end of the GSMF clearly dominate the uncertainty in the slope regardless of $\nabla_{\rm FMR0}$.

\section{Conclusions}

In this study we expand the framework introduced in \citet{ChruslinskaNelemans19} to 
construct the observation-based distribution of the cosmic SFRD at different metallicities and redshifts.
We evaluate its uncertainty in light of additional factors that were not discussed in the original study:
the redshift-invariant mass-metallicity-SFR relation (FMR), starburst galaxies and the slope of the local mass-metallicity relation.
We show that all of those factors have a strong impact on the low metallicity cut of the cosmic SFH -
crucial in the discussion of the origin of various energetic transients.
\\
Similarly to \citet{ChruslinskaNelemans19}, we consider different MZR and SFMR variations to cover the range of possibilities discussed in the literature and explore the realistic extremes of the f$_{SFR}$(Z,z).
The diversity of observational derivations of the MZR and SFMR stems from
 differences in the applied metallicity and SFR determination techniques - those differences in the methods also affect the FMR determinations.
To ensue the consistency of our method, we develop a phenomenological model to construct the FMR based on the local MZR and SFMR.
We introduce a parameter $\nabla_{\rm FMR0}$ to characterise the strength of the SFR-metallicity correlation in the high SFR part of the FMR. That is the only parameter which is not constrained by the choice of $z\sim0$ MZR and SFMR within our description and we discuss the sensitivity of our results to this factor.
Our description captures all characteristic features of the observationally inferred relation.
It reproduces the known systematic differences between the FMR derived with various metallicity and SFR determination techniques
and predicts dependence of the properties of the FMR on the parameters of the $z\sim0$ MZR and SFMR that can be verified with future observational studies.
\\
Furthermore, we use the results of \citet{Caputi17} and \citet{Bisigello18} to describe the properties of starburst galaxies 
and include their contribution in our f$_{SFR}$(Z,z) models. 
We contrast the f$_{SFR}$(Z,z) derived that way with the f$_{SFR}$(Z,z) obtained with the description of starbursts as recently used in \citet{Boco21} (based on \citet{Sargent12} and \citet{Bethermin12}, which predict a much smaller contribution of strabursts to the total cosmic SFRD budget).
In addition to full f$_{SFR}$(Z,z) distributions obtained for different variations of our model, we discuss the resulting cosmic SFH and its low and high metallicity cuts.
\\
Our main conclusions are the following:
\begin{itemize}
\item As recently discussed in \citet{Boco21}, 
     a redshift-invariant FMR can lead to drastically different conclusions about the metallicity at which the star formation occurs at $z\gtrsim$3 than the redshift-dependent MZR extrapolated to high redshifts (as used in \citet{ChruslinskaNelemans19}).
     However, while the high metallicity cut of the SFH at $z\gtrsim$3.5 is strongly affected by this assumption (Fig. \ref{fig: high Z SFRD}), 
     the low metallicity SFH estimate from \citet{ChruslinskaNelemans19} falls within the range covered by the variations explored in this study (Fig. \ref{fig: low Z SFRD}).
     In that case, the much shallower metallicity evolution implied by the non-evolving FMR is counterbalanced by the low metallicity contribution of starbursts (assuming that they follow the same FMR as regular star forming galaxies).
\item The low metallicity tail of the f$_{SFR}$(Z,z) is the most sensitive to variations in the model assumptions.
       We find a factor of $\sim$200 ($\sim$40) difference in the amount of SFRD occurring below 10\% solar metallicity
       at $z\sim$0 ($z\sim$2) between the extreme model variations considered in this study.
\item  At low redshifts, the uncertainty of the low metallicity SFH 
      is dominated by the assumptions about the slope and normalisation of the local MZR, while both MZR and starbursts strongly affect this quantity at $z>1$.
       However, the importance of starbursts for the low metallicity cut of the SFH is highly sensitive to the assumed strength of the SFR-metallicity correlation ($\nabla_{\rm FMR0}$).
\item  Beyond the peak of the cosmic SFH, the shape of the low metallicity cut of the SFH (peak location, slope)
       is set by the assumptions about the low mass end of the GSMF.
       Those assumptions also dominate the redshift dependence of the total SFRD at $z\gtrsim$3.
       Given this sensitivity, we propose that a tracer of the SFR happening at low metallicity available at high redshifts (e.g. volumetric rate of long gamma ray bursts, or potentially - stellar black hole mergers)
       could provide constraints on the contribution of low mass galaxies -difficult to constrain otherwise- to the SFRD at those redshifts.
\item  The uncertainty in the total SFRD in our models at $z\lesssim1$ is dominated by the uncertain shape of the high mass end of the SFMR. 
   Assumptions about the contribution of starbursts dominate the uncertainty of the cosmic SFH at $1<z\lesssim 3$.
        Model variations with SFMR with sharp flattening at high masses, non-evolving low mass slope of the GSMF and no/small fraction of starbursts may underpredict the total SFRD
        % Model variations with no/small fraction of starbursts may underpredict the total SFRD 
        at $z\sim$3.
        Model variations which account for the contribution of starbursts with B18/C17 based prescription
        lead to $\sim$2.5 times higher SFRD at $z\gtrsim$4 and shallower redshift evolution of the total SFRD at $z\gtrsim$2,
        more in line with the recent SFRD determinations at high redshifts.
        
\end{itemize}

\section*{Acknowledgements}
We  thank  Karina Caputi  and  Jarle Brinchmann  for  helpful discussions. 
We thank the anonymous referee for their thoughtful comments that helped to improve this paper.
MC  and  GN  acknowledge  support  from the  Netherlands  Organisation  for  Scientific  Research  (NWO).
AL and LB are supported by PRIN MIUR 2017 prot. 20173ML3WW,
"Opening the ALMA window on the cosmic evolution of gas, stars and
supermassive black holes", and by the EU H2020-MSCA-ITN-2019
Project 860744 "BiD4BESt: Big Data applications for black hole Evolution STudies."

%%%%%%%%%%%%%%%%%%%%%%%%%%%%%%%%%%%%%%%%%%%%%%%%%%
\section*{Data Availability}
The results of our calculations (allowing to reproduce all of the results shown in the paper) for all model variations discussed in this study will be made publicly available under this url: \url{https://ftp.science.ru.nl/astro/mchruslinska} upon the acceptance of this manuscript.
 
% The inclusion of a Data Availability Statement is a requirement for articles published in MNRAS. Data Availability Statements provide a standardised format for readers to understand the availability of data underlying the research results described in the article. The statement may refer to original data generated in the course of the study or to third-party data analysed in the article. The statement should describe and provide means of access, where possible, by linking to the data or providing the required accession numbers for the relevant databases or DOIs.



\bibliographystyle{mnras}
\bibliography{example} % if your bibtex file is called example.bib

\begin{thebibliography}{}
\makeatletter
\relax
\def\mn@urlcharsother{\let\do\@makeother \do\$\do\&\do\#\do\^\do\_\do\%\do\~}
\def\mn@doi{\begingroup\mn@urlcharsother \@ifnextchar [ {\mn@doi@}
  {\mn@doi@[]}}
\def\mn@doi@[#1]#2{\def\@tempa{#1}\ifx\@tempa\@empty \href
  {http://dx.doi.org/#2} {doi:#2}\else \href {http://dx.doi.org/#2} {#1}\fi
  \endgroup}
\def\mn@eprint#1#2{\mn@eprint@#1:#2::\@nil}
\def\mn@eprint@arXiv#1{\href {http://arxiv.org/abs/#1} {{\tt arXiv:#1}}}
\def\mn@eprint@dblp#1{\href {http://dblp.uni-trier.de/rec/bibtex/#1.xml}
  {dblp:#1}}
\def\mn@eprint@#1:#2:#3:#4\@nil{\def\@tempa {#1}\def\@tempb {#2}\def\@tempc
  {#3}\ifx \@tempc \@empty \let \@tempc \@tempb \let \@tempb \@tempa \fi \ifx
  \@tempb \@empty \def\@tempb {arXiv}\fi \@ifundefined
  {mn@eprint@\@tempb}{\@tempb:\@tempc}{\expandafter \expandafter \csname
  mn@eprint@\@tempb\endcsname \expandafter{\@tempc}}}

\bibitem[\protect\citeauthoryear{{Andrews} \& {Martini}}{{Andrews} \&
  {Martini}}{2013}]{AndrewsMartini13}
{Andrews} B.~H.,  {Martini} P.,  2013, \mn@doi [\apj]
  {10.1088/0004-637X/765/2/140}, \href
  {https://ui.adsabs.harvard.edu/abs/2013ApJ...765..140A} {765, 140}

\bibitem[\protect\citeauthoryear{{Belczynski}, {Holz}, {Bulik}  \&
  {O'Shaughnessy}}{{Belczynski} et~al.}{2016}]{Belczynski16}
{Belczynski} K.,  {Holz} D.~E.,  {Bulik} T.,   {O'Shaughnessy} R.,  2016,
  \mn@doi [\nat] {10.1038/nature18322}, \href
  {https://ui.adsabs.harvard.edu/abs/2016Natur.534..512B} {534, 512}

\bibitem[\protect\citeauthoryear{{Bensby}, {Feltzing}  \&
  {Lundstr{\"o}m}}{{Bensby} et~al.}{2004}]{Bensby04}
{Bensby} T.,  {Feltzing} S.,   {Lundstr{\"o}m} I.,  2004, \mn@doi [\aap]
  {10.1051/0004-6361:20031655}, \href
  {https://ui.adsabs.harvard.edu/abs/2004A&A...415..155B} {415, 155}

\bibitem[\protect\citeauthoryear{{B{\'e}thermin} et~al.,}{{B{\'e}thermin}
  et~al.}{2012}]{Bethermin12}
{B{\'e}thermin} M.,  et~al., 2012, \mn@doi [\apjl]
  {10.1088/2041-8205/757/2/L23}, \href
  {https://ui.adsabs.harvard.edu/abs/2012ApJ...757L..23B} {757, L23}

\bibitem[\protect\citeauthoryear{{Bisigello}, {Caputi}, {Grogin}  \&
  {Koekemoer}}{{Bisigello} et~al.}{2018}]{Bisigello18}
{Bisigello} L.,  {Caputi} K.~I.,  {Grogin} N.,   {Koekemoer} A.,  2018, \mn@doi
  [\aap] {10.1051/0004-6361/201731399}, \href
  {https://ui.adsabs.harvard.edu/abs/2018A&A...609A..82B} {609, A82}

\bibitem[\protect\citeauthoryear{{Boco}, {Lapi}, {Goswami}, {Perrotta},
  {Baccigalupi}  \& {Danese}}{{Boco} et~al.}{2019}]{Boco19}
{Boco} L.,  {Lapi} A.,  {Goswami} S.,  {Perrotta} F.,  {Baccigalupi} C.,
  {Danese} L.,  2019, \mn@doi [\apj] {10.3847/1538-4357/ab328e}, \href
  {https://ui.adsabs.harvard.edu/abs/2019ApJ...881..157B} {881, 157}

\bibitem[\protect\citeauthoryear{{Boco}, {Lapi}, {Chruslinska}, {Donevski},
  {Sicilia}  \& {Danese}}{{Boco} et~al.}{2021}]{Boco21}
{Boco} L.,  {Lapi} A.,  {Chruslinska} M.,  {Donevski} D.,  {Sicilia} A.,
  {Danese} L.,  2021, \mn@doi [\apj] {10.3847/1538-4357/abd3a0}, \href
  {https://ui.adsabs.harvard.edu/abs/2021ApJ...907..110B} {907, 110}

\bibitem[\protect\citeauthoryear{{Boogaard} et~al.,}{{Boogaard}
  et~al.}{2018}]{Boogaard18}
{Boogaard} L.~A.,  et~al., 2018, \mn@doi [\aap] {10.1051/0004-6361/201833136},
  \href {https://ui.adsabs.harvard.edu/abs/2018A&A...619A..27B} {619, A27}

\bibitem[\protect\citeauthoryear{{Bothwell}, {Maiolino}, {Kennicutt}, {Cresci},
  {Mannucci}, {Marconi}  \& {Cicone}}{{Bothwell} et~al.}{2013}]{Bothwell13}
{Bothwell} M.~S.,  {Maiolino} R.,  {Kennicutt} R.,  {Cresci} G.,  {Mannucci}
  F.,  {Marconi} A.,   {Cicone} C.,  2013, \mn@doi [\mnras]
  {10.1093/mnras/stt817}, \href
  {https://ui.adsabs.harvard.edu/abs/2013MNRAS.433.1425B} {433, 1425}

\bibitem[\protect\citeauthoryear{{Broekgaarden} et~al.,}{{Broekgaarden}
  et~al.}{2021}]{Broekgaarden21}
{Broekgaarden} F.~S.,  et~al., 2021, arXiv e-prints, \href
  {https://ui.adsabs.harvard.edu/abs/2021arXiv210302608B} {p. arXiv:2103.02608}

\bibitem[\protect\citeauthoryear{{Caputi} et~al.,}{{Caputi}
  et~al.}{2017}]{Caputi17}
{Caputi} K.~I.,  et~al., 2017, \mn@doi [\apj] {10.3847/1538-4357/aa901e}, \href
  {https://ui.adsabs.harvard.edu/abs/2017ApJ...849...45C} {849, 45}

\bibitem[\protect\citeauthoryear{{Casey} et~al.,}{{Casey}
  et~al.}{2018}]{Casey18}
{Casey} C.~M.,  et~al., 2018, \mn@doi [\apj] {10.3847/1538-4357/aac82d}, \href
  {https://ui.adsabs.harvard.edu/abs/2018ApJ...862...77C} {862, 77}

\bibitem[\protect\citeauthoryear{{Chruslinska} \& {Nelemans}}{{Chruslinska} \&
  {Nelemans}}{2019}]{ChruslinskaNelemans19}
{Chruslinska} M.,  {Nelemans} G.,  2019, \mn@doi [\mnras]
  {10.1093/mnras/stz2057}, \href
  {https://ui.adsabs.harvard.edu/abs/2019MNRAS.488.5300C} {488, 5300}

\bibitem[\protect\citeauthoryear{{Chruslinska}, {Nelemans}  \&
  {Belczynski}}{{Chruslinska} et~al.}{2019}]{ChruslinskaNelemansBelczynski19}
{Chruslinska} M.,  {Nelemans} G.,   {Belczynski} K.,  2019, \mn@doi [\mnras]
  {10.1093/mnras/sty3087}, \href
  {https://ui.adsabs.harvard.edu/abs/2019MNRAS.482.5012C} {482, 5012}

\bibitem[\protect\citeauthoryear{{Chru{\'s}li{\'n}ska},
  {Je{\v{r}}{\'a}bkov{\'a}}, {Nelemans}  \& {Yan}}{{Chru{\'s}li{\'n}ska}
  et~al.}{2020}]{Chruslinska20}
{Chru{\'s}li{\'n}ska} M.,  {Je{\v{r}}{\'a}bkov{\'a}} T.,  {Nelemans} G.,
  {Yan} Z.,  2020, \mn@doi [\aap] {10.1051/0004-6361/202037688}, \href
  {https://ui.adsabs.harvard.edu/abs/2020A&A...636A..10C} {636, A10}

\bibitem[\protect\citeauthoryear{{Cresci}, {Mannucci}  \& {Curti}}{{Cresci}
  et~al.}{2019}]{Cresci19}
{Cresci} G.,  {Mannucci} F.,   {Curti} M.,  2019, \mn@doi [\aap]
  {10.1051/0004-6361/201834637}, \href
  {https://ui.adsabs.harvard.edu/abs/2019A&A...627A..42C} {627, A42}

\bibitem[\protect\citeauthoryear{{Cullen} et~al.,}{{Cullen}
  et~al.}{2021}]{Cullen21}
{Cullen} F.,  et~al., 2021, \mn@doi [\mnras] {10.1093/mnras/stab1340}, \href
  {https://ui.adsabs.harvard.edu/abs/2021MNRAS.505..903C} {505, 903}

\bibitem[\protect\citeauthoryear{{Curti}, {Mannucci}, {Cresci}  \&
  {Maiolino}}{{Curti} et~al.}{2020}]{Curti20}
{Curti} M.,  {Mannucci} F.,  {Cresci} G.,   {Maiolino} R.,  2020, \mn@doi
  [\mnras] {10.1093/mnras/stz2910}, \href
  {https://ui.adsabs.harvard.edu/abs/2020MNRAS.491..944C} {491, 944}

\bibitem[\protect\citeauthoryear{{Dav{\'e}}, {Finlator}  \&
  {Oppenheimer}}{{Dav{\'e}} et~al.}{2011}]{Dave11}
{Dav{\'e}} R.,  {Finlator} K.,   {Oppenheimer} B.~D.,  2011, \mn@doi [\mnras]
  {10.1111/j.1365-2966.2011.19132.x}, \href
  {https://ui.adsabs.harvard.edu/abs/2011MNRAS.416.1354D} {416, 1354}

\bibitem[\protect\citeauthoryear{{De Lucia}, {Xie}, {Fontanot}  \&
  {Hirschmann}}{{De Lucia} et~al.}{2020}]{DeLucia20}
{De Lucia} G.,  {Xie} L.,  {Fontanot} F.,   {Hirschmann} M.,  2020, \mn@doi
  [\mnras] {10.1093/mnras/staa2556}, \href
  {https://ui.adsabs.harvard.edu/abs/2020MNRAS.498.3215D} {498, 3215}

\bibitem[\protect\citeauthoryear{{De Rossi}, {Bower}, {Font}, {Schaye}  \&
  {Theuns}}{{De Rossi} et~al.}{2017}]{DeRossi17}
{De Rossi} M.~E.,  {Bower} R.~G.,  {Font} A.~S.,  {Schaye} J.,   {Theuns} T.,
  2017, \mn@doi [\mnras] {10.1093/mnras/stx2158}, \href
  {https://ui.adsabs.harvard.edu/abs/2017MNRAS.472.3354D} {472, 3354}

\bibitem[\protect\citeauthoryear{{Ellison}, {Patton}, {Simard}  \&
  {McConnachie}}{{Ellison} et~al.}{2008}]{Ellison08}
{Ellison} S.~L.,  {Patton} D.~R.,  {Simard} L.,   {McConnachie} A.~W.,  2008,
  \mn@doi [\apjl] {10.1086/527296}, \href
  {https://ui.adsabs.harvard.edu/abs/2008ApJ...672L.107E} {672, L107}

\bibitem[\protect\citeauthoryear{{Giacobbo}, {Mapelli}  \& {Spera}}{{Giacobbo}
  et~al.}{2018}]{Giacobbo18}
{Giacobbo} N.,  {Mapelli} M.,   {Spera} M.,  2018, \mn@doi [\mnras]
  {10.1093/mnras/stx2933}, \href
  {https://ui.adsabs.harvard.edu/abs/2018MNRAS.474.2959G} {474, 2959}

\bibitem[\protect\citeauthoryear{{Gr{\"a}fener} \& {Hamann}}{{Gr{\"a}fener} \&
  {Hamann}}{2008}]{GrafenerHamann08}
{Gr{\"a}fener} G.,  {Hamann} W.~R.,  2008, \mn@doi [\aap]
  {10.1051/0004-6361:20066176}, \href
  {https://ui.adsabs.harvard.edu/abs/2008A&A...482..945G} {482, 945}

\bibitem[\protect\citeauthoryear{{Grevesse} \& {Sauval}}{{Grevesse} \&
  {Sauval}}{1998}]{GrevesseSauval98}
{Grevesse} N.,  {Sauval} A.~J.,  1998, \mn@doi [\ssr]
  {10.1023/A:1005161325181}, \href
  {https://ui.adsabs.harvard.edu/abs/1998SSRv...85..161G} {85, 161}

\bibitem[\protect\citeauthoryear{{Gruppioni} et~al.,}{{Gruppioni}
  et~al.}{2020}]{Gruppioni20}
{Gruppioni} C.,  et~al., 2020, \mn@doi [\aap] {10.1051/0004-6361/202038487},
  \href {https://ui.adsabs.harvard.edu/abs/2020A&A...643A...8G} {643, A8}

\bibitem[\protect\citeauthoryear{{Hunt} et~al.,}{{Hunt} et~al.}{2012}]{Hunt12}
{Hunt} L.,  et~al., 2012, \mn@doi [\mnras] {10.1111/j.1365-2966.2012.21761.x},
  \href {https://ui.adsabs.harvard.edu/abs/2012MNRAS.427..906H} {427, 906}

\bibitem[\protect\citeauthoryear{{Hunt}, {Dayal}, {Magrini}  \&
  {Ferrara}}{{Hunt} et~al.}{2016}]{Hunt16_obs}
{Hunt} L.,  {Dayal} P.,  {Magrini} L.,   {Ferrara} A.,  2016, \mn@doi [\mnras]
  {10.1093/mnras/stw1993}, \href
  {https://ui.adsabs.harvard.edu/abs/2016MNRAS.463.2002H} {463, 2002}

\bibitem[\protect\citeauthoryear{{Ilbert} et~al.,}{{Ilbert}
  et~al.}{2015}]{Ilbert15}
{Ilbert} O.,  et~al., 2015, \mn@doi [\aap] {10.1051/0004-6361/201425176}, \href
  {https://ui.adsabs.harvard.edu/abs/2015A&A...579A...2I} {579, A2}

\bibitem[\protect\citeauthoryear{{Kashino}, {Renzini}, {Silverman}  \&
  {Daddi}}{{Kashino} et~al.}{2016}]{Kashino16}
{Kashino} D.,  {Renzini} A.,  {Silverman} J.~D.,   {Daddi} E.,  2016, \mn@doi
  [\apjl] {10.3847/2041-8205/823/2/L24}, \href
  {https://ui.adsabs.harvard.edu/abs/2016ApJ...823L..24K} {823, L24}

\bibitem[\protect\citeauthoryear{{Kewley} \& {Ellison}}{{Kewley} \&
  {Ellison}}{2008}]{KewleyEllison08}
{Kewley} L.~J.,  {Ellison} S.~L.,  2008, \mn@doi [\apj] {10.1086/587500}, \href
  {https://ui.adsabs.harvard.edu/abs/2008ApJ...681.1183K} {681, 1183}

\bibitem[\protect\citeauthoryear{{Khusanova} et~al.,}{{Khusanova}
  et~al.}{2021}]{Khusanova20}
{Khusanova} Y.,  et~al., 2021, \mn@doi [\aap] {10.1051/0004-6361/202038944},
  \href {https://ui.adsabs.harvard.edu/abs/2021A&A...649A.152K} {649, A152}

\bibitem[\protect\citeauthoryear{{Klencki}, {Moe}, {Gladysz}, {Chruslinska},
  {Holz}  \& {Belczynski}}{{Klencki} et~al.}{2018}]{Klencki18}
{Klencki} J.,  {Moe} M.,  {Gladysz} W.,  {Chruslinska} M.,  {Holz} D.~E.,
  {Belczynski} K.,  2018, \mn@doi [\aap] {10.1051/0004-6361/201833025}, \href
  {https://ui.adsabs.harvard.edu/abs/2018A&A...619A..77K} {619, A77}

\bibitem[\protect\citeauthoryear{{Kobulnicky} \& {Kewley}}{{Kobulnicky} \&
  {Kewley}}{2004}]{KobulnickyKeweley04}
{Kobulnicky} H.~A.,  {Kewley} L.~J.,  2004, \mn@doi [\apj] {10.1086/425299},
  \href {https://ui.adsabs.harvard.edu/abs/2004ApJ...617..240K} {617, 240}

\bibitem[\protect\citeauthoryear{{Kroupa}}{{Kroupa}}{2001}]{Kroupa01}
{Kroupa} P.,  2001, \mn@doi [\mnras] {10.1046/j.1365-8711.2001.04022.x}, \href
  {https://ui.adsabs.harvard.edu/abs/2001MNRAS.322..231K} {322, 231}

\bibitem[\protect\citeauthoryear{{Lagos} et~al.,}{{Lagos}
  et~al.}{2016}]{Lagos16}
{Lagos} C. d.~P.,  et~al., 2016, \mn@doi [\mnras] {10.1093/mnras/stw717}, \href
  {https://ui.adsabs.harvard.edu/abs/2016MNRAS.459.2632L} {459, 2632}

\bibitem[\protect\citeauthoryear{{Lara-L{\'o}pez} et~al.,}{{Lara-L{\'o}pez}
  et~al.}{2010}]{Lara-Lopez13}
{Lara-L{\'o}pez} M.~A.,  et~al., 2010, \mn@doi [\aap]
  {10.1051/0004-6361/201014803}, \href
  {https://ui.adsabs.harvard.edu/abs/2010A&A...521L..53L} {521, L53}

\bibitem[\protect\citeauthoryear{{Loiacono} et~al.,}{{Loiacono}
  et~al.}{2021}]{Loiacono21}
{Loiacono} F.,  et~al., 2021, \mn@doi [\aap] {10.1051/0004-6361/202038607},
  \href {https://ui.adsabs.harvard.edu/abs/2021A&A...646A..76L} {646, A76}

\bibitem[\protect\citeauthoryear{{Madau} \& {Dickinson}}{{Madau} \&
  {Dickinson}}{2014}]{MadauDickinson14}
{Madau} P.,  {Dickinson} M.,  2014, \mn@doi [\araa]
  {10.1146/annurev-astro-081811-125615}, \href
  {https://ui.adsabs.harvard.edu/abs/2014ARA&A..52..415M} {52, 415}

\bibitem[\protect\citeauthoryear{{Maiolino} \& {Mannucci}}{{Maiolino} \&
  {Mannucci}}{2019}]{MaiolinoMannucci19}
{Maiolino} R.,  {Mannucci} F.,  2019, \mn@doi [\aapr]
  {10.1007/s00159-018-0112-2}, \href
  {https://ui.adsabs.harvard.edu/abs/2019A&ARv..27....3M} {27, 3}

\bibitem[\protect\citeauthoryear{{Maiolino} et~al.,}{{Maiolino}
  et~al.}{2008}]{Maiolino08}
{Maiolino} R.,  et~al., 2008, \mn@doi [\aap] {10.1051/0004-6361:200809678},
  \href {https://ui.adsabs.harvard.edu/abs/2008A&A...488..463M} {488, 463}

\bibitem[\protect\citeauthoryear{{Mannucci} et~al.,}{{Mannucci}
  et~al.}{2009}]{Mannucci09}
{Mannucci} F.,  et~al., 2009, \mn@doi [\mnras]
  {10.1111/j.1365-2966.2009.15185.x}, \href
  {https://ui.adsabs.harvard.edu/abs/2009MNRAS.398.1915M} {398, 1915}

\bibitem[\protect\citeauthoryear{{Mannucci}, {Cresci}, {Maiolino}, {Marconi}
  \& {Gnerucci}}{{Mannucci} et~al.}{2010}]{Mannucci10}
{Mannucci} F.,  {Cresci} G.,  {Maiolino} R.,  {Marconi} A.,   {Gnerucci} A.,
  2010, \mn@doi [\mnras] {10.1111/j.1365-2966.2010.17291.x}, \href
  {https://ui.adsabs.harvard.edu/abs/2010MNRAS.408.2115M} {408, 2115}

\bibitem[\protect\citeauthoryear{{Mannucci}, {Salvaterra}  \&
  {Campisi}}{{Mannucci} et~al.}{2011}]{Mannucci11}
{Mannucci} F.,  {Salvaterra} R.,   {Campisi} M.~A.,  2011, \mn@doi [\mnras]
  {10.1111/j.1365-2966.2011.18459.x}, \href
  {https://ui.adsabs.harvard.edu/abs/2011MNRAS.414.1263M} {414, 1263}

\bibitem[\protect\citeauthoryear{{Neijssel} et~al.,}{{Neijssel}
  et~al.}{2019}]{Neijssel19}
{Neijssel} C.~J.,  et~al., 2019, \mn@doi [\mnras] {10.1093/mnras/stz2840},
  \href {https://ui.adsabs.harvard.edu/abs/2019MNRAS.490.3740N} {490, 3740}

\bibitem[\protect\citeauthoryear{{Orlitova}}{{Orlitova}}{2020}]{Orlitova20}
{Orlitova} I.,  2020, arXiv e-prints, \href
  {https://ui.adsabs.harvard.edu/abs/2020arXiv201212378O} {p. arXiv:2012.12378}

\bibitem[\protect\citeauthoryear{{Pearson} et~al.,}{{Pearson}
  et~al.}{2018}]{Pearson18}
{Pearson} W.~J.,  et~al., 2018, \mn@doi [\aap] {10.1051/0004-6361/201832821},
  \href {https://ui.adsabs.harvard.edu/abs/2018A&A...615A.146P} {615, A146}

\bibitem[\protect\citeauthoryear{{Pettini} \& {Pagel}}{{Pettini} \&
  {Pagel}}{2004}]{PettiniPagel04}
{Pettini} M.,  {Pagel} B. E.~J.,  2004, \mn@doi [\mnras]
  {10.1111/j.1365-2966.2004.07591.x}, \href
  {https://ui.adsabs.harvard.edu/abs/2004MNRAS.348L..59P} {348, L59}

\bibitem[\protect\citeauthoryear{{Renzini} \& {Peng}}{{Renzini} \&
  {Peng}}{2015}]{RenziniPeng15}
{Renzini} A.,  {Peng} Y.-j.,  2015, \mn@doi [\apjl]
  {10.1088/2041-8205/801/2/L29}, \href
  {https://ui.adsabs.harvard.edu/abs/2015ApJ...801L..29R} {801, L29}

\bibitem[\protect\citeauthoryear{{Rodighiero} et~al.,}{{Rodighiero}
  et~al.}{2011}]{Rodighiero11}
{Rodighiero} G.,  et~al., 2011, \mn@doi [\apjl] {10.1088/2041-8205/739/2/L40},
  \href {https://ui.adsabs.harvard.edu/abs/2011ApJ...739L..40R} {739, L40}

\bibitem[\protect\citeauthoryear{{Salim}, {Lee}, {Ly}, {Brinchmann},
  {Dav{\'e}}, {Dickinson}, {Salzer}  \& {Charlot}}{{Salim}
  et~al.}{2014}]{Salim14}
{Salim} S.,  {Lee} J.~C.,  {Ly} C.,  {Brinchmann} J.,  {Dav{\'e}} R.,
  {Dickinson} M.,  {Salzer} J.~J.,   {Charlot} S.,  2014, \mn@doi [\apj]
  {10.1088/0004-637X/797/2/126}, \href
  {https://ui.adsabs.harvard.edu/abs/2014ApJ...797..126S} {797, 126}

\bibitem[\protect\citeauthoryear{{Sanders} et~al.,}{{Sanders}
  et~al.}{2018}]{Sanders18}
{Sanders} R.~L.,  et~al., 2018, \mn@doi [\apj] {10.3847/1538-4357/aabcbd},
  \href {https://ui.adsabs.harvard.edu/abs/2018ApJ...858...99S} {858, 99}

\bibitem[\protect\citeauthoryear{{Sanders} et~al.,}{{Sanders}
  et~al.}{2020}]{Sanders20}
{Sanders} R.~L.,  et~al., 2020, arXiv e-prints, \href
  {https://ui.adsabs.harvard.edu/abs/2020arXiv200907292S} {p. arXiv:2009.07292}

\bibitem[\protect\citeauthoryear{{Sargent}, {B{\'e}thermin}, {Daddi}  \&
  {Elbaz}}{{Sargent} et~al.}{2012}]{Sargent12}
{Sargent} M.~T.,  {B{\'e}thermin} M.,  {Daddi} E.,   {Elbaz} D.,  2012, \mn@doi
  [\apjl] {10.1088/2041-8205/747/2/L31}, \href
  {https://ui.adsabs.harvard.edu/abs/2012ApJ...747L..31S} {747, L31}

\bibitem[\protect\citeauthoryear{{Schreiber} et~al.,}{{Schreiber}
  et~al.}{2015}]{Schreiber15}
{Schreiber} C.,  et~al., 2015, \mn@doi [\aap] {10.1051/0004-6361/201425017},
  \href {https://ui.adsabs.harvard.edu/abs/2015A&A...575A..74S} {575, A74}

\bibitem[\protect\citeauthoryear{{Speagle}, {Steinhardt}, {Capak}  \&
  {Silverman}}{{Speagle} et~al.}{2014}]{Speagle14}
{Speagle} J.~S.,  {Steinhardt} C.~L.,  {Capak} P.~L.,   {Silverman} J.~D.,
  2014, \mn@doi [\apjs] {10.1088/0067-0049/214/2/15}, \href
  {https://ui.adsabs.harvard.edu/abs/2014ApJS..214...15S} {214, 15}

\bibitem[\protect\citeauthoryear{{Steidel}, {Strom}, {Pettini}, {Rudie},
  {Reddy}  \& {Trainor}}{{Steidel} et~al.}{2016}]{Steidel16}
{Steidel} C.~C.,  {Strom} A.~L.,  {Pettini} M.,  {Rudie} G.~C.,  {Reddy} N.~A.,
    {Trainor} R.~F.,  2016, \mn@doi [\apj] {10.3847/0004-637X/826/2/159}, \href
  {https://ui.adsabs.harvard.edu/abs/2016ApJ...826..159S} {826, 159}

\bibitem[\protect\citeauthoryear{{Telford}, {Dalcanton}, {Skillman}  \&
  {Conroy}}{{Telford} et~al.}{2016}]{Telford16}
{Telford} O.~G.,  {Dalcanton} J.~J.,  {Skillman} E.~D.,   {Conroy} C.,  2016,
  \mn@doi [\apj] {10.3847/0004-637X/827/1/35}, \href
  {https://ui.adsabs.harvard.edu/abs/2016ApJ...827...35T} {827, 35}

\bibitem[\protect\citeauthoryear{{Tolstoy}, {Hill}  \& {Tosi}}{{Tolstoy}
  et~al.}{2009}]{Tolstoy09}
{Tolstoy} E.,  {Hill} V.,   {Tosi} M.,  2009, \mn@doi [\araa]
  {10.1146/annurev-astro-082708-101650}, \href
  {https://ui.adsabs.harvard.edu/abs/2009ARA&A..47..371T} {47, 371}

\bibitem[\protect\citeauthoryear{{Tomczak} et~al.,}{{Tomczak}
  et~al.}{2016}]{Tomczak16}
{Tomczak} A.~R.,  et~al., 2016, \mn@doi [\apj] {10.3847/0004-637X/817/2/118},
  \href {https://ui.adsabs.harvard.edu/abs/2016ApJ...817..118T} {817, 118}

\bibitem[\protect\citeauthoryear{{Topping}, {Shapley}, {Reddy}, {Sanders},
  {Coil}, {Kriek}, {Mobasher}  \& {Siana}}{{Topping} et~al.}{2020}]{Topping20}
{Topping} M.~W.,  {Shapley} A.~E.,  {Reddy} N.~A.,  {Sanders} R.~L.,  {Coil}
  A.~L.,  {Kriek} M.,  {Mobasher} B.,   {Siana} B.,  2020, \mn@doi [\mnras]
  {10.1093/mnras/staa2941}, \href
  {https://ui.adsabs.harvard.edu/abs/2020MNRAS.499.1652T} {499, 1652}

\bibitem[\protect\citeauthoryear{{Torrey} et~al.,}{{Torrey}
  et~al.}{2018}]{Torrey18}
{Torrey} P.,  et~al., 2018, \mn@doi [\mnras] {10.1093/mnrasl/sly031}, \href
  {https://ui.adsabs.harvard.edu/abs/2018MNRAS.477L..16T} {477, L16}

\bibitem[\protect\citeauthoryear{{Torrey} et~al.,}{{Torrey}
  et~al.}{2019}]{Torrey19}
{Torrey} P.,  et~al., 2019, \mn@doi [\mnras] {10.1093/mnras/stz243}, \href
  {https://ui.adsabs.harvard.edu/abs/2019MNRAS.484.5587T} {484, 5587}

\bibitem[\protect\citeauthoryear{{Vink} \& {de Koter}}{{Vink} \& {de
  Koter}}{2005}]{VinkdeKoter05}
{Vink} J.~S.,  {de Koter} A.,  2005, \mn@doi [\aap]
  {10.1051/0004-6361:20052862}, \href
  {https://ui.adsabs.harvard.edu/abs/2005A&A...442..587V} {442, 587}

\bibitem[\protect\citeauthoryear{{Vink}, {de Koter}  \& {Lamers}}{{Vink}
  et~al.}{2001}]{Vink01}
{Vink} J.~S.,  {de Koter} A.,   {Lamers} H.~J.~G.~L.~M.,  2001, \mn@doi [\aap]
  {10.1051/0004-6361:20010127}, \href
  {https://ui.adsabs.harvard.edu/abs/2001A&A...369..574V} {369, 574}

\bibitem[\protect\citeauthoryear{{Wheeler}, {Sneden}  \& {Truran}}{{Wheeler}
  et~al.}{1989}]{Wheeler89}
{Wheeler} J.~C.,  {Sneden} C.,   {Truran} Jr. J.~W.,  1989, \mn@doi [\araa]
  {10.1146/annurev.aa.27.090189.001431}, \href
  {http://adsabs.harvard.edu/abs/1989ARA%26A..27..279W} {27, 279}

\bibitem[\protect\citeauthoryear{{Yabe} et~al.,}{{Yabe} et~al.}{2015}]{Yabe15}
{Yabe} K.,  et~al., 2015, \mn@doi [\pasj] {10.1093/pasj/psv079}, \href
  {https://ui.adsabs.harvard.edu/abs/2015PASJ...67..102Y} {67, 102}

\bibitem[\protect\citeauthoryear{{Yates}, {Kauffmann}  \& {Guo}}{{Yates}
  et~al.}{2012}]{Yates12}
{Yates} R.~M.,  {Kauffmann} G.,   {Guo} Q.,  2012, \mn@doi [\mnras]
  {10.1111/j.1365-2966.2012.20595.x}, \href
  {https://ui.adsabs.harvard.edu/abs/2012MNRAS.422..215Y} {422, 215}

\bibitem[\protect\citeauthoryear{{Zahid}, {Dima}, {Kudritzki}, {Kewley},
  {Geller}, {Hwang}, {Silverman}  \& {Kashino}}{{Zahid}
  et~al.}{2014a}]{Zahid14a}
{Zahid} H.~J.,  {Dima} G.~I.,  {Kudritzki} R.-P.,  {Kewley} L.~J.,  {Geller}
  M.~J.,  {Hwang} H.~S.,  {Silverman} J.~D.,   {Kashino} D.,  2014a, \mn@doi
  [\apj] {10.1088/0004-637X/791/2/130}, \href
  {https://ui.adsabs.harvard.edu/abs/2014ApJ...791..130Z} {791, 130}

\bibitem[\protect\citeauthoryear{{Zahid} et~al.,}{{Zahid}
  et~al.}{2014b}]{Zahid14b}
{Zahid} H.~J.,  et~al., 2014b, \mn@doi [\apj] {10.1088/0004-637X/792/1/75},
  \href {https://ui.adsabs.harvard.edu/abs/2014ApJ...792...75Z} {792, 75}

\makeatother
\end{thebibliography}

%\begin{thebibliography}{99}
%\end{thebibliography}

%%%%%%%%%%%%%%%%%%%%%%%%%%%%%%%%%%%%%%%%%%%%%%%%%%

%%%%%%%%%%%%%%%%% APPENDICES %%%%%%%%%%%%%%%%%%%%%

\appendix

\newpage

\section{FMR variations - additional figures} \label{app: FMR variations}
\begin{figure}
\vspace*{-0.5cm}
\includegraphics[width=1\columnwidth]{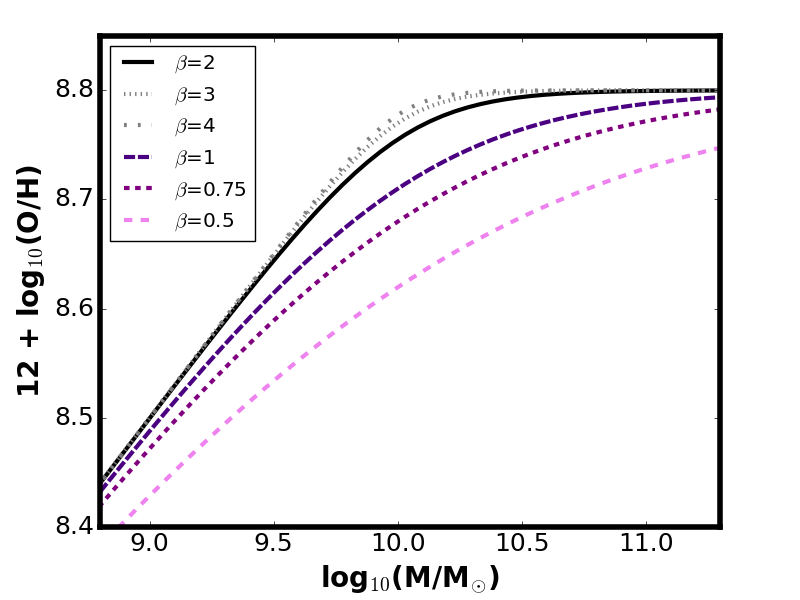}
\vspace*{-0.3cm}
\caption{
Dependence on $\beta$ parameter in eq. \ref{eq: Curti parametrisation},
illustrated using the best fit MZR z$\sim$0 parameters from Curti et al. (2020)
and various choices of $\beta$.
If $\beta$ can be constrained to values $\beta\gtrsim$1, 
 the $Z_{O/H}$ uncertainty induced by the choice of this parameter is relatively small 
 and affects a relatively small range of galaxy masses.
 However, if $\beta\lesssim$1, small variation in this parameter can add a appreciable uncertainty to $Z_{O/H}$ .
 %test_C20_parametrisation.py
}
\label{fig: beta dependence}
\end{figure}

\begin{figure*}
\vspace*{-0.5cm}
\includegraphics[scale=0.48]{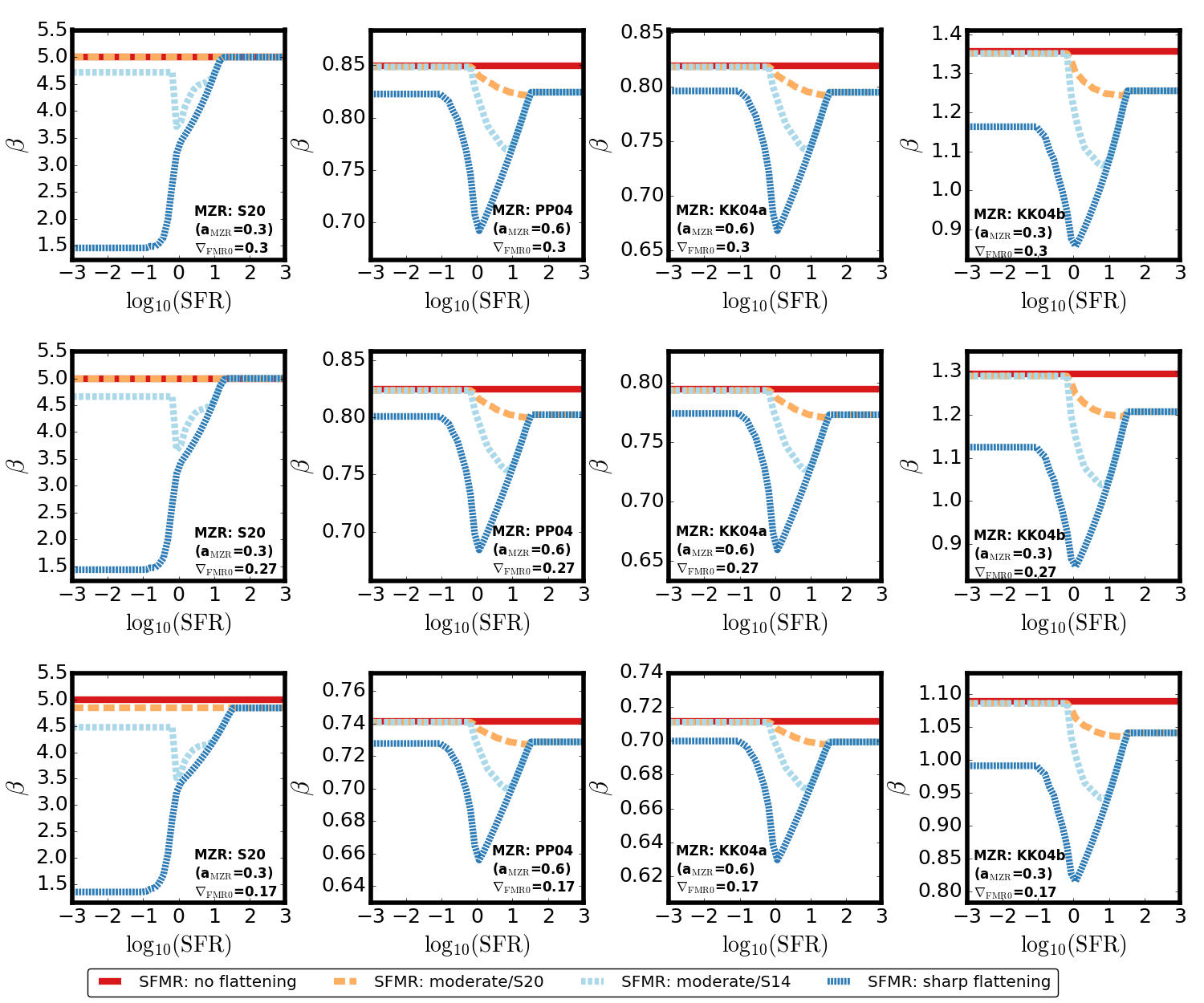}
\vspace*{-0.3cm}
\caption{
Parameter $\beta$ as a function of log$_{10}$(SFR) (see eq. \ref{eq: Curti parametrisation}), depending on the choice of $z\sim0$ MZR (different columns), $\nabla_{\rm FMR0}$ (different rows) and the shape of the high mass end of the $z\sim0$ SFMR (different lines).
Single value of $\beta$ provides a good description if the SFMR does not significantly depart from a single power law. 
See Sec \ref{sec: model variations} for the description of different model variations.
%FMR - figure_beta_multipanel.py
}
\label{fig: beta summary}
\end{figure*}

\begin{figure*}
\vspace*{-0.5cm}
\includegraphics[scale=0.48]{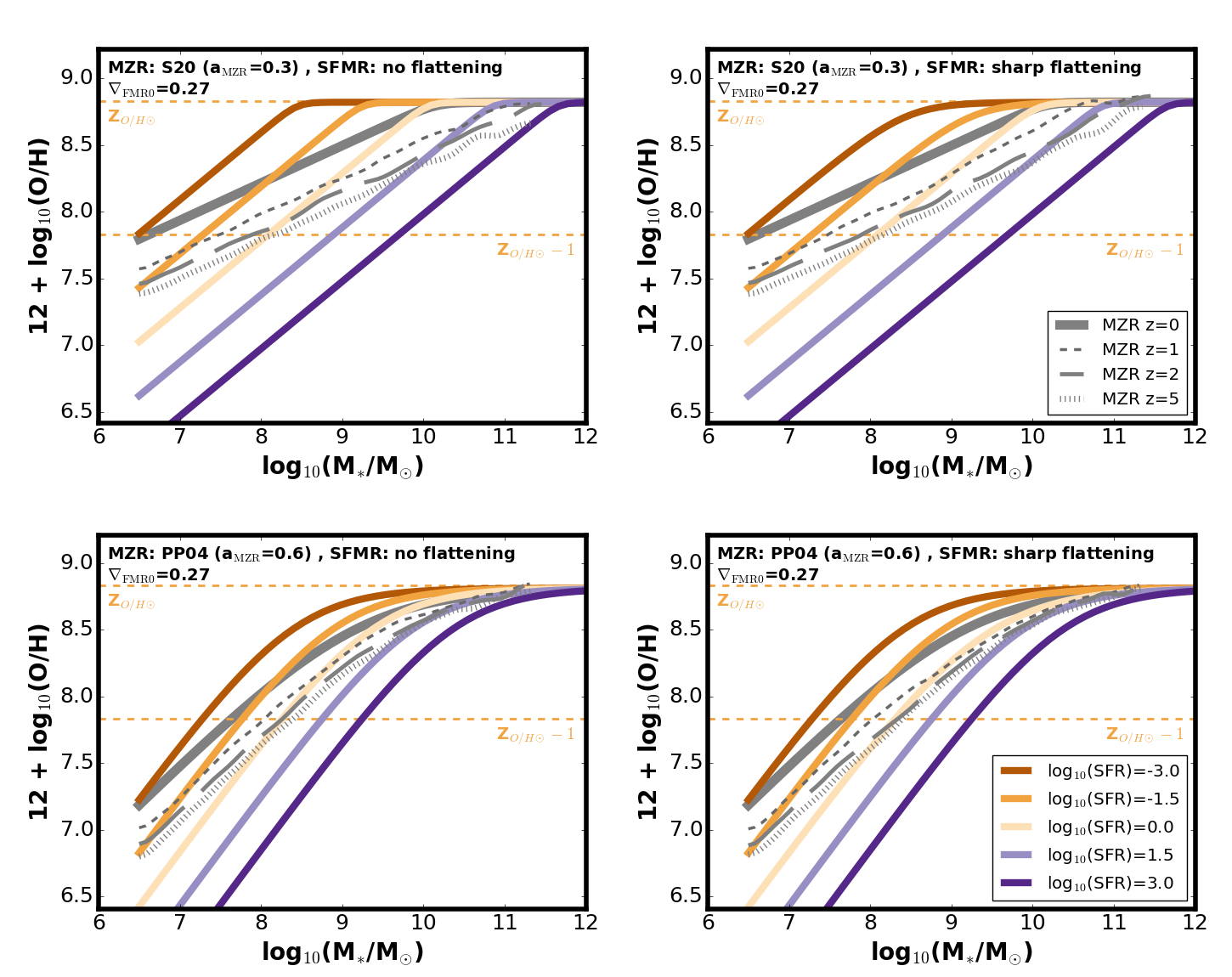}
\vspace*{-0.3cm}
\caption{
Same as Fig. \ref{fig: FMR MZR choice}, but for different choices of the $z\sim 0$ SFMR and MZR.
Left (right) panels assume SFMR with no (sharp) flattening at the high mass end.
Top (bottom) panels assume S20 (PP04) $z\sim0$ MZR.
All panels assume $\nabla_{\rm FMR0}$=0.27.
}
\label{fig: FMR SFMR choice}
\end{figure*}

\begin{figure*}
\vspace*{-0.5cm}
\includegraphics[scale=0.48]{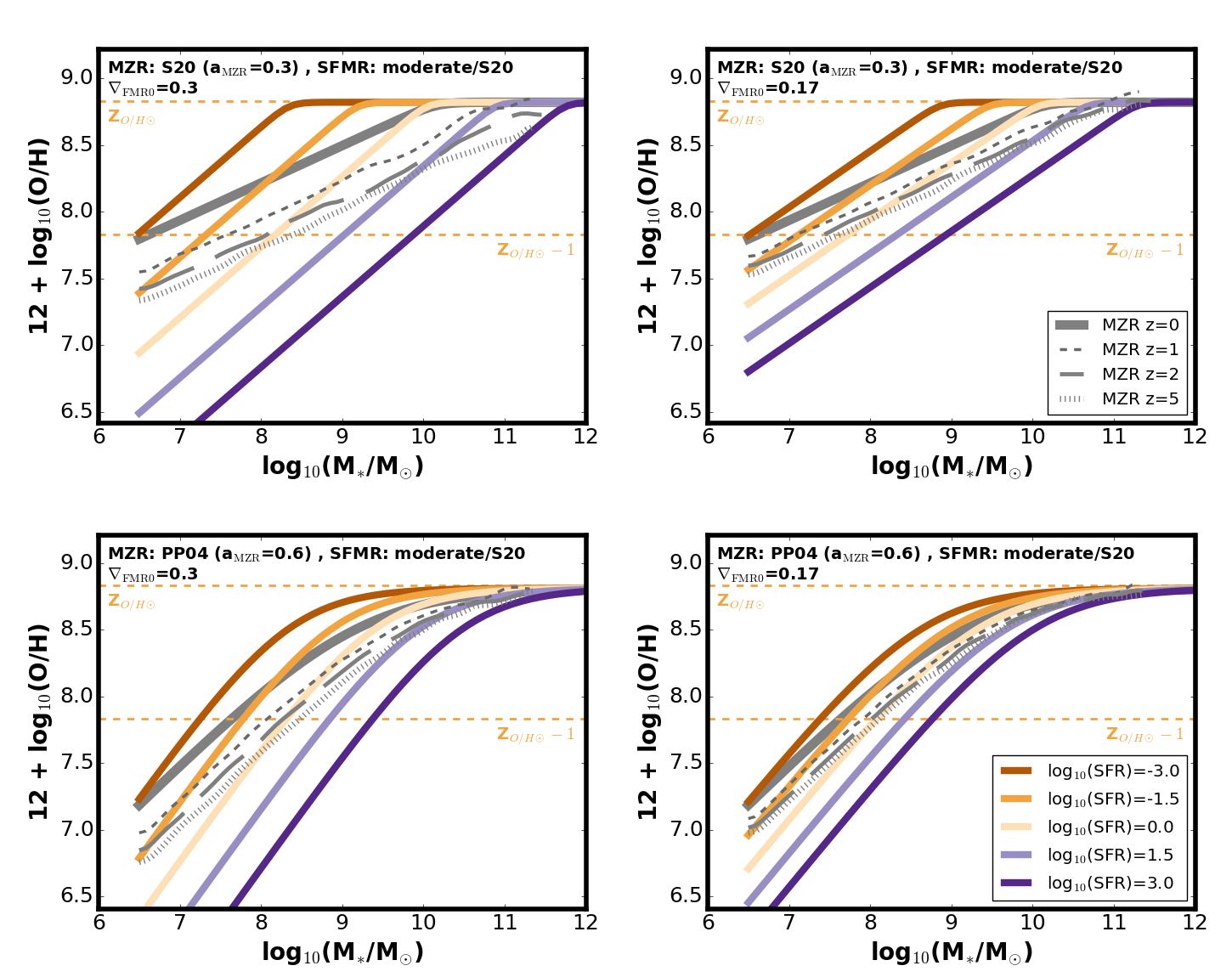}
\vspace*{-0.3cm}
\caption{
Same as Fig. \ref{fig: FMR MZR choice}, but for different choices of $\nabla_{\rm FMR0}$.
Top (bottom) panels assume S20 (PP04) $z\sim0$ MZR.
All panels assume SFMR with a moderate flattening at high masses (a$_{SFMR}$=0.72 at the high mass end).
}
\label{fig: FMR FMR slope choice}
\end{figure*}

In this Section we show additional figures illustrating the sensitivity of Z$_{O/H}$(SFR,M$_{*}$) to the parameter $\beta$ (Figure \ref{fig: beta dependence}) and the Z$_{O/H}$(SFR,M$_{*}$) dependence on various choices of the parameters of our observation-based model. 
\\
Figure \ref{fig: beta summary} shows the values of $\beta$ as a function of log$_{10}$(SFR) obtained  as described in Sec. \ref{sec: model: beta} for different choices of the $z\sim0$ MZR, $\nabla_{\rm FMR0}$ and SFMR used in this study.
\\
Figures \ref{fig: FMR SFMR choice} and \ref{fig: FMR FMR slope choice} show the comparison between the Z$_{O/H}$(SFR,M$_{*}$) obtained for different SFMR and $\nabla_{\rm FMR0}$ choices, respectively.

\section{Cosmic SFH - additional materials}\label{app: cosmic SFH appendix}

\begin{figure*}
\vspace*{-0.5cm}
\includegraphics[width=1.7\columnwidth]{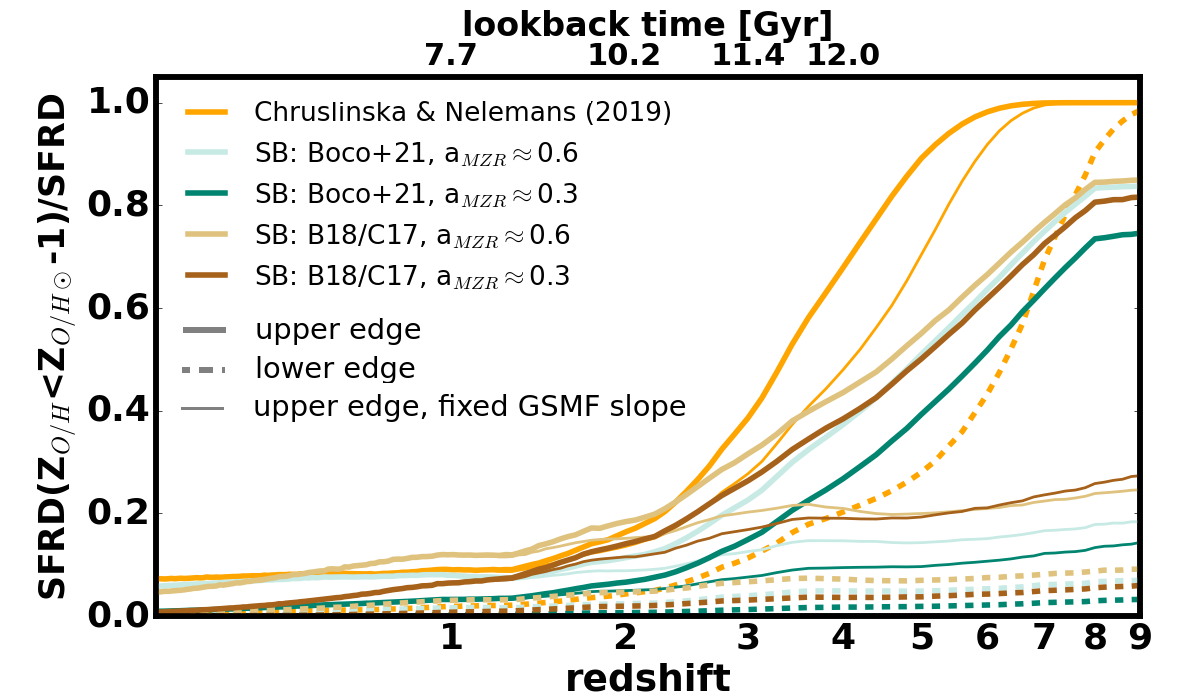}
\vspace*{-0.3cm}
\caption{
Fraction of the total SFRD that occurs below Z$_{O/H\odot}$-1 as a function of redshift/lookback time, plotted for the model variations shown in Fig. \ref{fig: low Z SFRD}.
Different colours correspond to different assumptions about $a_{MZR}$ and starburst galaxies (see legend).
Thick solid (dashed) lines mark the upper (lower) edges of the corresponding ranges for each of the model variations. Thin solid lines show the location of the upper edge obtained under the assumption of a non-evolving low mass end of the GSMF.
}
\label{fig: low Z SFRD - fraction}
\end{figure*}

\begin{figure*}
\vspace*{-0.5cm}
\includegraphics[width=1.7\columnwidth]{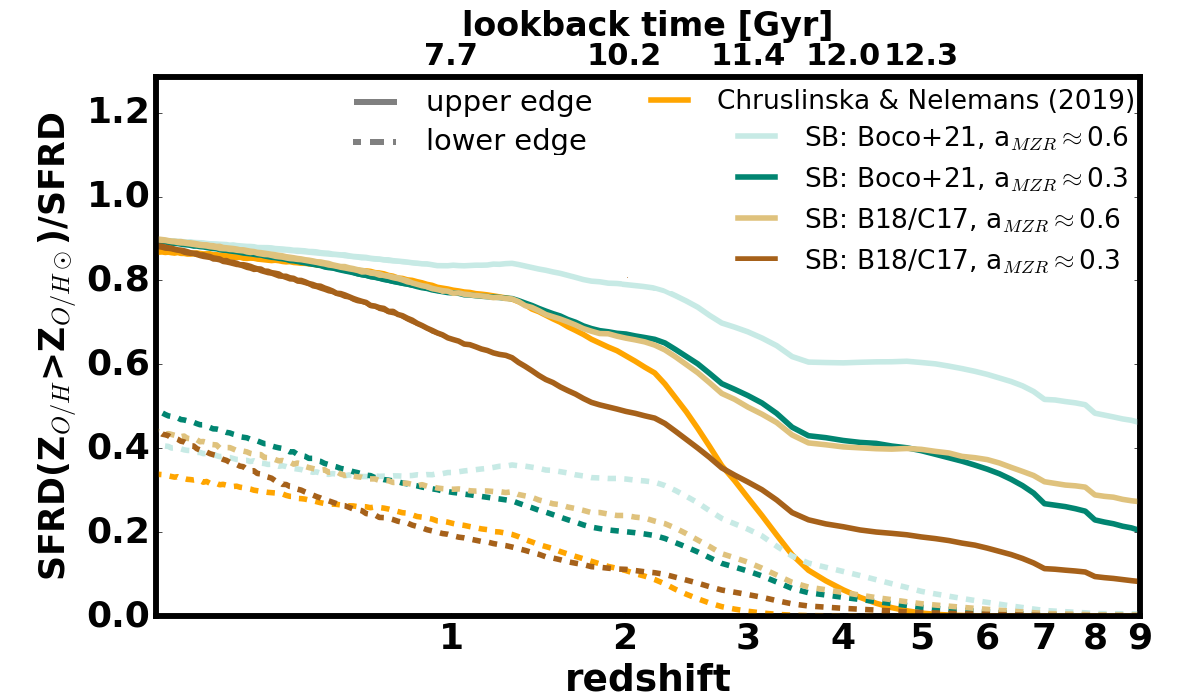}
\vspace*{-0.3cm}
\caption{
Fraction of the total SFRD that occurs above Z$_{O/H\odot}$ as a function of redshift/lookback time, plotted for the model variations shown in Fig. \ref{fig: high Z SFRD}.
Different colours correspond to different assumptions about $a_{MZR}$ and starburst galaxies (see legend).
Thick solid (dashed) lines mark the upper (lower) edges of the corresponding ranges for each of the model variations.
}
\label{fig: high Z SFRD - fraction}
\end{figure*}

Table \ref{tab: total SFRD, alpha GSMF fixed} provides the coefficients of the broken power law fits 
to the total cosmic SFH for the variations with non-evolving low mass end of the GSMF.
Table \ref{tab: low Z SFH} (\ref{tab: low Z SFH GSMF fixed}) provides the corresponding coefficients 
obtained for the low metallicity cut of the cosmic SFH, fitted for the variations with redshift-dependent (non evolving)
 low mass end of the GSMF.
 \\
%  Figure \ref{fig: total SFRD - lMmin=8} is analogous to Fig. \ref{fig: total SFRD} discussed in Sec. \ref{sec: results - total SFH}, but obtained for the scaling relations and GSMF integrated down to log$_{10}$(M$_{*}$/M$_{\odot}$)$=$8 instead of log$_{10}$(M$_{*}$/M$_{\odot}$)$=$6.
%  \\
 Figure \ref{fig: low Z SFRD - fraction} (\ref{fig: high Z SFRD - fraction}) shows the fraction of the total SFRD that happens below 10\% solar metallicity, i.e $Z_{O/H} < Z_{O/H\odot}$-1 (above solar metallicity, i.e. $Z>Z_{O/H}$)
as a function of redshift/lookback time, obtained for the different variations of our model discussed in Sec. \ref{sec: results - low metallicity SFH} (\ref{sec: results - high metallicity SFH}). Note that those variations in general lead to different total SFRD at different redshift, which also affects the spread between the estimates of the low (high) metallicity part of the cosmic SFH shown in Fig. \ref{fig: low Z SFRD} (\ref{fig: high Z SFRD}).
Fig. \ref{fig: low Z SFRD - fraction} allows to directly compare the differences in the low (high) metallicity tail of the f$_{\rm SFR}$(Z,z) distribution.

%  \begin{figure*}
% \vspace*{-0.5cm}
% \includegraphics[width=1.7\columnwidth]{./figures/total_t_SFRD_cut_fixedFMRslope_Mmin8.png}
% \vspace*{-0.3cm}
% \caption{
% Same as Fig. \ref{fig: total SFRD}, but considering only galaxies
% with log$_{10}$(M$_{*}/\Msun$)$>$8.
% }
% \label{fig: total SFRD - lMmin=8}
% \end{figure*}

\begin{table*}
\centering
\small
\caption{Same as Table \ref{tab: total SFRD}, but considering only variations with non-evolving low mass end of the GSMF}
\begin{tabular}{c c c c c c c }
\hline
\multicolumn{7}{c}{ \textbf{cosmic SFH - lower edge ($\alpha_{\rm GSMF}$ fixed)} }\\ variation:  & \multicolumn{2}{c}{no SB} & \multicolumn{2}{c}{SB: Boco+21} & \multicolumn{2}{c}{ SB: B18/C17}\\
$z$ range & $\kappa$ & A & $\kappa$ & A & $\kappa$ & A  \\ \hline
0<$z\leqslant$1.0 & 2.6307 & 0.00779 & 2.5875 & 0.00924 & 2.7706 & 0.0125\\ \hline
1.0<$z\leqslant$1.8 & 2.3888 & 0.00922 & 2.3265 & 0.0111 & 2.6969 & 0.0131\\ \hline
1.8<$z\leqslant$3.0 & -1.0535 & 0.319 & -1.1284 & 0.388 & -0.7558 & 0.458\\ \hline
3.0<$z\leqslant$7.0 & -2.2771 & 1.74 & -2.2909 & 1.95 & -1.9628 & 2.44\\ \hline
7.0<$z\leqslant$8.8 & -12.5425 & 3.24$\cdot10^{9}$ & -12.5526 & 3.6$\cdot10^{9}$ & -12.4895 & 7.85$\cdot10^{9}$\\ \hline
8.8<$z\leqslant$10 & 0 & 0.0012 & 0 & 0.0013 & 0 & 0.00327\\ \hline \hline
\multicolumn{7}{c}{ \textbf{cosmic SFH - upper edge ($\alpha_{\rm GSMF}$ fixed)} }\\ variation:  & \multicolumn{2}{c}{no SB} & \multicolumn{2}{c}{SB: Boco+21} & \multicolumn{2}{c}{ SB: B18/C17}\\
$z$ range & $\kappa$ & A & $\kappa$ & A & $\kappa$ & A  \\ \hline
0<$z\leqslant$1.0 & 2.2211 & 0.0194 & 2.2208 & 0.0209 & 2.3822 & 0.0248\\ \hline
1.0<$z\leqslant$1.8 & 1.8466 & 0.0251 & 1.8463 & 0.027 & 2.2105 & 0.028\\ \hline
1.8<$z\leqslant$4.0 & -1.7884 & 1.06 & -1.7847 & 1.14 & -1.2278 & 0.964\\ \hline
4.0<$z\leqslant$7.0 & -2.6385 & 4.16 & -2.6376 & 4.49 & -2.4769 & 7.2\\ \hline
7.0<$z\leqslant$8.8 & -12.6417 & 4.5$\cdot10^{9}$ & -12.6485 & 4.93$\cdot10^{9}$ & -12.5280 & 8.6$\cdot10^{9}$\\ \hline
8.8<$z\leqslant$10 & 0 & 0.00133 & 0 & 0.00143 & 0 & 0.00328\\ \hline \hline
\end{tabular}
\label{tab: total SFRD, alpha GSMF fixed}
\end{table*}

\begin{table*}
\centering
\small
\caption{
Coefficients of the power law fits to the low metallicity cut of the cosmic SFH (lower and upper edges of the ranges shown in  Fig. \ref{fig: low Z SFRD}) obtained for different variations of our observation-based model.
Within each redshift range indicated in the first column total SFRD can be approximated with the dependence: SFRD=A$\cdot(1+z)^{\kappa} [\Msun/yr]$.
}
\begin{tabular}{c c c c c c c c c}
\hline
%Boco, a=0.3  Boco a=06, BiC a=0.3 BiC a=0.6 -- kappa A
%LOWER EDGE
\multicolumn{9}{c}{ \textbf{low metallicity SFH - lower edge} }\\
 variation:  & \multicolumn{2}{c}{SB: Boco+21, a$_{MZR}\sim$0.3} & \multicolumn{2}{c}{SB: Boco+21, a$_{MZR}\sim$0.6} & \multicolumn{2}{c}{SB: B18/C17, a$_{MZR}\sim$0.3} & \multicolumn{2}{c}{SB: B18/C17, a$_{MZR}\sim$0.6} \\ 
$z$ range & $\kappa$ & A & $\kappa$ & A & $\kappa$ & A  & $\kappa$ & A \\ \hline

0$<z\leqslant$0.5  & 6.4712  & 2.912$\cdot 10^{-6}$ & 3.8153  & 1$\cdot 10^{-4}$    & 8.3323 & 4.519$\cdot 10^{-6}$ & 4.8660  & 1$\cdot 10^{-4}$      \\ \hline
0.5$<z\leqslant$1  & 6.1722  & 3.287$\cdot 10^{-6}$ & 3.9229  & 1.036$\cdot 10^{-4}$& 6.8234 & 8.331$\cdot 10^{-6}$ & 4.7246  & 1.5941$\cdot 10^{-4}$ \\ \hline
1$<z\leqslant$1.8  & 4.7030  & 9.101$\cdot 10^{-6}$ & 2.9237  & 2.071$\cdot 10^{-4}$& 4.8817 & 3.2$\cdot 10^{-5}$   & 3.4034  & 3.9821$\cdot 10^{-4}$ \\ \hline
1.8$<z\leqslant$7  & -0.6499 & 2.253$\cdot 10^{-3}$ & -1.0032 & 0.01181             &-0.6698 & 9.719$\cdot 10^{-3}$ & -0.9736 & 0.3608                \\ \hline
7$<z\leqslant$8.8  & -11.888 & 3.172$\cdot 10^{7}$  & -12.176 & 1.455$\cdot 10^{8}$ &-11.931 & 1.437$\cdot 10^{8}$  &-12.1147 & 4.1571$\cdot 10^{8}$  \\ \hline
8.8$<z\leqslant$10 & 0       & 5.225$\cdot 10^{-5}$ & 0       & 1.239$\cdot 10^{-4}$& 0      & 2.144$\cdot 10^{-4}$ & 0       & 4.0771$\cdot 10^{-4}$ \\ \hline
\hline
\multicolumn{9}{c}{ \textbf{low metallicity SFH - upper edge} }\\
 variation:  & \multicolumn{2}{c}{SB: Boco+21, a$_{MZR}\sim$0.3} & \multicolumn{2}{c}{SB: Boco+21, a$_{MZR}\sim$0.6} & \multicolumn{2}{c}{SB: B18/C17, a$_{MZR}\sim$0.3} & \multicolumn{2}{c}{SB: B18/C17, a$_{MZR}\sim$0.6} \\ 
$z$ range & $\kappa$ & A & $\kappa$ & A & $\kappa$ & A  & $\kappa$ & A \\ \hline
%UPPER EDGE
0$<z\leqslant$0.5  & 4.4163  & 9.410$\cdot 10^{-5}$ & 2.9836 & 6$\cdot 10^{-4}$     & 5.9088 & 1.149$\cdot 10^{-4}$ & 4.0787  & 6$\cdot 10^{-4}$      \\ \hline
0.5$<z\leqslant$1  & 4.2247  & 1.017$\cdot 10^{-4}$ & 3.0826 & 5.389$\cdot 10^{-4}$ & 5.3151 & 1.462$\cdot 10^{-4}$ & 4.1063  & 6.262$\cdot 10^{-4}$  \\ \hline
1$<z\leqslant$1.5  & 3.9141  & 1.261$\cdot 10^{-4}$ & 2.7905 & 6.599$\cdot 10^{-4}$ & 4.8008 & 2.088$\cdot 10^{-4}$ & 3.7273  & 8.143$\cdot 10^{-4}$  \\ \hline
1.5$<z\leqslant$7  & 2.4934  & 4.636$\cdot 10^{-4}$ & 2.0535 & 1.296$\cdot 10^{-3}$ & 2.0328 & 2.638$\cdot 10^{-3}$ & 1.8127  & 4.707$\cdot 10^{-3}$  \\ \hline
7$<z\leqslant$8.8  & -8.5779 & 4.619$\cdot 10^{6}$  &-8.7763 & 7.816$\cdot 10^{6}$  & -8.896 & 1.954$\cdot 10^{7}$  & -8.9519 & 2.478$\cdot 10^{7}$   \\ \hline
8.8$<z\leqslant$10 & 0       & 0.01452              & 0      & 0.01562              & 0      & 0.02972              & 0       & 0.03317               \\ \hline
\hline
\end{tabular}
\label{tab: low Z SFH}
\end{table*}

\begin{table*}
\centering
\small
\caption{
Coefficients of the power law fits to the low metallicity cut of the cosmic SFH (lower and upper edges of the ranges as in  Fig. \ref{fig: low Z SFRD}) obtained for different variations of our observation-based model under the assumption of a non-evolving low mass end of the GSMF.
Within each redshift range indicated in the first column total SFRD can be approximated with the dependence: SFRD=A$\cdot(1+z)^{\kappa} [\Msun/yr]$.
}
\begin{tabular}{c c c c c c c c c}
\hline
%Boco, a=0.3  Boco a=06, BiC a=0.3 BiC a=0.6 -- kappa A
%LOWER EDGE
\multicolumn{9}{c}{ \textbf{low metallicity SFH - lower edge} }\\
 variation:  & \multicolumn{2}{c}{SB: Boco+21, a$_{MZR}\sim$0.3} & \multicolumn{2}{c}{SB: Boco+21, a$_{MZR}\sim$0.6} & \multicolumn{2}{c}{SB: B18/C17, a$_{MZR}\sim$0.3} & \multicolumn{2}{c}{SB: B18/C17, a$_{MZR}\sim$0.6} \\ 
$z$ range & $\kappa$ & A & $\kappa$ & A & $\kappa$ & A  & $\kappa$ & A \\ \hline

0$<z\leqslant$0.5  & 6.0172  & 4.393$\cdot 10^{-6}$ & 3.3024 & 1.667$\cdot 10^{-4}$ & 7.9464 & 6.561$\cdot 10^{-6}$ &  4.3940 & 2.085$\cdot 10^{-4}$\\ \hline
0.5$<z\leqslant$1  & 5.5287  & 5.355$\cdot 10^{-6}$ & 3.3306 & 1.635$\cdot 10^{-4}$ & 6.1805 & 1.343$\cdot 10^{-5}$ &4.2022 & 2.381$\cdot 10^{-4}$  \\ \hline
1$<z\leqslant$1.7  & 4.7328  & 9.297$\cdot 10^{-6}$ & 2.8406 & 2.297$\cdot 10^{-4}$ & 5.0089 & 3.024$\cdot 10^{-5}$ &3.4101 & 4.123$\cdot 10^{-4}$ \\ \hline
1.7$<z\leqslant$3.5& 0.4510  & 6.537$\cdot 10^{-4}$ & -0.0495& 0.004053             & 0.4466 & 0.002809             & 0.0103 & 0.01207 \\ \hline
3.5$<z\leqslant$7  & -1.8116 & 0.01965              & -2.2411& 0.1095               & -1.8350& 0.08689              &-2.2043& 0.3376  \\ \hline
7$<z\leqslant$8.8  & -11.8876& 2.471$\cdot 10^{7}$  & -12.177& 1.028$\cdot 10^{8}$  &-11.9308& 1.139$\cdot 10^{8}$  & -12.115& 3.009$\cdot 10^{8}$ \\ \hline
8.8$<z\leqslant$10 & 0       & 4.069$\cdot 10^{-5}$ & 0      & 8.754$\cdot 10^{-5}$ & 0      & 1.699$\cdot 10^{-4}$ & 0      & 2.951$\cdot 10^{-4}$ \\ \hline
\hline
\multicolumn{9}{c}{ \textbf{low metallicity SFH - upper edge} }\\
 variation:  & \multicolumn{2}{c}{SB: Boco+21, a$_{MZR}\sim$0.3} & \multicolumn{2}{c}{SB: Boco+21, a$_{MZR}\sim$0.6} & \multicolumn{2}{c}{SB: B18/C17, a$_{MZR}\sim$0.3} & \multicolumn{2}{c}{SB: B18/C17, a$_{MZR}\sim$0.6} \\ 
$z$ range & $\kappa$ & A & $\kappa$ & A & $\kappa$ & A  & $\kappa$ & A \\ \hline
%UPPER EDGE
0$<z\leqslant$0.5  & 4.4163  & 9.410$\cdot 10^{-5}$ & 2.9836 & 5.772$\cdot 10^{-4}$ & 5.9088 & 1.149$\cdot 10^{-4}$& 4.0787 & 6.383$\cdot 10^{-4}$  \\ \hline
0.5$<z\leqslant$1  & 4.2247  & 1.017$\cdot 10^{-4}$ & 3.0826 & 5.389$\cdot 10^{-4}$ & 5.3151 & 1.462$\cdot 10^{-4}$& 4.1063 & 6.262$\cdot 10^{-4}$  \\ \hline
1$<z\leqslant$1.7  & 3.6167  &  1.55$\cdot 10^{-4}$ & 2.6761 & 7.143$\cdot 10^{-4}$ & 4.4986 & 2.575$\cdot 10^{-4}$& 3.5242 & 9.374$\cdot 10^{-4}$  \\ \hline
1.7$<z\leqslant$3.5& 0.1790  & 0.004712             & -0.1177& 0.01146              & 0.4267 & 0.0147              & 0.0785 & 0.02873               \\ \hline
3.5$<z\leqslant$7  & -2.0303 &  0.1307              & -2.3001& 0.3052               &-1.9322 & 0.5106              &-2.2568 & 0.9633  \\ \hline
7$<z\leqslant$8.8  & -12.0388& 1.429$\cdot 10^{8}$  &-12.1984& 2.653$\cdot 10^{8}$  &-12.0147&6.509$\cdot 10^{8}$  &-12.1707& 8.648$\cdot 10^{8}$ \\ \hline
8.8$<z\leqslant$10 & 0       & 1.666$\cdot 10^{-4}$ & 0      & 2.149$\cdot 10^{-4}$ & 0      & 8.02$\cdot 10^{-4}$ &0      & 7.464$\cdot 10^{-4}$ \\ \hline
\hline
\end{tabular}
\label{tab: low Z SFH GSMF fixed}
\end{table*}

\section{Z$_{O/H}$(SFR, M$_{*}$) implementation with SFR-dependent $\nabla_{\rm FMR}$ (and $\gamma$)}\label{app: SFR dependent nabla}

\begin{figure*}
\vspace*{-0.5cm}
\includegraphics[width=1.8\columnwidth]{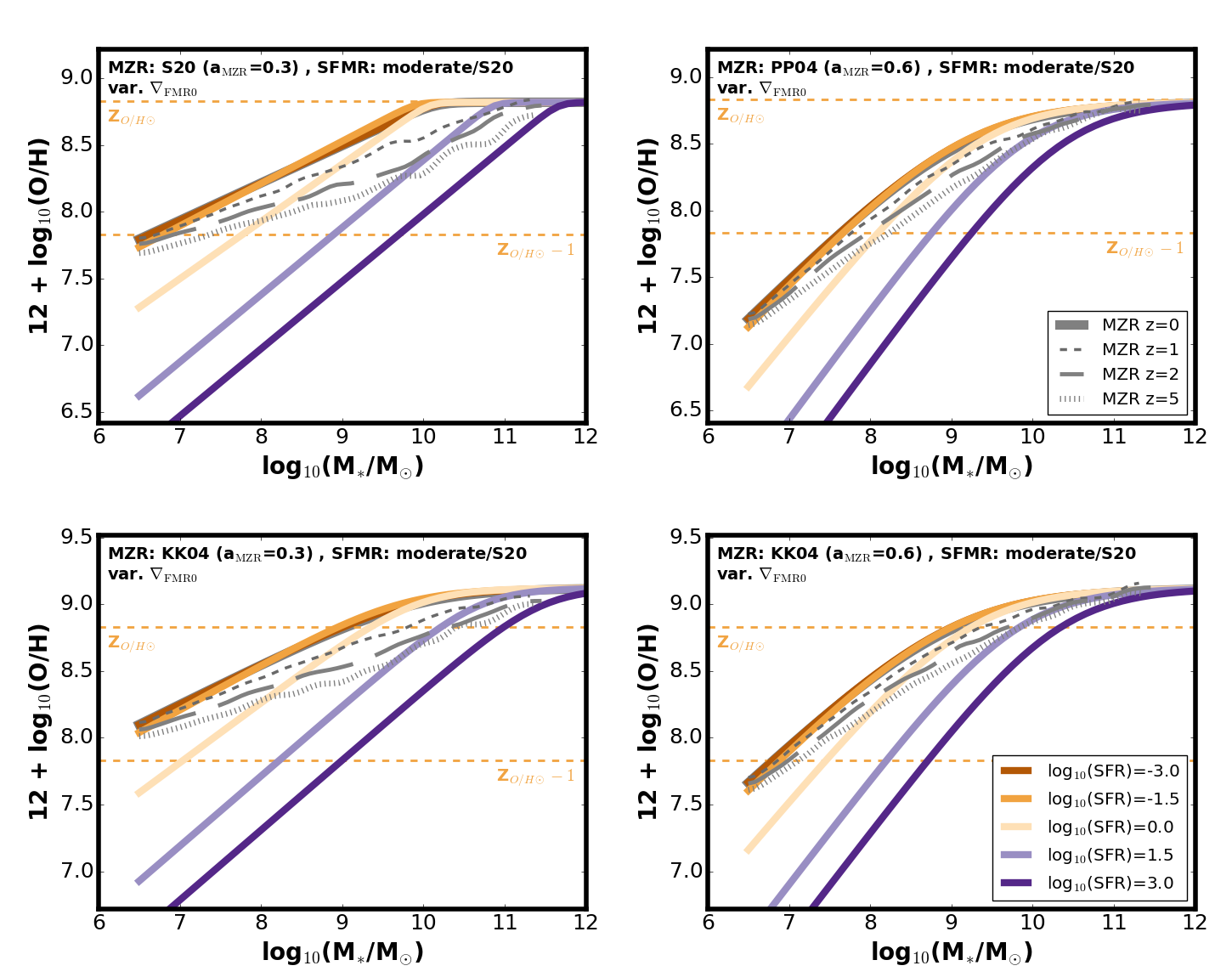}
\vspace*{-0.3cm}
\caption{
Same as Fig. \ref{fig: FMR MZR choice}, but assuming a SFR-dependent $\nabla_{\rm FMR}$ (leading to SFR-dependent $\gamma$).
Top (bottom) panels assume S20 (PP04) $z\sim0$ MZR.
All panels assume SFMR with a moderate flattening at high masses (a$_{SFMR}$=0.72 at the high mass end), log$_{10}$ (SFR$_{\nabla0}$)=1 and $\nabla_{\rm FMR0}$=0.27.
}
\label{fig: FMR varN}
\end{figure*}

\begin{figure*}
\vspace*{-0.5cm}
\includegraphics[width=1.7\columnwidth]{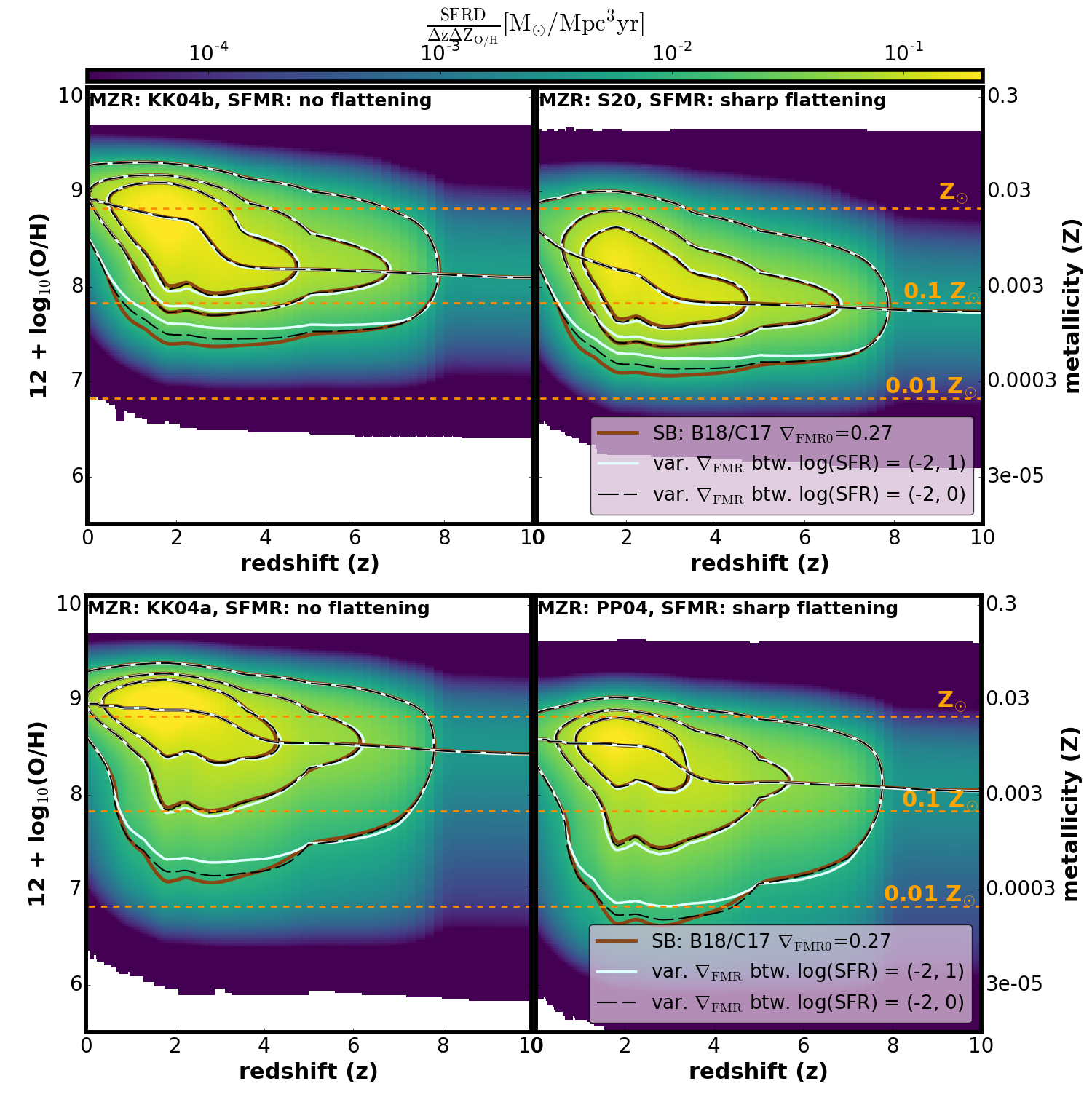}
\vspace*{-0.3cm}
\caption{
Distribution of the cosmic SFRD at different metallicities and redshift for different Z$_{O/H}$(SFR, M$_{*}$) implementations, shown for B18/C17 starbursts implementation (effect is analogous for the starburst implementation following Boco et al. (2021)).
Brown contours and background colours correspond to Z$_{O/H}$(SFR, M$_{*}$) with a fixed $\nabla_{\rm FMR0}=0.27$.
Solid white (dashed black) contours correspond to Z$_{O/H}$(SFR, M$_{*}$) modelled with $\nabla_{\rm FMR}$ varying between log$_{10}$ (SFR)=-2 and log$_{10}$ (SFR)=1 (log$_{10}$ (SFR)=0) as described in  Sec. \ref{app: SFR dependent nabla}.\\
Different panels correspond to different z$\sim$0 MZR and SFMR variations (same as in Fig. \ref{fig: SFRD(Z,z) - MZR and FMR slope}). Contours indicate constant SFRD and are plotted for 0.01,0.05 and 0.1 [ $\Msun$/Mpc$^{3}$ yr] (with the highest value corresponding to the innermost contour).
% Horizontal violet long-dashed line indicates the metallicity of log$_{10}$(M$_{*}$/$\Msun$)=8 galaxies that follow the $z\sim0$ SFMR.
The alternative Z$_{O/H}$(SFR, M$_{*}$) implementation has a relatively small effect on the overall distribution, but leads to a less extended low metallicity tail.
}
\label{fig: fSFRD varN}
\end{figure*}

As discussed in Sec. \ref{Sec: FMR - model}, Z$_{O/H}$(SFR, M$_{*}$) at fixed SFR and M$_{*}<M_{\rm 0;SFR}$ appear as lines roughly parallel to each other (see examples in, e.g. \citealt{Mannucci10}, \citealt{Telford16}, \citealt{Cresci19}) and we assume that the Z$_{O/H}$(SFR, M$_{*}$) slope $\gamma$ is independent of SFR/M$_{*}$.
However, in their recent study \citet{Curti20} fit the slopes of  Z$_{O/H}$(SFR, M$_{*}$) in individual log$_{10}$(SFR) bins and indicate a possible dependence of $\gamma$ on SFR. They find that $\gamma$ decreases with decreasing SFR, reaching the value close to that of $\sim$0 MZR slope at low SFR (see figure 9 therein).
When Z$_{O/H}$(SFR, M$_{*}$) is extrapolated down to low (specific) SFR values, such dependence leads to considerably higher metallicity of galaxies with the lowest masses (M$_{*}\lesssim 10^{8}\Msun$), and therefore can affect the low metallicity tail of the f$_{\rm SFR}$(Z,z) distribution obtained with our model.
\\
We explore additional variation of our model to illustrate the effect of such a dependence on our results.
Within our description we can model a similar dependence of $\gamma$ on SFR to that found by \citet{Curti20} by implementing a SFR-dependent $\nabla_{\rm FMR}$  (and therefore of $\gamma$, see e.q. \ref{eq: gamma}) and requiring that the strength of the correlation weakens towards low SFRs, eventually disappearing at a certain low SFR (where $\gamma$ approaches the slope of the z$\sim$0 MZR). 
As discussed in Sec. \ref{sec: FMR - nabla}, there is some observational evidence for the weakening of the SFR-metallicity correlation below a certain specific SFR threshold ($\rm \lesssim 10^{-10}/yr$, but it should be noted that this behaviour is difficult to constrain at low M$_{*}\lesssim10^{8.5}\Msun$).
For simplicity, we assume that there is a linear dependence of $\nabla_{\rm FMR}$  on log$_{10}$(SFR), such that at log$_{10}$(SFR)$\geqslant$log$_{10}$ (SFR$_{\nabla0}$) $\nabla_{\rm FMR}$  = $\nabla_{\rm FMR0}$  and at log$_{10}$(SFR)$\leqslant$ log$_{10}$(SFR$_{\rm no \ corr.}$) $\nabla_{\rm FMR}$=0.
We set log$_{10}$(SFR$_{\rm no \ corr.}$)=-2, which roughly corresponds to the SFR of M$_{*}\sim10^{8}\Msun$ galaxies according to our z$\sim$0 SFMR.
We note that below that mass and SFR the observational FMR is not constrained \footnote{But note that galaxies with $M_{*}\lesssim 10^{8}\Msun$ and high specific SFR (i.e. SFMR outliers with high SFR) were found to follow the FMR \citep[e.g.][]{Mannucci11,Hunt16_obs}},
and by assuming that for those galaxies the correlation disappears we obtain a conservative estimate of the low metallicity contribution of low mass galaxies.
Results of our calculations are also sensitive to the value of
the SFR threshold above which $\gamma$ is fixed: the higher the value of log$_{10}$ (SFR$_{\nabla0}$), the lower the amount of the SFRD happening at low metallicity.
The choice of this value is unclear and for the illustrative purposes we show our results using two example thresholds:  log$_{10}$ (SFR$_{\nabla0}$)=0 and log$_{10}$ (SFR$_{\nabla0}$)=1.
The resulting  Z$_{O/H}$(SFR, M$_{*}$) is illustrated in Fig. \ref{fig: FMR varN} using the latter threshold and for the different z$\sim$0 MZR choices considered in our study.
\\
The overall f$_{\rm SFR}$(Z,z) distribution is only mildly affected by the explored variations with SFR-dependent $\nabla_{\rm FMR}$ (see Fig. \ref{fig: fSFRD varN}, the effect is analogous for the variations with the starburst implementation following Boco et al.).
The redshift evolution of the peak metallicity is indistinguishable from that estimated with a fixed $\nabla_{\rm FMR}=\nabla_{\rm FMR0}$ case and the only visible difference is in the extent of the low metallicity tail of the distribution (the fixed $\nabla_{\rm FMR}$ case leads to more extended low metallicity tail).
\\
With the low metallicity threshold of 10\% solar metallicity (oxygen abundance) used in our study, the difference in the amount of the total SFRD happening at low metallicity estimated in the model variations with fixed and SFR-dependent $\nabla_{\rm FMR}$ is amplified in the cases with flat z$\sim$0 MZR (a$_{\rm MZR}\sim$0.3) and high MZR normalisation (compare lower edge of the dark brown range and the lower thick dark brown lines in Fig. \ref{fig: lowZ cut varN} - note that the strength of this effect is very sensitive to the exact choice of the low metallicity threshold). In those cases the result is particularly sensitive to the assumed low mass/SFR FMR extrapolation, as at low redshifts only galaxies that are MZR outliers have metallicities falling below $Z_{O/H} < Z_{O/H\odot}$-1 (see top left panel in Fig. \ref{fig: FMR varN}). If there is no correlation between the SFR and metallicity at low SFRs, this leads to a negligible amount of SFRD happening below 10\% solar metallicity at low to intermediate redshifts.

% We show the resulting $\gamma$ as a function of SFR in Fig. \ref{fig: gamma varN}.

% \begin{figure}
% \vspace*{-0.5cm}
% \includegraphics[width=1\columnwidth]{./figures/gamma_varN.png}
% \vspace*{-0.3cm}
% \caption{
% FMR slope $\gamma$ for different z$\sim$0 MZR variations as a function of log$_{10}$(SFR) (bottom axis) or $\nabla_{\rm FMR}$ (top axis), assuming log$_{10}$ (SFR$_{\nabla0}$)=1 and $\nabla_{\rm FMR0}$=0.27.
% }
% \label{fig: gamma varN}
% \end{figure}

\begin{figure*}
\vspace*{-0.5cm}
\includegraphics[width=1.8\columnwidth]{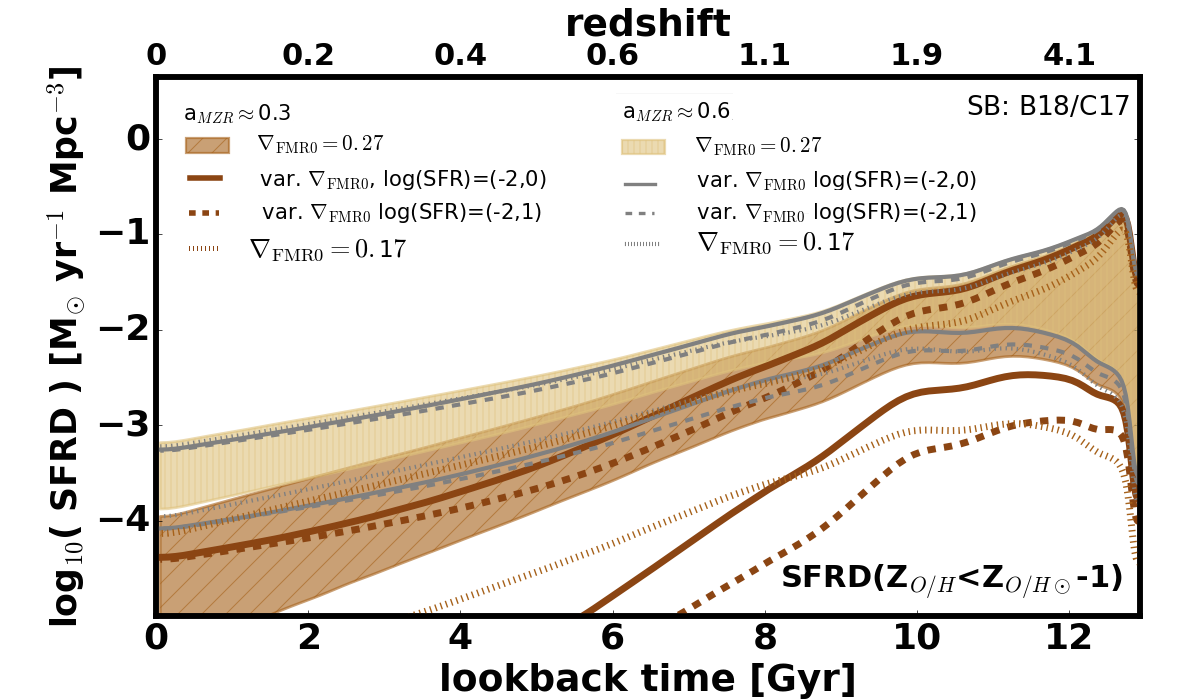}
\vspace*{-0.15cm}
\caption{
SFRD happening below 10\% solar metallicity ($Z_{O/H}<Z_{O/H \odot}$-1) as a function of lookback time/redshift estimated with different Z$_{O/H}$(SFR, M$_{*}$) implementations, shown for B18/C17 starbursts implementation (effect is analogous for the starburst implementation following Boco et al. (2021)).
The light and dark brown coloured ranges are the same as in Fig. \ref{fig: low Z SFRD} and correspond to a fixed $\nabla_{\rm FMR0}=0.27$ and a$_{\rm MZR}\approx$0.3 and a$_{\rm MZR}\approx$0.6 respectively.
Dotted gray/brown lines indicate upper and lower edges of the corresponding ranges obtained with $\nabla_{\rm FMR0}=0.17$.
Solid (log$_{10}$ (SFR$_{\nabla0}$)=0) and dashed (log$_{10}$ (SFR$_{\nabla0}$)=1) lines indicate the upper and lower edges of those ranges obtained with a SFR-dependent $\nabla_{\rm FMR}$ (leading to SFR-dependent $\gamma$, see Sec. \ref{app: SFR dependent nabla}).
}
\label{fig: lowZ cut varN}
\end{figure*}

\begin{figure*}
\vspace*{-0.3cm}
\includegraphics[width=1.8\columnwidth]{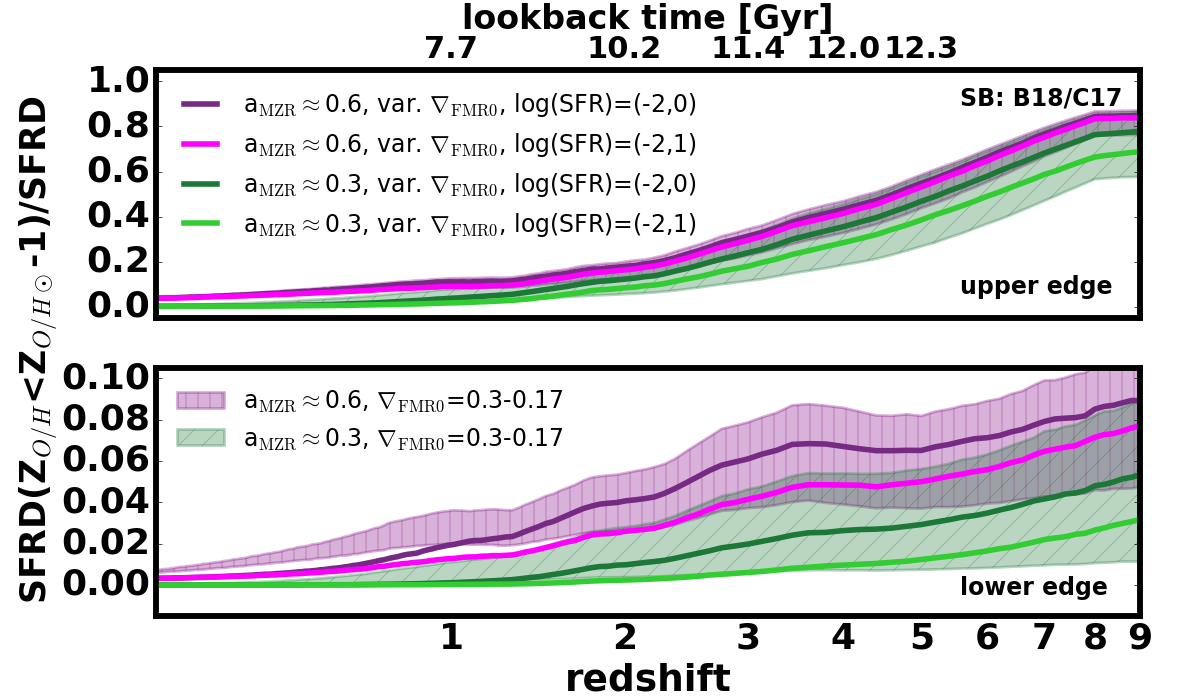}
\vspace*{-0.15cm}
\caption{Fraction of the total
SFRD happening below 10\% solar metallicity ($Z_{O/H}<Z_{O/H \odot}$-1) as a function of lookback time/redshift estimated with different Z$_{O/H}$(SFR, M$_{*}$) implementations, shown for B18/C17 starbursts implementation (effect is analogous for the starburst implementation following Boco et al. (2021)).
Top panel shows the estimates obtained for the low metallicity variations (i.e. corresponding to upper edges in Fig. \ref{fig: lowZ cut varN}),
bottom panel shows the estimates obtained for the high metallicity variations.
The green (purple) ranges span between the estimates obtained with Z$_{O/H}$(SFR, M$_{*}$) with a fixed $\nabla_{\rm FMR0}=0.3$ and 0.17 and a$_{\rm MZR}\approx$0.3 (a$_{\rm MZR}\approx$0.6).
Different lines indicate the corresponding estimate obtained with Z$_{O/H}$(SFR, M$_{*}$) with a SFR-dependent $\nabla_{\rm FMR}$, varying between log$_{10}$ (SFR)=-2 and log$_{10}$(SFR)=0 or 1 (see legend and  Sec. \ref{app: SFR dependent nabla}).
}
\label{fig: lowZ cut varN}
\end{figure*}
% With this SFR threshold we also reproduce the disappearance of the FMR at log$_{10}$(sSFR)$\sim$-10 suggested by observations.

% \section{Some extra material}

% If you want to present additional material which would interrupt the flow of the main paper,
% it can be placed in an Appendix which appears after the list of references.

%%%%%%%%%%%%%%%%%%%%%%%%%%%%%%%%%%%%%%%%%%%%%%%%%%

% Don't change these lines
\bsp	% typesetting comment
\label{lastpage}
\end{document}